\documentclass[
 reprint,
 superscriptaddress,
preprintnumbers,
 amsmath,amssymb,
 aps,
 prd
]{revtex4-2}

\usepackage{tabularx}     
\usepackage{multirow}     
\usepackage{ragged2e}     
\usepackage{booktabs}
\usepackage{multirow}
\usepackage{array}
\usepackage{amsmath,amssymb}
\usepackage{lineno}
\usepackage{bm}
\usepackage{upgreek}
\usepackage{comment}
\usepackage[dvipsnames]{xcolor}
\usepackage{gensymb}
\usepackage{placeins}
\usepackage{svg}
\usepackage[caption=false]{subfig}
\usepackage{graphicx}
\usepackage{dcolumn}
\usepackage{tikz}
\usepackage{hyperref}

\hypersetup{
  colorlinks=true,
  linkcolor=purple,
  filecolor=purple,
  urlcolor=purple,
  citecolor=purple
}

\newcommand{\code}[1]{\texttt{#1}}
\newcommand{\ppfx}{\code{PPFX}}
\newcommand{\pizero}{$\pi^0$}
\newcommand{\nue}{$\nu_{e}$}

\newcommand{\nuebar}{$\bar{\nu}_{e}$}

\newcommand{\numu}{$\nu_{\mu}$}

\newcommand{\numubar}{$\bar{\nu}_{\mu}$}

\newcommand{\npsel}{CC 1e0$\pi$Np}

\begin{document}

\widetext
\hspace{4.84in} \mbox{-    }

\title{Measurements of differential charged-current cross sections on argon for electron neutrinos with final-state protons in MicroBooNE}

\newcommand{\ANL}{Argonne National Laboratory (ANL), Lemont, IL, 60439, USA}
\newcommand{\Bern}{Universit{\"a}t Bern, Bern CH-3012, Switzerland}
\newcommand{\BNL}{Brookhaven National Laboratory (BNL), Upton, NY, 11973, USA}
\newcommand{\UCSB}{University of California, Santa Barbara, CA, 93106, USA}
\newcommand{\Cambridge}{University of Cambridge, Cambridge CB3 0HE, United Kingdom}
\newcommand{\CIEMAT}{Centro de Investigaciones Energ\'{e}ticas, Medioambientales y Tecnol\'{o}gicas (CIEMAT), Madrid E-28040, Spain}
\newcommand{\Chicago}{University of Chicago, Chicago, IL, 60637, USA}
\newcommand{\Cincinnati}{University of Cincinnati, Cincinnati, OH, 45221, USA}
\newcommand{\CSU}{Colorado State University, Fort Collins, CO, 80523, USA}
\newcommand{\Columbia}{Columbia University, New York, NY, 10027, USA}
\newcommand{\Edinburgh}{University of Edinburgh, Edinburgh EH9 3FD, United Kingdom}
\newcommand{\FNAL}{Fermi National Accelerator Laboratory (FNAL), Batavia, IL 60510, USA}
\newcommand{\Granada}{Universidad de Granada, Granada E-18071, Spain}
\newcommand{\IIT}{Illinois Institute of Technology (IIT), Chicago, IL 60616, USA}
\newcommand{\ICL}{Imperial College London, London SW7 2AZ, United Kingdom}
\newcommand{\Indiana}{Indiana University, Bloomington, IN 47405, USA}
\newcommand{\Kansas}{The University of Kansas, Lawrence, KS, 66045, USA}
\newcommand{\KSU}{Kansas State University (KSU), Manhattan, KS, 66506, USA}
\newcommand{\Lancaster}{Lancaster University, Lancaster LA1 4YW, United Kingdom}
\newcommand{\LANL}{Los Alamos National Laboratory (LANL), Los Alamos, NM, 87545, USA}
\newcommand{\Louisiana}{Louisiana State University, Baton Rouge, LA, 70803, USA}
\newcommand{\Manchester}{The University of Manchester, Manchester M13 9PL, United Kingdom}
\newcommand{\MIT}{Massachusetts Institute of Technology (MIT), Cambridge, MA, 02139, USA}
\newcommand{\Michigan}{University of Michigan, Ann Arbor, MI, 48109, USA}
\newcommand{\MSU}{Michigan State University, East Lansing, MI 48824, USA}
\newcommand{\Minnesota}{University of Minnesota, Minneapolis, MN, 55455, USA}
\newcommand{\Nankai}{Nankai University, Nankai District, Tianjin 300071, China}
\newcommand{\NMSU}{New Mexico State University (NMSU), Las Cruces, NM, 88003, USA}
\newcommand{\Oxford}{University of Oxford, Oxford OX1 3RH, United Kingdom}
\newcommand{\Pitt}{University of Pittsburgh, Pittsburgh, PA, 15260, USA}
\newcommand{\QMUL}{Queen Mary University of London, London E1 4NS, United Kingdom}
\newcommand{\Rutgers}{Rutgers University, Piscataway, NJ, 08854, USA}
\newcommand{\SLAC}{SLAC National Accelerator Laboratory, Menlo Park, CA, 94025, USA}
\newcommand{\SDSMT}{South Dakota School of Mines and Technology (SDSMT), Rapid City, SD, 57701, USA}
\newcommand{\Maine}{University of Southern Maine, Portland, ME, 04104, USA}
\newcommand{\TelAviv}{Tel Aviv University, Tel Aviv, Israel, 69978}
\newcommand{\UTA}{University of Texas, Arlington, TX, 76019, USA}
\newcommand{\Tufts}{Tufts University, Medford, MA, 02155, USA}
\newcommand{\VTech}{Center for Neutrino Physics, Virginia Tech, Blacksburg, VA, 24061, USA}
\newcommand{\Warwick}{University of Warwick, Coventry CV4 7AL, United Kingdom}

\affiliation{\ANL}
\affiliation{\Bern}
\affiliation{\BNL}
\affiliation{\UCSB}
\affiliation{\Cambridge}
\affiliation{\CIEMAT}
\affiliation{\Chicago}
\affiliation{\Cincinnati}
\affiliation{\CSU}
\affiliation{\Columbia}
\affiliation{\Edinburgh}
\affiliation{\FNAL}
\affiliation{\Granada}
\affiliation{\IIT}
\affiliation{\ICL}
\affiliation{\Indiana}
\affiliation{\Kansas}
\affiliation{\KSU}
\affiliation{\Lancaster}
\affiliation{\LANL}
\affiliation{\Louisiana}
\affiliation{\Manchester}
\affiliation{\MIT}
\affiliation{\Michigan}
\affiliation{\MSU}
\affiliation{\Minnesota}
\affiliation{\Nankai}
\affiliation{\NMSU}
\affiliation{\Oxford}
\affiliation{\Pitt}
\affiliation{\QMUL}
\affiliation{\Rutgers}
\affiliation{\SLAC}
\affiliation{\SDSMT}
\affiliation{\Maine}
\affiliation{\TelAviv}
\affiliation{\UTA}
\affiliation{\Tufts}
\affiliation{\VTech}
\affiliation{\Warwick}

\author{P.~Abratenko} \affiliation{\Tufts}
\author{D.~Andrade~Aldana} \affiliation{\IIT}
\author{L.~Arellano} \affiliation{\Manchester}
\author{J.~Asaadi} \affiliation{\UTA}
\author{A.~Ashkenazi}\affiliation{\TelAviv}
\author{S.~Balasubramanian}\affiliation{\FNAL}
\author{B.~Baller} \affiliation{\FNAL}
\author{A.~Barnard} \affiliation{\Oxford}
\author{G.~Barr} \affiliation{\Oxford}
\author{D.~Barrow} \affiliation{\Oxford}
% ...

\collaboration{The MicroBooNE Collaboration}
\thanks{microboone\_info@fnal.gov}\noaffiliation      
\date{\today}

\begin{abstract}
This work presents single-differential electron-neutrino charged-current cross sections on argon measured using the MicroBooNE detector at the Fermi National Accelerator Laboratory. The analysis uses data recorded when the Neutrinos at the Main Injector beam was operating in both neutrino and antineutrino modes, with exposures of $2 \times 10^{20}$ and $5 \times 10^{20}$ protons on target, respectively. A selection algorithm targeting electron-neutrino charged-current interactions with at least one proton, one electron, and no pions in the final topology is used to measure differential cross sections as a function of outgoing electron energy, total visible energy, and the cosine of the opening angle between the electron and the most energetic proton. The interaction rate as a function of proton multiplicity is also reported. The total cross section is measured as [4.1 $\pm$ 0.3 (stat.) $\pm$ 1.1 (syst.)]\,$\times \,10^{-39} \mathrm{cm}^{2}/\,\mathrm{nucleon}$. The unfolded cross-section measurements are compared to predictions from neutrino event generators commonly employed in the field. Good agreement is seen across all variables within uncertainties.
\end{abstract}

\maketitle

\section{Introduction}
\label{introduction}

Precise measurements of differential cross sections for electron-neutrino interactions on nuclei are vital to search for sterile neutrino flavors \cite{steriles1, steriles2} and for performing precision oscillation measurements with long-baseline accelerator experiments \cite{t2k-deltaCP, hyperk, dune-potential}. To enable the upcoming era of precision measurements, a robust understanding of neutrino-nucleus interactions is critical to ensure the accuracy of the models used to interpret data \cite{snowmass}. Upgraded neutrino-nucleus interaction models will improve measurements of the CP-violating phase at long baselines \cite{dune1} and allow for a vigorous exploration of new flavors at short baselines \cite{sbn-proposal, sbn}.  

Over the past several decades, a suite of measurements have been performed to constrain theoretical models of \numu{} charged-current (CC) interactions \cite{numu-xsecs}. Typically, these measurements are extrapolated to make predictions of \nue{} cross sections without direct experimental verification. Only in the past decade have direct \nue{} CC measurements begun to appear in the literature. Differential \nue{} and joint \nue{} + \nuebar{} cross sections on carbon have been reported using both inclusive and exclusive signal definitions \cite{t2k-2014, minerva-2016, t2k-2020, nova-2023, t2knuepi}. More recently, experimental \nue{} cross sections on argon have been measured \cite{argoneut-2020, numi-singlebin, bnb-exclusive, numi-inclusive, patrick}. These interactions are of particular importance to the forthcoming Deep Underground Neutrino Experiment \cite{dunetdr1}, which will make use of the liquid argon time projection chamber (LArTPC) technology \cite{rubbia-lartpc}. 

MicroBooNE \cite{uboone_det}, the first multi-ton detector in the LArTPC program at the Fermi National Accelerator Laboratory (Fermilab) exposed to a neutrino beam, is well-poised to explore electron-neutrino interactions on argon at energies relevant to accelerator-based neutrino experiments. The first \nue{} CC cross section on argon with no pions in the final state was reported as functions of the electron and leading proton energies and angles with respect to the neutrino beam using data collected at the MicroBooNE detector from the Booster Neutrino Beam (BNB) \cite{bnb-exclusive}. A significant flux of electron neutrinos has also been observed at MicroBooNE from a higher proton energy beamline called Neutrinos at the Main Injector (NuMI) \cite{numi_beam}, which operates in neutrino and antineutrino modes. The first MicroBooNE \nue{} + \nuebar{} inclusive cross sections on argon were extracted as a function of lepton energy and scattering angle using NuMI neutrino-mode data \cite{numi-inclusive}. More recently, the first \nue{} + \nuebar{} CC single charged pion differential cross sections on argon were extracted as a function of electron energy, electron and pion angles, and electron-pion opening angle using the full MicroBooNE NuMI dataset \cite{patrick}. 

These earlier results established benchmarks for \nue{} interactions on argon, but measurements of exclusive final states remain limited. The \nue{} CC channel with no pions and one or more protons in the final state $\left(\nu_{e}\,\mathrm{CC}0\pi Np\right)$ is a dominant contribution to electron neutrino event rates at $\mathcal{O}(1$\,GeV) and probes quasi-elastic-like processes and nuclear effects that alter proton kinematics and multiplicities. Requiring protons in the final state reduces \nuebar{} contamination, enabling a higher-purity \nue{} measurement. While MicroBooNE has measured this topology before, the present analysis uses higher-statistics NuMI data and reports flux-integrated differential cross sections in complementary observables. 

Using NuMI data, this work presents measurements of \nue{} CC cross sections for events characterized by one electron with greater than 20\,MeV kinetic energy, at least one proton with greater than 40\,MeV kinetic energy, zero charged pions with greater than 40\,MeV kinetic energy, and zero neutral pions in the final topology. Only \nue{} are part of the signal definition; \nuebar{} are not included. Flux-integrated differential cross sections are extracted as a function of outgoing electron energy, total visible energy, and the cosine of the opening angle between the electron and the leading proton. The total cross section is also reported, as well as interaction rates as a function of proton multiplicity. 

\section{Experimental Setup}
\label{uboone}

The MicroBooNE experiment \cite{uboone_det} is a surface LArTPC detector situated on the axis of the BNB beamline and off-axis to the NuMI beamline. The NuMI beamline operates with higher-energy protons incident on the target compared to the BNB beamline, resulting in an increased production of electron neutrinos. This larger flux of \nue{} from NuMI at MicroBooNE makes it ideal for performing measurements of \nue{} interactions. This section summarizes the derivation of the NuMI flux prediction at MicroBooNE, the detector design, and event simulation, readout, and reconstruction. 

 
\begin{figure*}
\centering

\includegraphics[width=\textwidth]{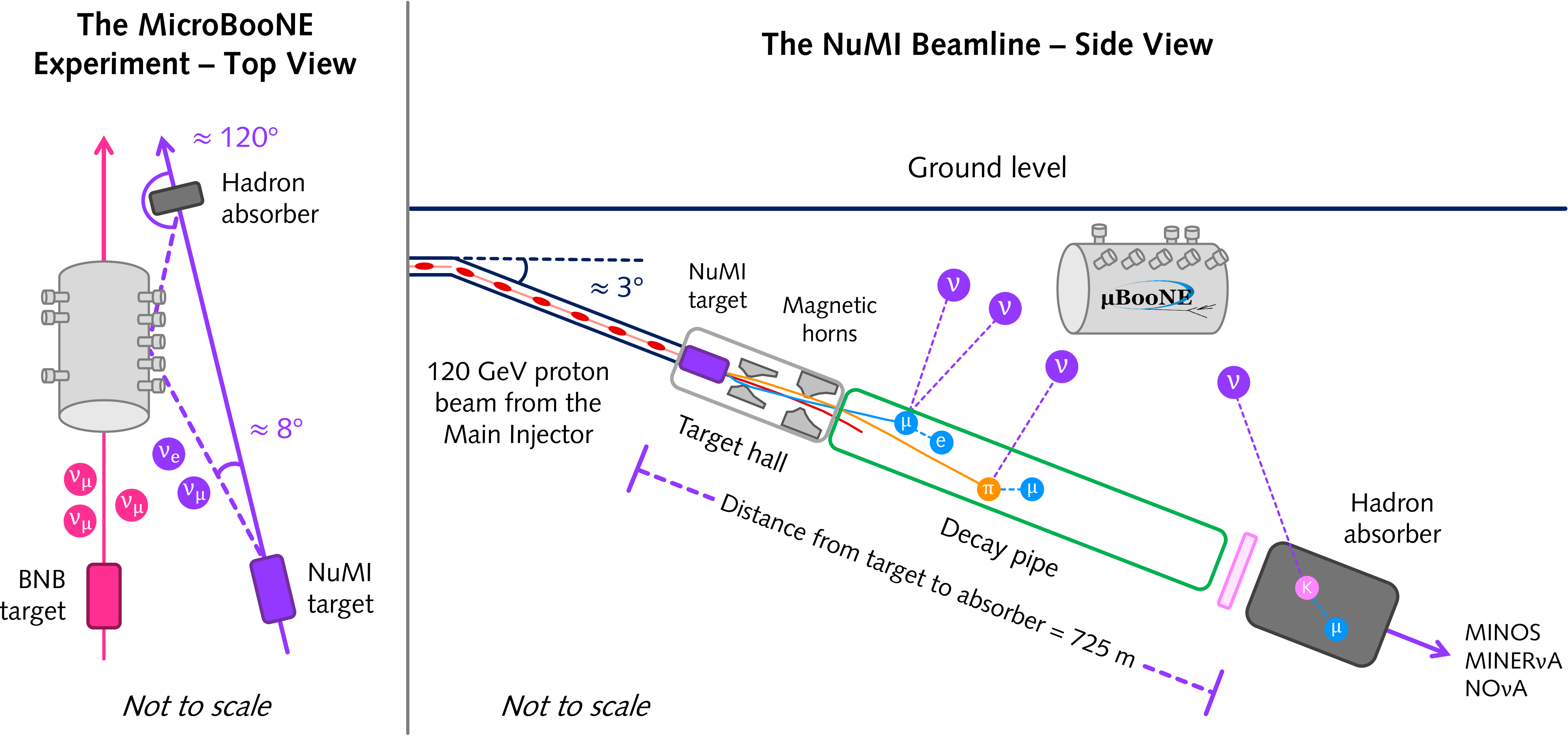}

\caption{\label{fig:numi_uboone}{The position of MicroBooNE relative to the NuMI beamline from top (left) and side (right) views.  The NuMI beamline is angled $3\degree$ downward. The distance from the NuMI target to MicroBooNE is approximately 675 meters. Neutrinos enter MicroBooNE at angles ${\approx}\,8-120\degree$ off the beamline. Most of the flux arriving at MicroBooNE originates from kaon decays in-flight on or near the target.}}
\end{figure*}


\subsection{The NuMI Flux Prediction}

The NuMI beam \cite{numi_beam} is generated by the accelerator complex at Fermilab. Protons with an energy of 120\,GeV are extracted from the Main Injector to collide with a fixed graphite target. The collisions generate cascades of hadrons that are focused in or deflected from the forward direction based on the hadron's charge and the polarity of the current in a pair of magnetic horns located immediately beyond the target. Focused hadrons travel down a 675-meter pipe where they decay and produce neutrinos. Charged particles surviving the decay pipe are stopped at the beam absorber. NuMI is designed to operate in one of two beam modes. These are referred to as forward horn current (FHC) and reverse horn current (RHC). In FHC (RHC) mode, the magnetic horn current is  $+200$\,kA ($-200$\,kA), and parent particles with positive (negative) charge are focused into the decay pipe to produce a primarily neutrino (antineutrino) beam. 

A schematic of NuMI's geometry, as it pertains to MicroBooNE, is shown in Fig. \ref{fig:numi_uboone}. The beam points into the Earth at an angle of approximately $3\degree$ from the horizontal. MicroBooNE is situated ${\approx}\,675$ meters from the NuMI target and is at an angle of about $8\degree$ off the beamline. The majority of the flux arriving at MicroBooNE originates from on or near the target from kaon decays in-flight. 


\begin{figure}[h]
\centering
\begin{tikzpicture} \draw (0, 0) node[inner sep=0] {
\includegraphics[width=.5\textwidth]{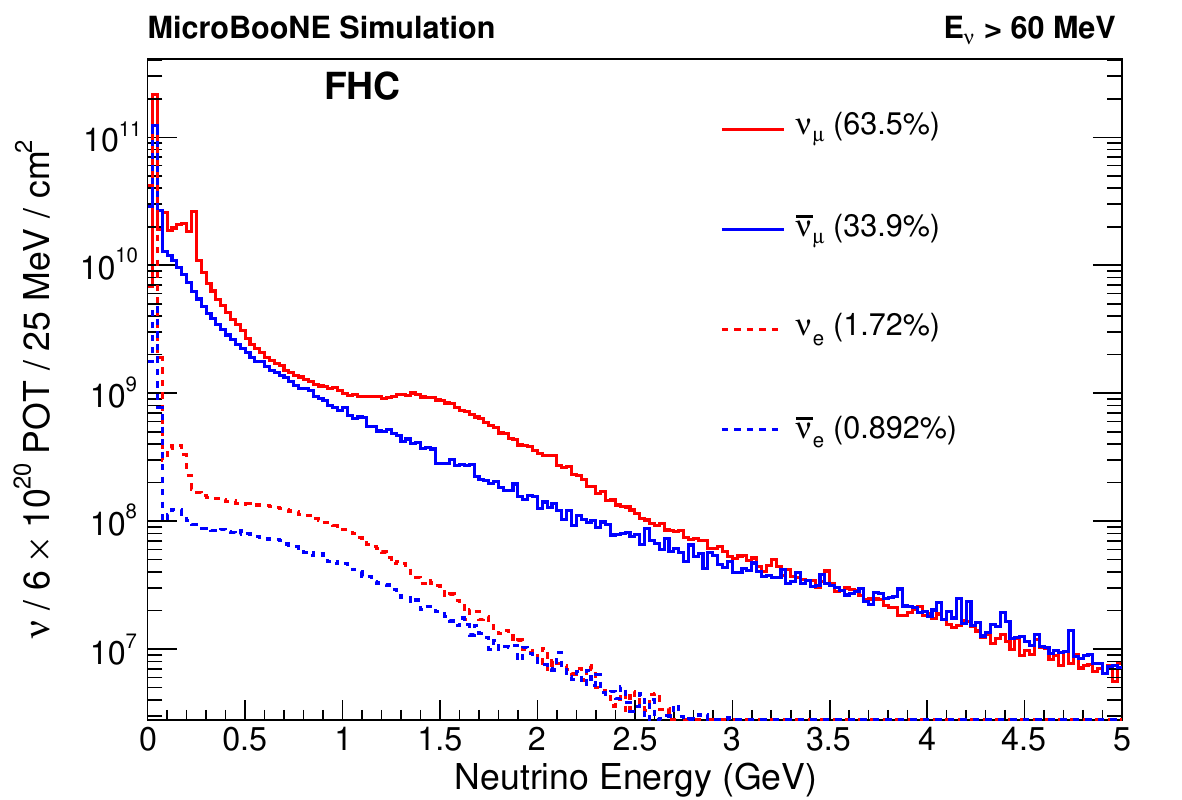}
		};
  \draw (-2.8, -2.0) node {\textbf{(a)}};
\end{tikzpicture} 

\begin{tikzpicture} \draw (0, 0) node[inner sep=0] {
\includegraphics[width=.5\textwidth]{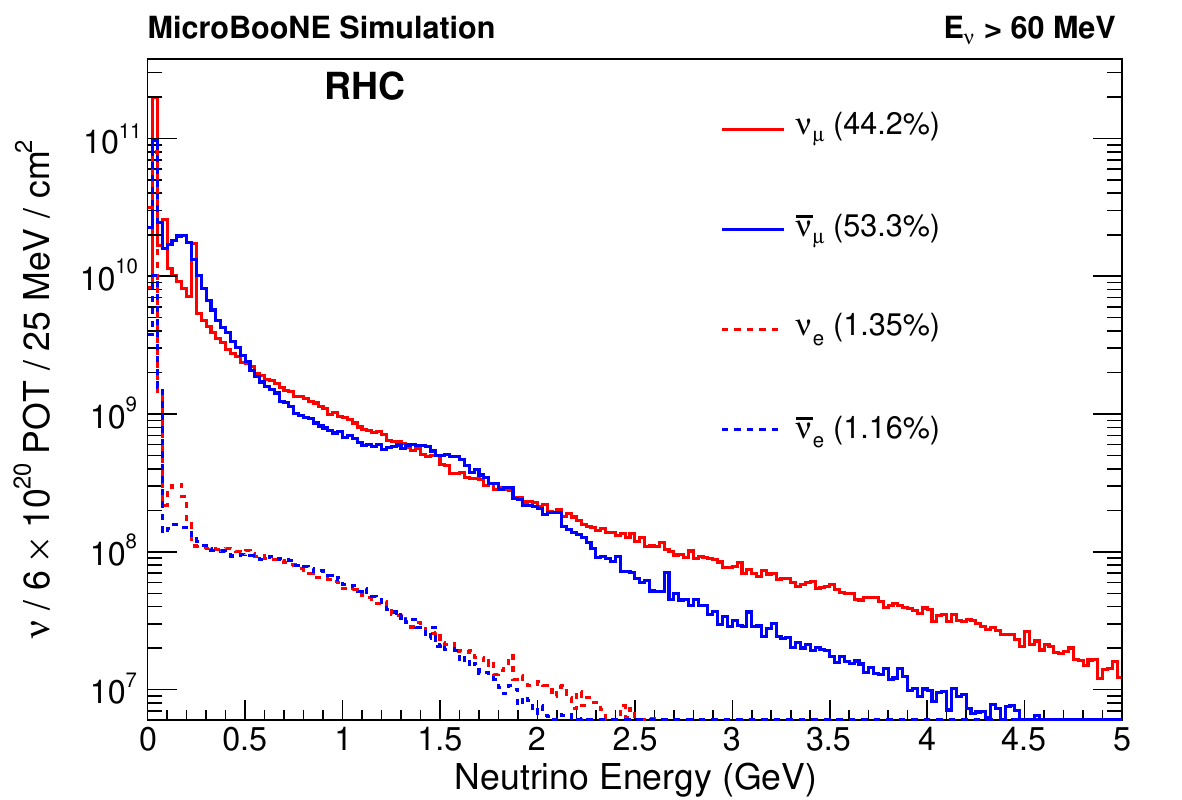}
		};
  \draw (-2.8, -2.0) node {\textbf{(b)}};
\end{tikzpicture}

\caption{\label{fig:flux} {The neutrino flux prediction at MicroBooNE for the NuMI beam operating in FHC mode (a) and RHC mode (b) \cite{flux_public_note}. Shown is the flux of \nue{} (dashed red), \nuebar{} (dashed blue), \numu{} (solid red), and \numubar{} (solid blue).}}
\end{figure}


The NuMI flux prediction is generated using \code{Geant4} \cite{geant4, geant4-dev, flux_public_note, steven-geant4} and a detailed geometric model of the beamline. The simulation, which accounts for particle interactions and their propagation, begins with proton collisions at the graphite target and ends with secondary charged particles that decay and produce neutrinos. Precise knowledge of pion, kaon, and nucleon production at the target, as well as the focusing properties of the beamline, should yield an accurate neutrino flux. However, the flux is more complicated to predict due to hadronic interactions with other beamline components---including the magnetic horns, decay pipe walls, shielding, and air in the target hall---which also create neutrinos \cite{numi_flux}. Data-driven corrections to the initial NuMI flux prediction are applied using \code{PPFX}, an experiment-agnostic software package originally developed by the MINER$\upnu$A Collaboration \cite{leo-thesis}. \ppfx{} constrains the \code{Geant4} simulation using external measurements of hadron production and absorption cross sections \cite{na49-pion, na49-nucleon, na49-kaon, barton, mipp-thin, bellettini, na61, denisov, carroll, abrams, allaby, cronin, allardyce, longo, bobchenko, fedorov, roberts}. This work also incorporates recent updates to the NuMI flux prediction, including a refined beamline geometry description, a more modern \code{Geant4} version (\code{v4.9.2} to \code{v4.10.4}), and \code{PPFX} changes accounting for the underlying model updates  \cite{flux_public_note}.

The \ppfx-constrained NuMI flux prediction at MicroBooNE as a function of true neutrino energy is shown in Fig. \ref{fig:flux}. Decay-at-rest muons contribute to a low energy ($<$ 60\,MeV) peak in the \nue{} and \nuebar{} flux predictions for both FHC and RHC \cite{numi_beam, numi_flux}. Higher energy \nue{} and \nuebar{} primarily result from charged and long-lived neutral kaon decays, 

\begin{equation}
    K^{\pm} \rightarrow \nu_{e} + e^{\pm} + \pi^{0},
\end{equation}
\begin{equation}
    K^{0}_{L} \rightarrow \nu_{e} + e^{\pm} + \pi^{\mp}.
\end{equation}
\noindent In both FHC and RHC modes, the NuMI flux contains comparable amounts of \nue{} and \nuebar{}. 

MicroBooNE collected data for a total of five run periods. Measurements described in this work make use of the NuMI Run~1 FHC exposure of $2.0 \times10^{20}$ protons-on-target (POT), which occurred between October 23, 2015, and May 2, 2016, as well as the Run~3 RHC exposure of $5.0 \times10^{20}$\,POT that occurred between November 7, 2017, and July 6, 2018. For neutrinos that interact in the detector, the Run~1 FHC sample consists of 84.09\% \nue{} and 15.91\% \nuebar{}, while the Run~3 RHC sample consists of 80.79\% \nue{} and 19.21\% \nuebar{}. The signal topology, $\mathrm{CC}0\pi Np$, is dominated by $\nu_{e}$ events. 


\subsection{Detector Design}

The MicroBooNE LArTPC is housed inside a cylindrical, stainless steel cryostat filled with 170 tonnes of liquid argon. Inside the cryostat is a field cage comprised of rectangular loops connected via a voltage divider chain linking a cathode and an anode on opposite-facing sides. The field cage encases 85 metric tons of liquid argon, which defines the active volume. Behind the anode is an array of 32 8-inch Hamamatsu photomultiplier tubes (PMTs) that act as MicroBooNE's light collection system. 

The anode consists of three wire planes: two induction planes, oriented at $\pm\,60\degree$ from the vertical, and one collection plane with vertically oriented wires. The planes are supplied with bias voltages of $-110$\,V, 0\,V, and 230\,V, respectively. The cathode is operated at $-70$\,kV, resulting in a uniform electric field of 273\,V/cm within the active volume. More information on the design of the detector can be found in \cite{uboone_det}. 

Charged particles created in neutrino interactions excite and ionize nearby argon atoms. The applied electric field drifts liberated electrons to the anode planes, where they induce signals on the wires. Excited argon atoms will bond with nearby ground-state atoms to form excimers, which quickly decay via isotropic de-excitation. This produces scintillation photons detectable by the MicroBooNE PMT array. 

Information about the signal formation on the anode planes and the electronics response is deconvolved from recorded wire waveforms \cite{charge-processing, charge-processing-2}. Each pulse in the deconvolved charge waveform is fit with a Gaussian distribution to form a ``hit'', which quantifies the number of ionization electrons detected on a single wire at a definite drift time.

Time-coincident waveforms from neighboring PMTs are combined to reconstruct flashes in the detector and determine the observed number of photoelectrons associated with an event. A PMT-based trigger selects events in the beam window with light levels above a certain threshold. An external (EXT) trigger is also implemented to collect data when there is no beam, using identical trigger conditions as the beam-on NuMI data stream to enable a standalone sample for data-driven estimations of cosmic-triggered events. An additional data stream with no trigger requirement (EXTUnbiased) yields a high-statistics sample of random, unbiased events. When combined with MC, this provides data-driven estimations of cosmic-ray background and modeling of detector noise occurring in time with beam-induced neutrino interactions.


\begin{figure}
\centering
\includegraphics[width=.47\textwidth]{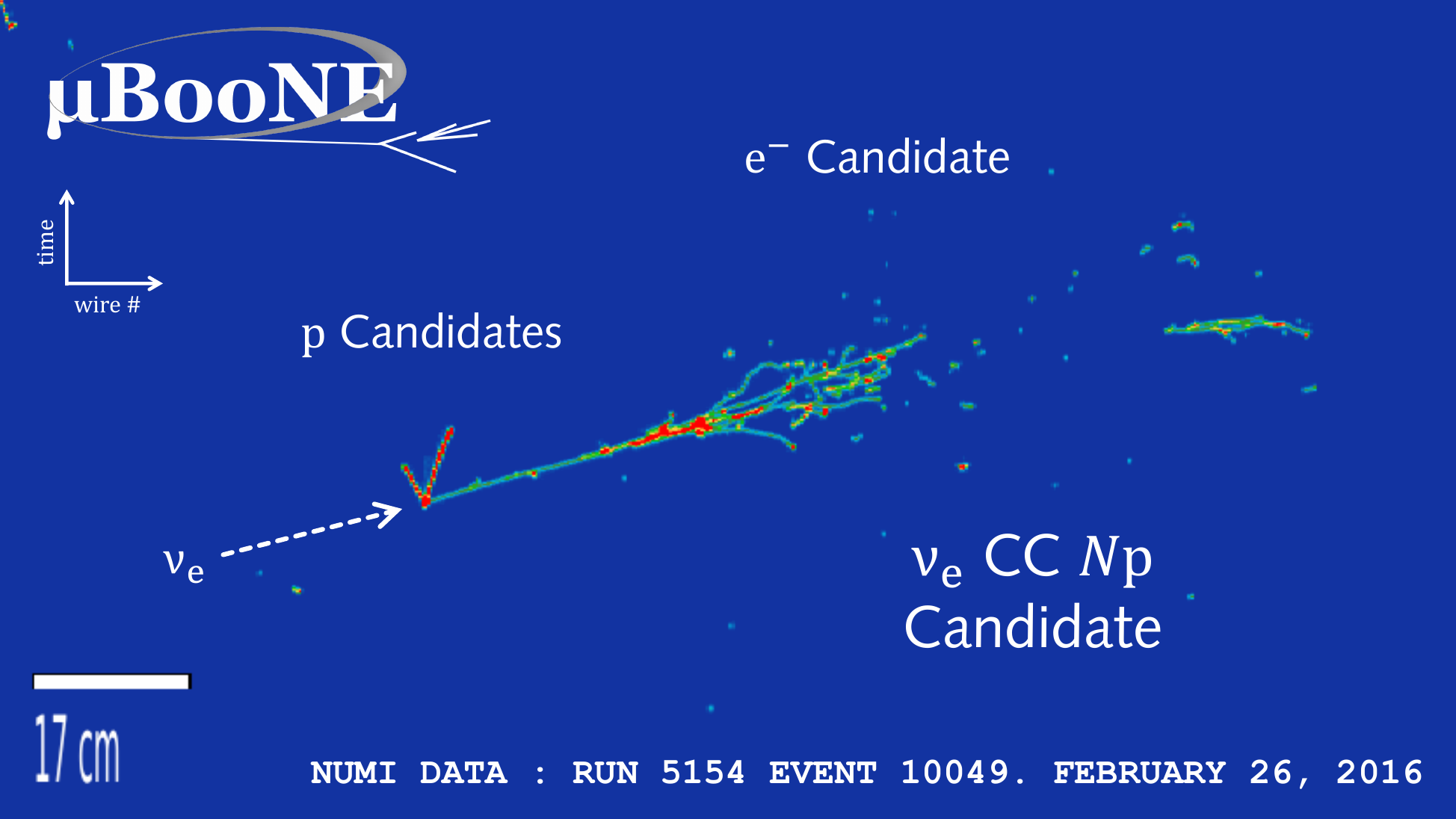}
\caption{\label{fig:event_display}{A selected signal candidate in the MicroBooNE NuMI Run~1 FHC dataset, as viewed from the collection plane. The event is characterized by a single electron-like electromagnetic shower and two proton-like tracks emanating from a common interaction vertex.}}
\end{figure}


One of the main advantages of LArTPCs for neutrino studies is the ability to discriminate among particles, particularly electrons and photons, based on calorimetric and topological information. In MicroBooNE, electrons and photons can be distinguished based on the characteristics of the electromagnetic showers they produce. Electrons have a characteristic energy loss ($\mathrm{d}E/\mathrm{d}x$) profile that extends back to the original neutrino interaction vertex. In contrast, photonic showers begin after a characteristic gap from the neutrino interaction vertex and deposit approximately twice the initial d$E$/d$x$ \cite{argoneut-dedx, pi0-gammagamma}. A common source of these photons is from \pizero decay, which produces photon pairs that travel some distance from the original neutrino interaction vertex before initiating showers. This production of two photons provides an additional handle for distinguishing electrons from photons. An example of an interaction signature is shown in Fig. \ref{fig:event_display}, which displays a \nue{} CC event candidate from the MicroBooNE NuMI dataset characterized by an electron-like electromagnetic shower and two proton-like tracks emanating from a common interaction vertex. 


\subsection{Event Simulation \& Reconstruction}

Data analysis in MicroBooNE relies on simulations of neutrino interactions and the reconstruction of events in the detector to interpret experimental observations. Simulated event samples are used to characterize the detector response and understand the expected signal and background distributions of collected data. The MicroBooNE event simulation and reconstruction are performed within the \code{LArSoft} framework \cite{larsoft}. 

The simulation of NuMI neutrino interactions in MicroBooNE begins with a prediction of the flux (Fig. \ref{fig:flux}), which serves as input to the \code{GENIE} Monte Carlo (MC) event generator \cite{genie}. Specifically, MicroBooNE employs \code{GENIE v3.0.6 G18\_10a\_02\_11a}. Based on this model configuration, a MicroBooNE-specific tune was developed \cite{genie_tune} using independent measurements collected by the T2K experiment \cite{t2k-experiment}. The modified cross section implemented through this tune affects the \nue{} CC simulation as well since the same underlying models are used in \code{GENIE} for both flavors. 

Custom algorithms in \code{LArSoft} simulate the detector response, including the readout of scintillation and ionization signals, accounting for detector effects that can attenuate light and ionization electrons and alter their trajectories as they traverse the detector. To provide a complete prediction of MicroBooNE data, raw EXTUnbiased waveforms are overlaid onto simulated neutrino interactions to form combined ``MC+EXT'' samples. The combined waveforms are processed through the same reconstruction chain as data, from which hits are extracted from the wire waveforms and used as input to high-level event reconstruction. 

The purpose of reconstruction is to build physics-relevant quantities from MicroBooNE's charge and light readout. Several published results from MicroBooNE \cite{numi-singlebin, numi-inclusive, bnb-exclusive, pelee-prd, numu-0piNp, numu-ccqelike, numu-inclusive, numu-pi0, numu-mult, higgs-portal-scalar, hnl-higgs} employ an automated, multi-algorithm approach to pattern recognition called \code{Pandora} \cite{pandora} to reconstruct interactions in the LArTPC. This toolkit uses topological and calorimetric information to assemble three-dimensional particles traversing the detector and identify candidate neutrino interactions. The output includes a reconstructed vertex and a hierarchy of particles produced by the event. Charged particles are also assigned a score by \code{Pandora}, indicating their consistency with a shower (more probable for scores between 0 and 0.5) or track (more probable for scores between 0.5 and 1) topology. 

Calorimetric energy reconstruction for showers is performed by summing the charge associated with all hits in the shower and converting these hits to the total shower energy using a constant recombination factor of 0.62. This conversion assumes that the electrons and positrons creating the shower are minimally ionizing and lose energy at a fixed rate of 2.3 MeV/cm \cite{michel-reco}.  The total shower energy is corrected to account for charge clustering inefficiencies in \code{Pandora} \cite{pi0-gammagamma} via a constant factor of 1.2.

For tracks, the reconstructed track length is measured geometrically. Under an assigned particle hypothesis, the kinetic energy is estimated from the track range \cite{berger2017stopping}, and the corresponding momentum is calculated from the kinetic energy.

Both shower and track energies are corrected for position- and time-dependent detector effects \cite{dedx-cal}. These corrections account for spatial and temporal variations in the detector response arising from space charge effects, electron attachment to impurities, diffusion, cross-connected TPC channels, and other nonuniformities such as changes in argon purity. The fully calibrated output of the \code{Pandora} reconstruction provides the basis for all subsequent data analysis. 

\section{Event Selection}
\label{selection}

Candidate events are identified based on reconstructed observables and must meet several pre-selection requirements to raise the likelihood that they are well reconstructed and consistent with the signal definition. Interactions must be identified as neutrino candidates by \code{Pandora} with an interaction vertex within the fiducial volume, defined as 10\,cm inward from any side of the LArTPC active volume. In addition, at least 90\% of the hits that make up reconstructed showers and tracks must be contained within the fiducial volume. Neutrino candidate events must have one reconstructed electromagnetic shower (assumed to be an electron candidate) with kinetic energy greater than 20\,MeV and one or more reconstructed tracks with kinetic energy greater than 40\,MeV. The track requirement removes most $\bar\nu_e$ events.

After the preselection, a set of quality cuts and loose constraints is applied to remove obvious backgrounds while retaining sufficiently high statistics in the event samples. To reject \numu{} CC interactions, the longest track in each event must pass a proton log-likelihood test based on its measured $\mathrm{d}E/\mathrm{d}x$ profile \cite{llr-pid}. The test assigns a score between $-1$ (proton-like) and $+1$ (muon-like) to each track object. Events with the longest track score greater than 0.35 are removed. To reject events misreconstructed with a candidate shower object, events are required to have a \code{Pandora} shower score of less than 0.3. Interactions involving a \pizero{} are removed by requiring the three-dimensional distance between the reconstructed interaction vertex and the start of the electromagnetic shower to be less than 12\,cm. Electron-induced showers originate at the vertex, while photons typically convert after a characteristic gap, with a conversion distance of approximately 25\,cm \cite{numu-pi0}. The chosen threshold accounts for shower-vertex misreconstruction. In addition, the $\mathrm{d}E/\mathrm{d}x$ on the collection plane is required to be less than 7\,MeV/cm. This loose cut helps to remove events with obvious reconstruction failures and enables the BDT to discriminate between electron and photon showers. Finally, the average angle between the shower's momentum vector and its associated hits (a proxy for the Moli\`ere angle) is required to be less than 15$\degree$. See Supplemental Material for event distributions as a function of the selection parameters \cite{SupplementalMaterial}.

Next, a Boosted Decision Tree (BDT) is used to discriminate $1e+Np$ interactions from background via a multivariate assessment of reconstructed parameters related to the event topology and energy profile. Separate BDT models are developed for the Run~1 FHC and Run~3 RHC event rates using the gradient boosting framework \code{XGBoost} \cite{xgboost} to account for differences in the rates and energies of the neutrinos produced in the different beam configurations.

MC+EXT event samples that pass the preceding requirements are used to train the BDT. These samples are split equally, with one half reserved for training and the other for testing. To constrain the training phase space, only events with a reconstructed shower energy greater than 70\,MeV and a reconstructed cosine of the opening angle, defined as the cosine of the angle between the shower momentum and the longest track in the interaction, between $\pm\,0.9$ are included. These criteria apply only to the training phase space and are not part of the event selection or measurement. The training phase space is intentionally restricted to reduce bias from training in poorly reconstructed regions. The shower energy constraint suppresses the population of delta rays and low-energy Michel electrons in the samples used to train the BDT model, while the opening-angle cosine constraint removes events with poorly reconstructed topologies. Both criteria improve the signal-to-background ratio for optimal training. 


\begin{figure}
\centering
\begin{tikzpicture} \draw (0, 0) node[inner sep=0] {
\includegraphics[width=.48\textwidth]{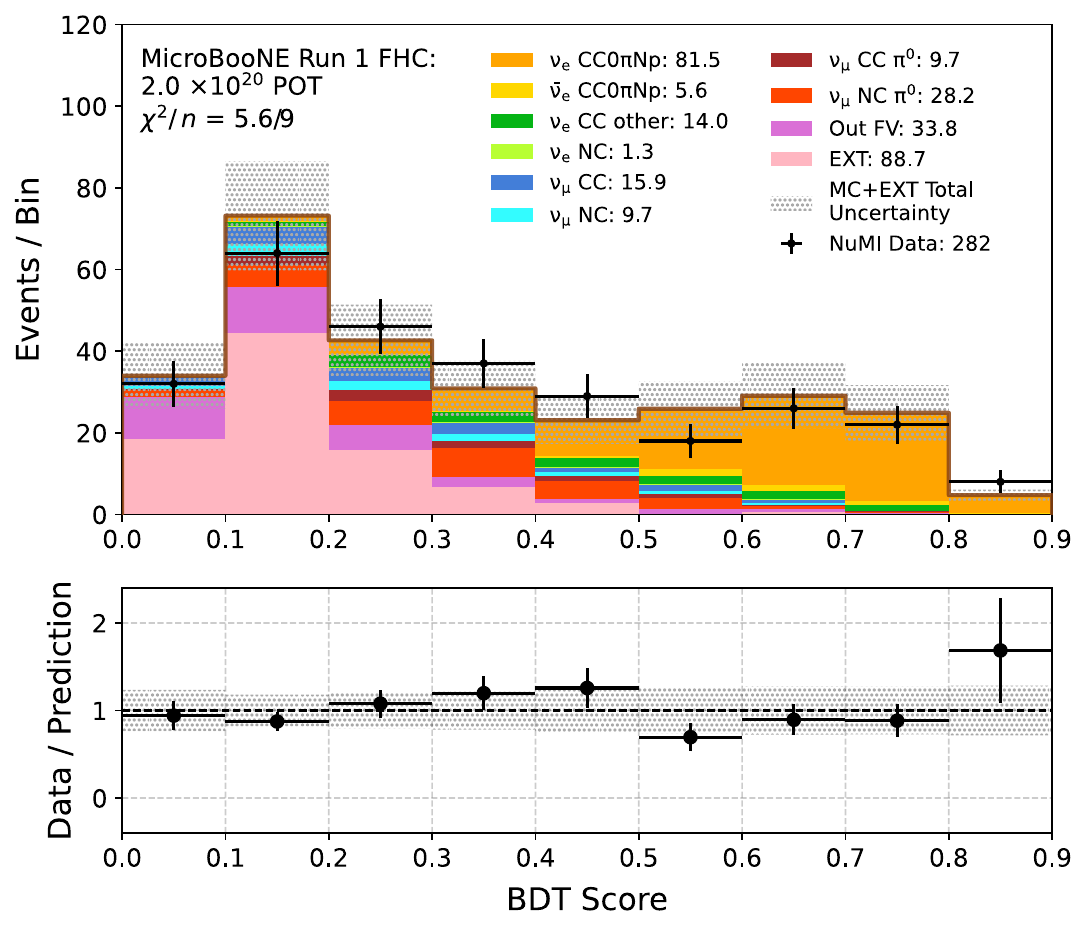}
		};
  \draw (-3.0, -1.25) node {\textbf{(a)}};
\end{tikzpicture}

\begin{tikzpicture} \draw (0, 0) node[inner sep=0] {
\includegraphics[width=.48\textwidth]{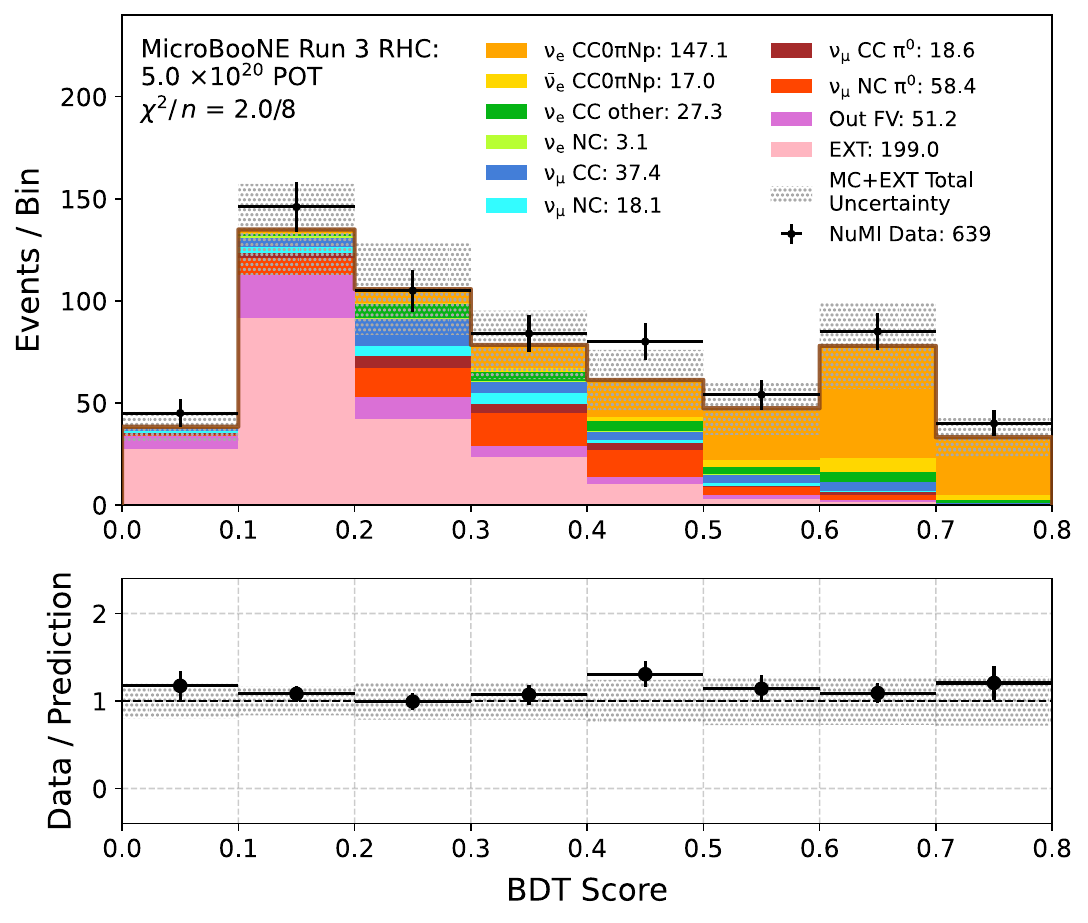}
		};
  \draw (-3.0, -1.25) node {\textbf{(b)}};
\end{tikzpicture}

\caption{\label{fig:bdtscore}{Estimated event rates for Run~1 FHC (a) and Run~3 RHC (b) as a function of the output of the BDT trained to discriminate signal (orange) from various sources of background. The gray band represents the total MC+EXT uncertainty, as described in Sec. \ref{uncertainties}. Data is overlaid for comparison, shown in black with associated statistical uncertainties.}}
\end{figure}


\begin{figure*}[t]
\centering

\begin{tikzpicture} \draw (0, 0) node[inner sep=0] {
\includegraphics[width=.48\textwidth]{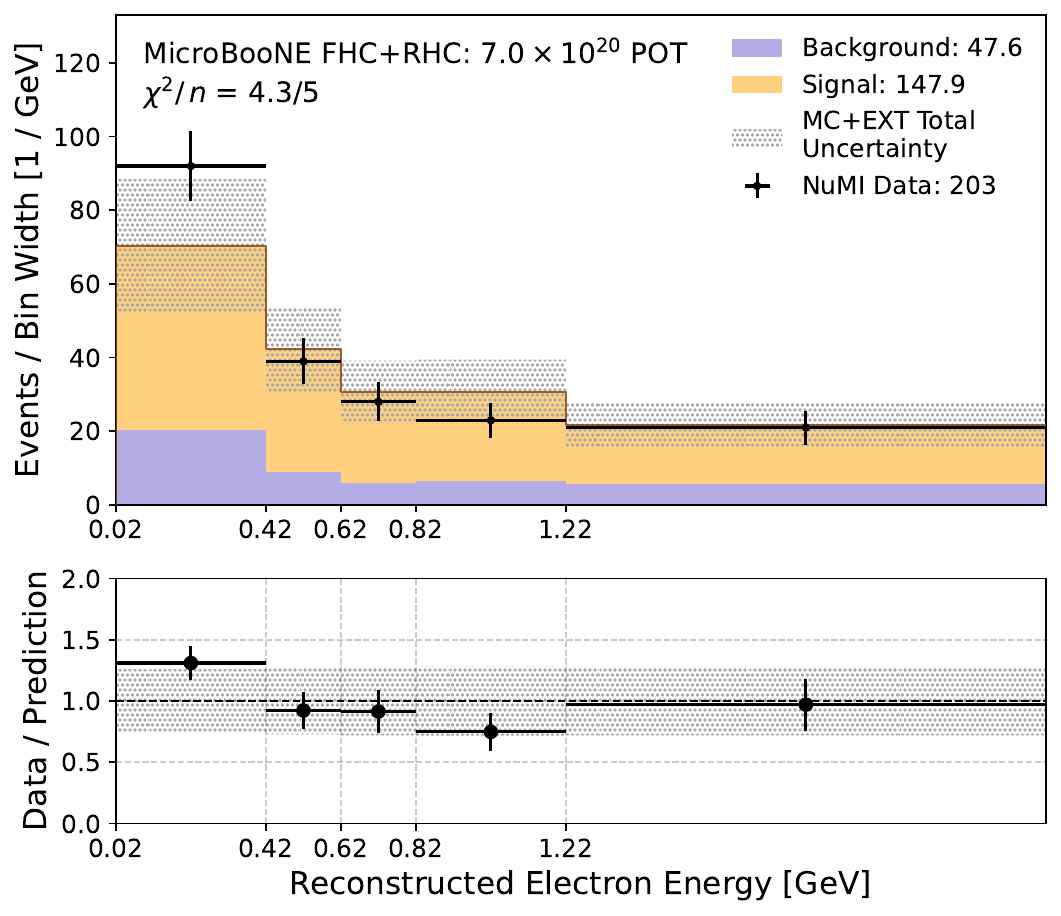}
		};
  \draw (3.85, -1.3) node {\textbf{(a)}};
\end{tikzpicture}
\hfill
\begin{tikzpicture} \draw (0, 0) node[inner sep=0] {
\includegraphics[width=.48\textwidth]{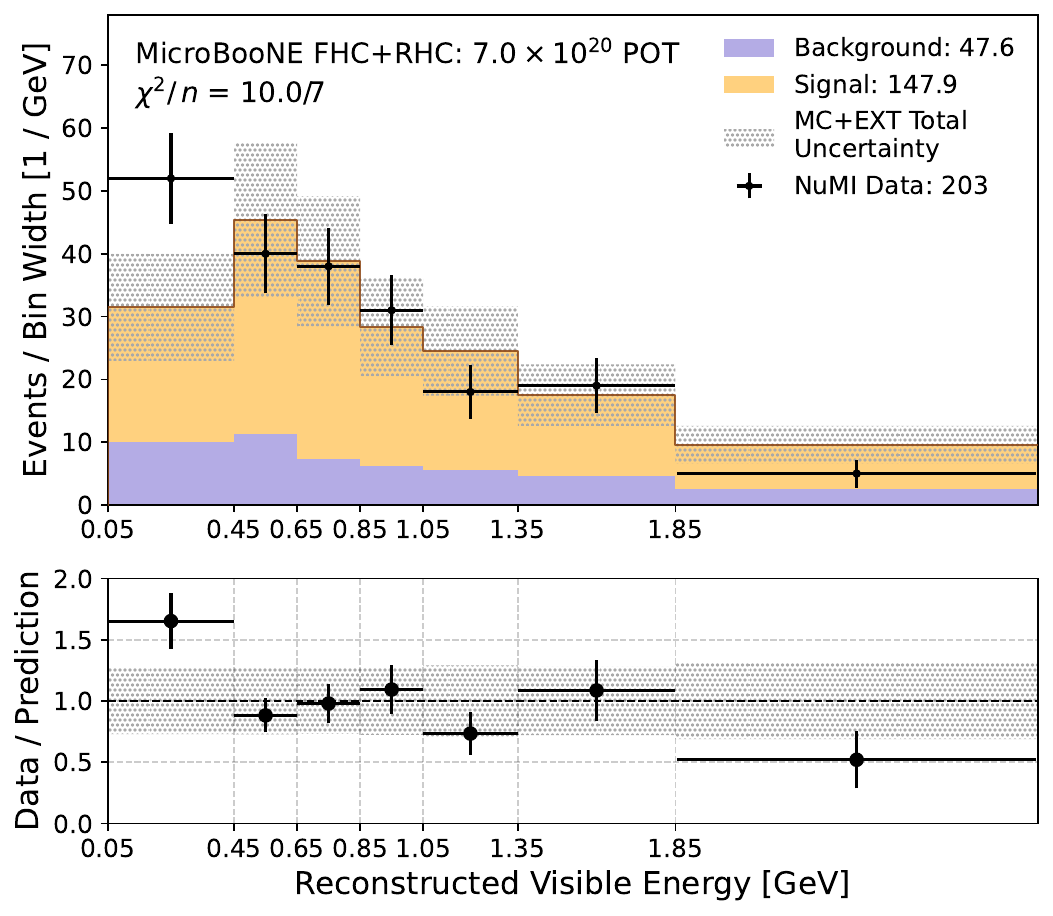}
		};
  \draw (3.85, -1.265) node {\textbf{(b)}};
\end{tikzpicture}
\hfill
\begin{tikzpicture} \draw (0, 0) node[inner sep=0] {
\includegraphics[width=.495\textwidth]{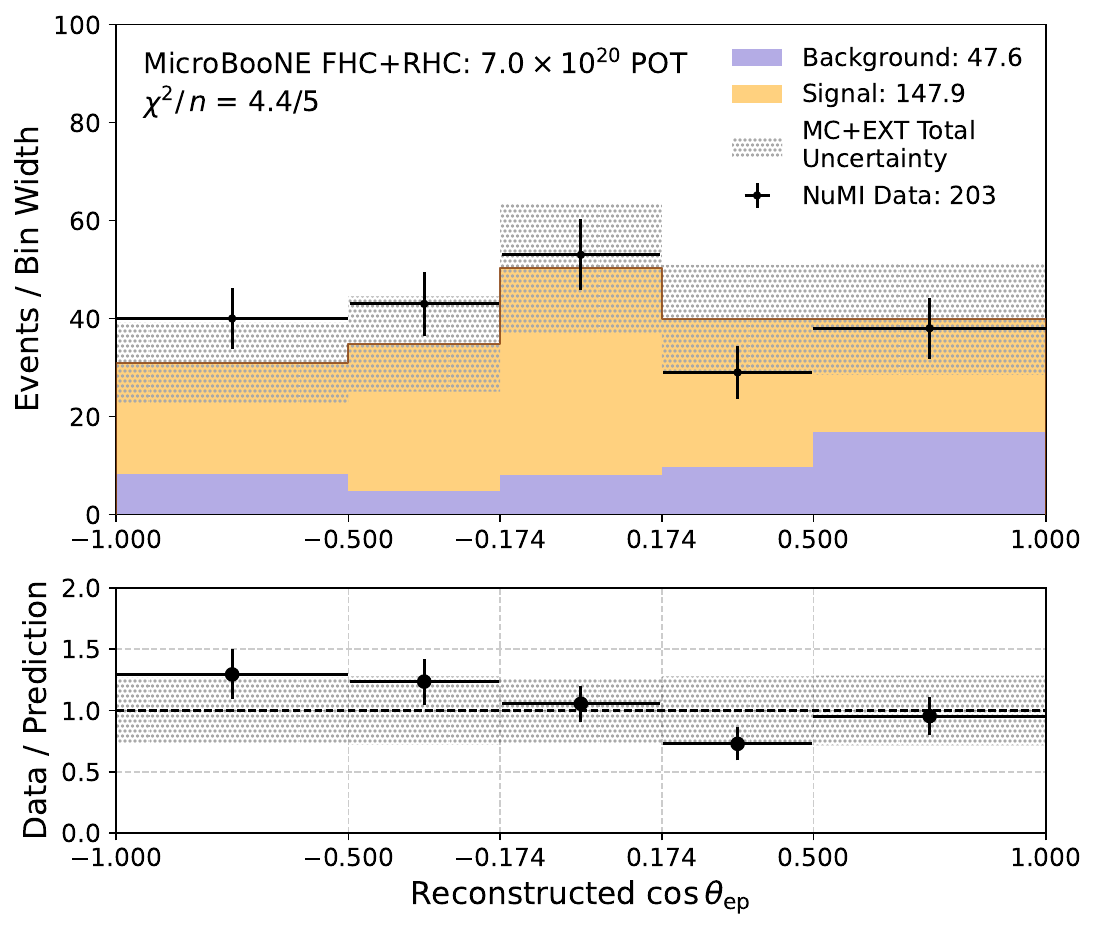}
		};
  \draw (3.75, -1.25) node {\textbf{(c)}};
\end{tikzpicture}
\hfill
\begin{tikzpicture} \draw (0, 0) node[inner sep=0] {
\includegraphics[width=.48\textwidth]{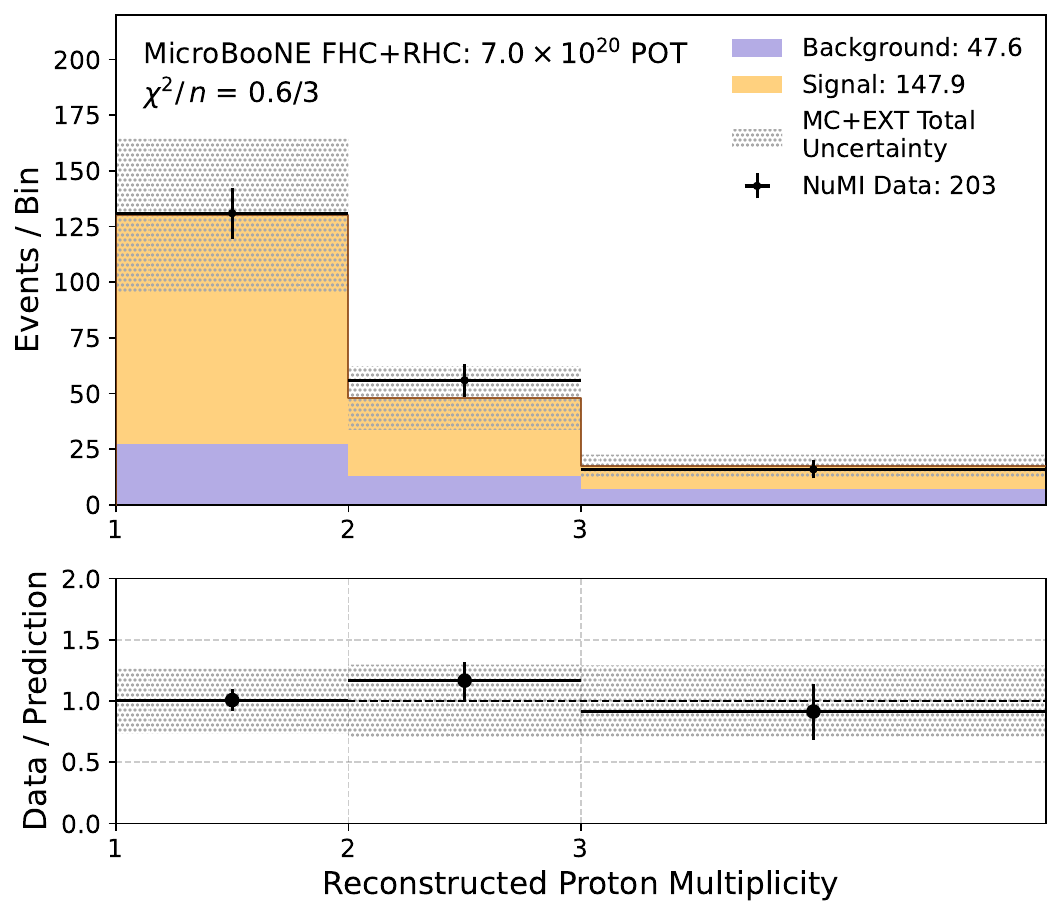}
		};
  \draw (3.85, -1.25) node {\textbf{(d)}};
\end{tikzpicture}

\caption{\label{fig:selected_full}{Estimated signal (orange) and background (purple) predictions of the FHC+RHC selected events after the full selection (including the BDT threshold requirement) for reconstructed electron energy (a), visible energy (b), $\cos{\theta_{ep}}$ (c), and proton multiplicity (d). The rightmost bin for the electron energy, visible energy, and proton multiplicity distributions is overflow. The gray band represents the total MC+EXT uncertainty, as described in Sec. \ref{uncertainties}. Data is overlaid for comparison, shown in black with associated statistical uncertainties. The data/MC ratios are also shown, where again the gray band represents the total MC+EXT uncertainty.}}
\end{figure*}
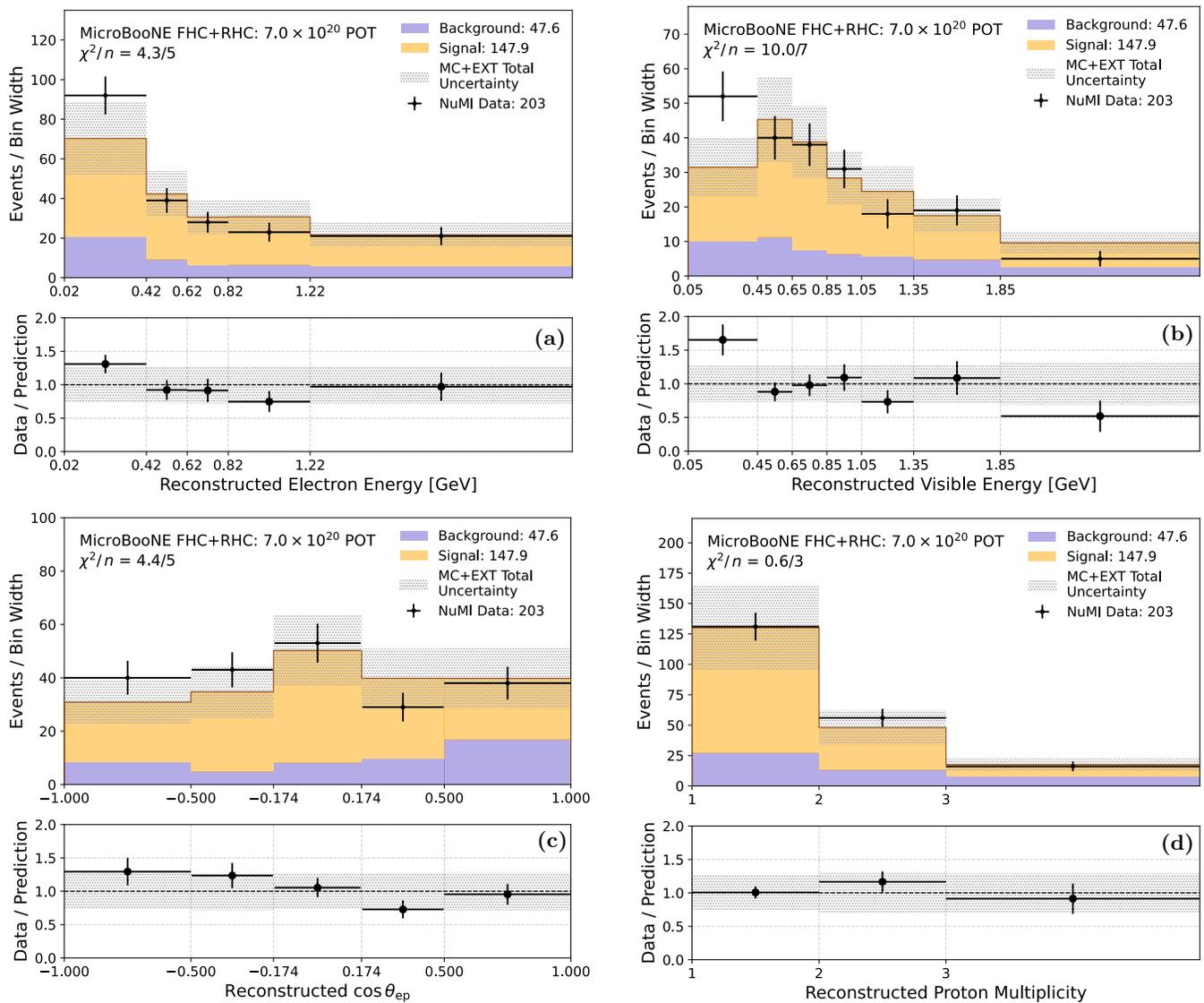


\begin{figure*}[t]
\centering

\begin{tikzpicture} \draw (0, 0) node[inner sep=0] {
\includegraphics[width=.48\textwidth]{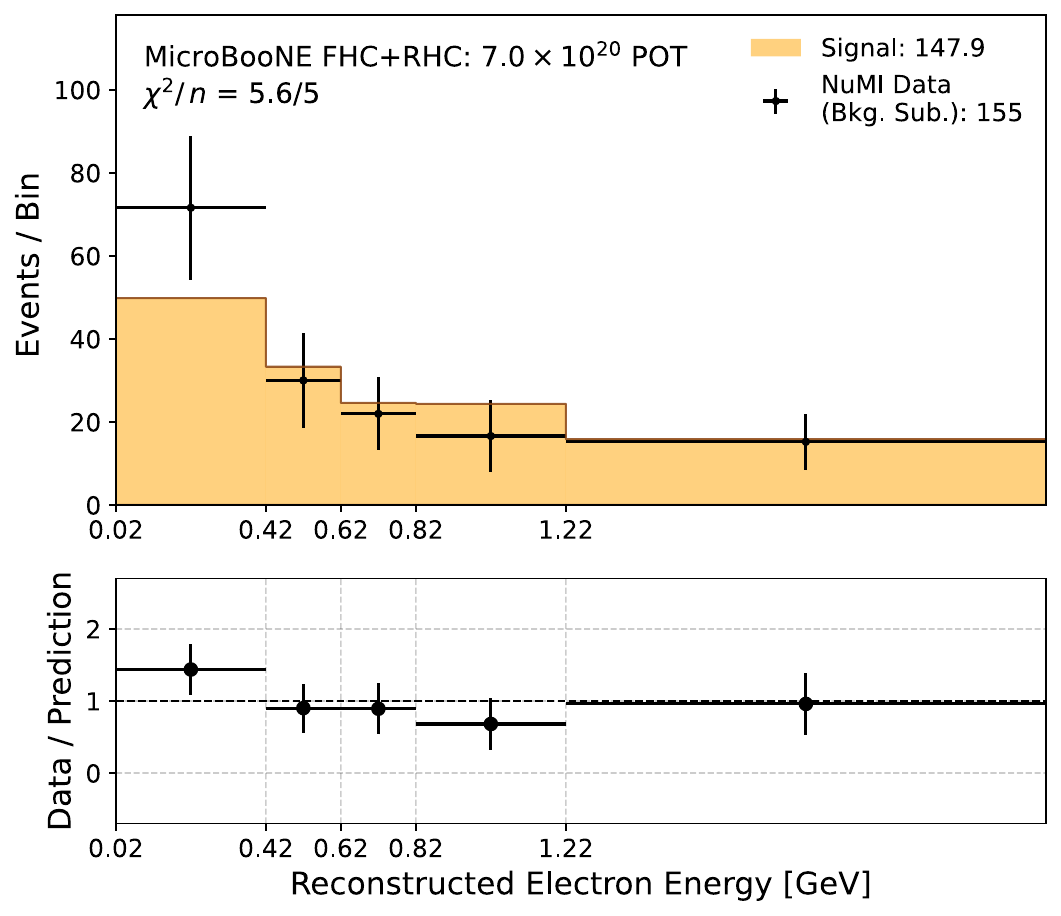}
		};
  \draw (3.85, -1.3) node {\textbf{(a)}};
\end{tikzpicture}
\hfill
\begin{tikzpicture} \draw (0, 0) node[inner sep=0] {
\includegraphics[width=.48\textwidth]{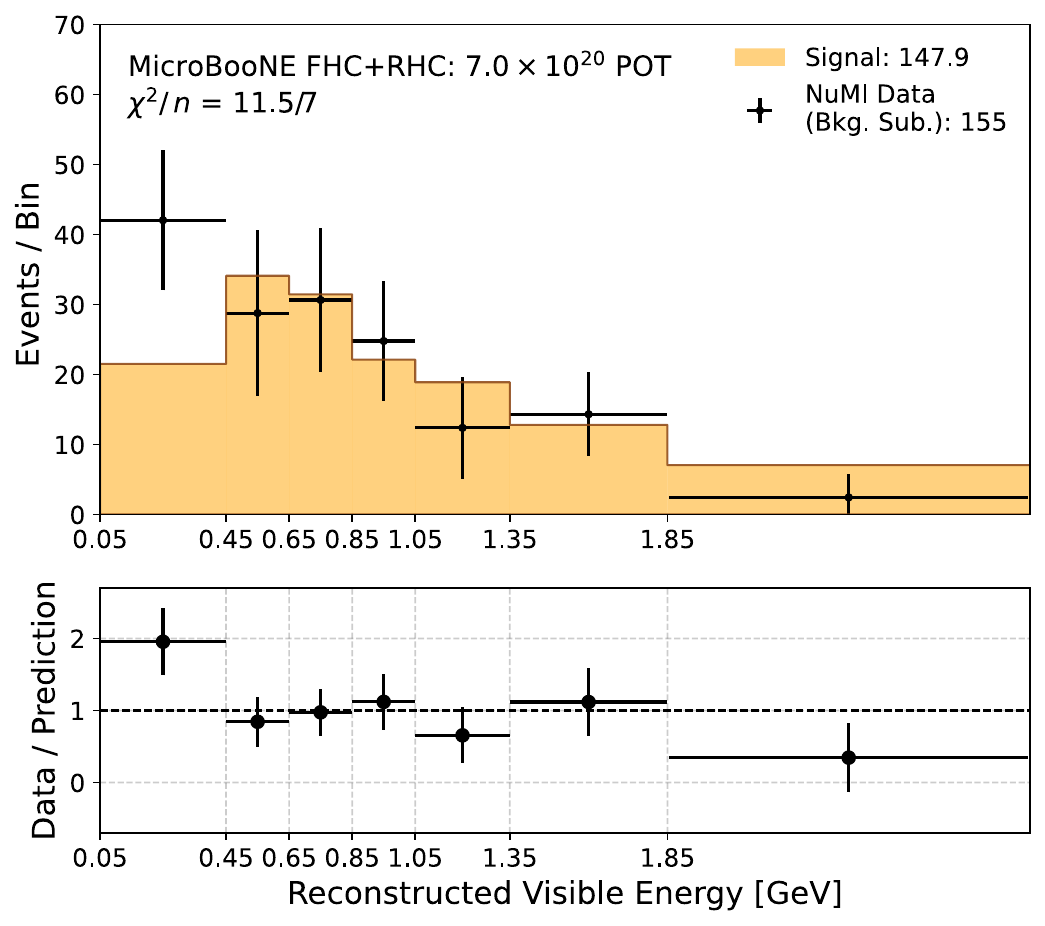}
		};
  \draw (3.85, -1.265) node {\textbf{(b)}};
\end{tikzpicture}
\hfill
\begin{tikzpicture} \draw (0, 0) node[inner sep=0] {
\includegraphics[width=.495\textwidth]{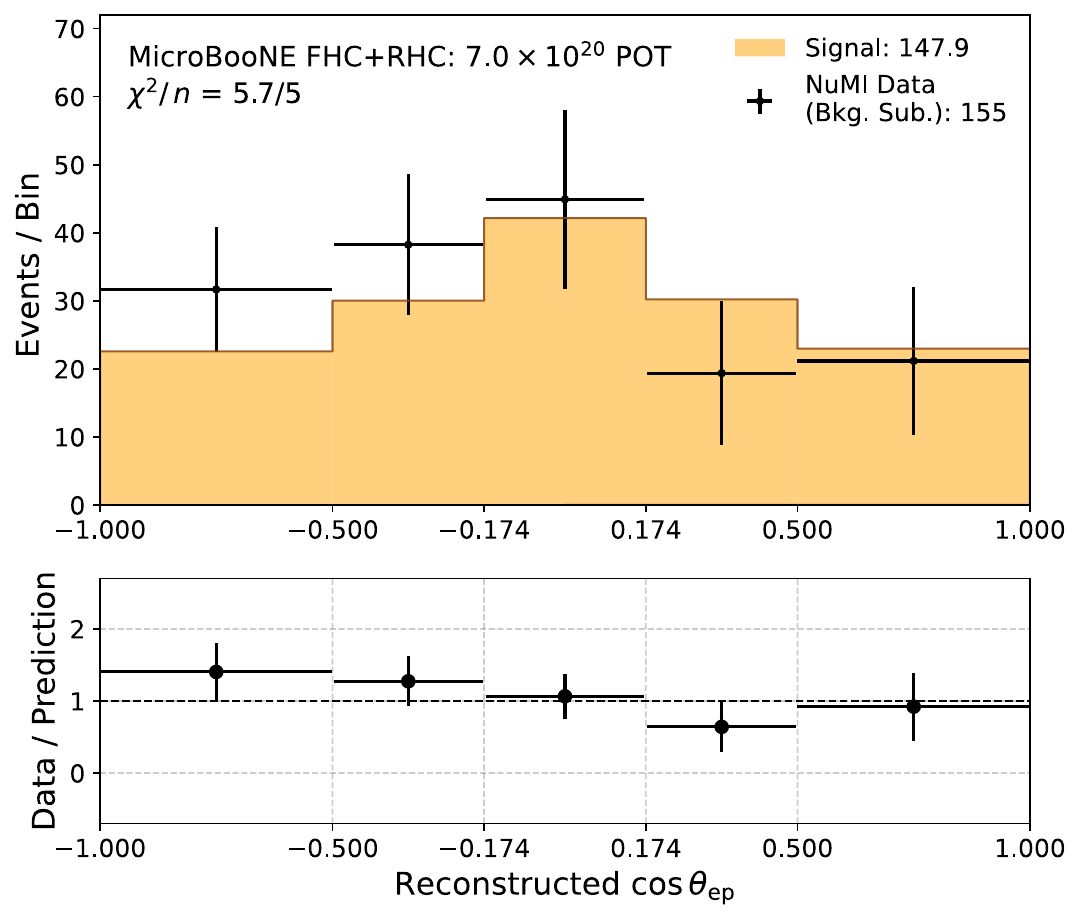}
		};
  \draw (3.75, -1.25) node {\textbf{(c)}};
\end{tikzpicture}
\hfill
\begin{tikzpicture} \draw (0, 0) node[inner sep=0] {
\includegraphics[width=.48\textwidth]{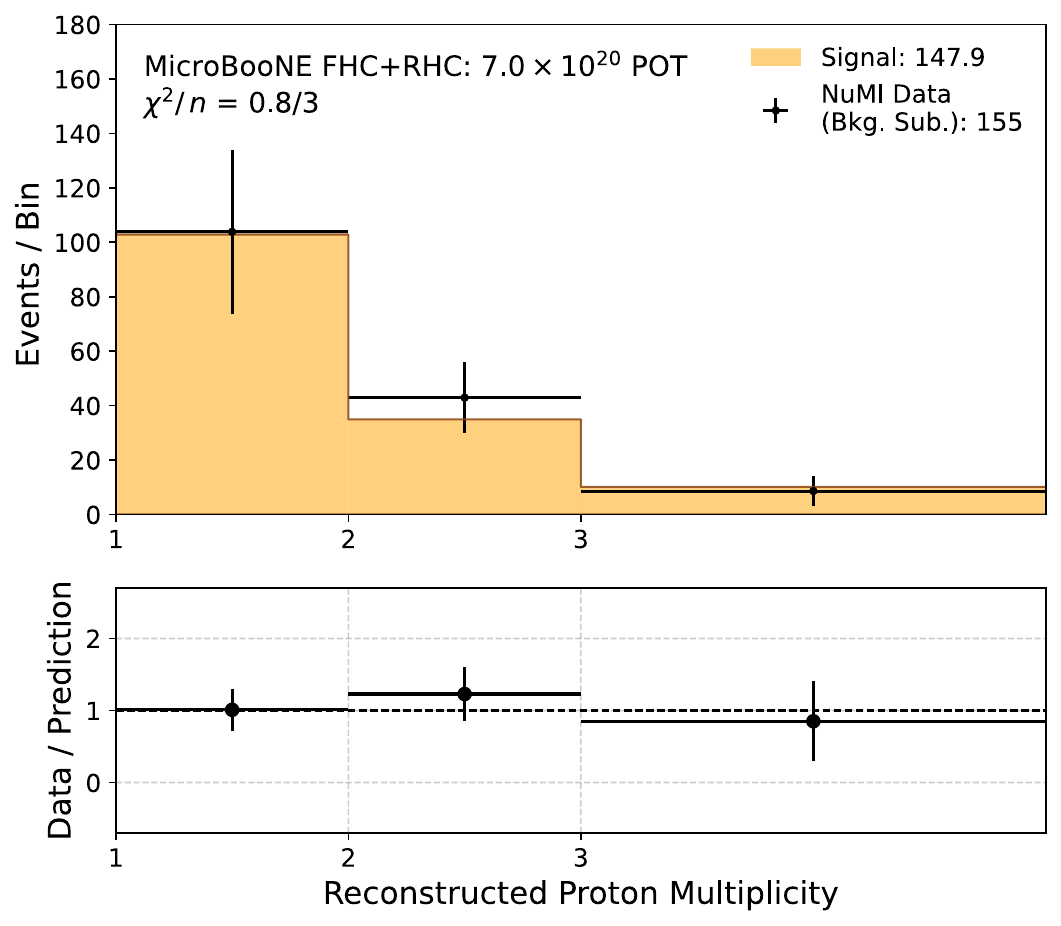}
		};
  \draw (3.85, -1.25) node {\textbf{(d)}};
\end{tikzpicture}

\caption{\label{fig:selected}{Estimated signal (orange) predictions of the FHC+RHC selected events after background subtraction for reconstructed electron energy (a), visible energy (b), $\cos{\theta_{ep}}$ (c), and proton multiplicity (d). The rightmost bin for the electron energy, visible energy, and proton multiplicity distributions is overflow. Background-subtracted NuMI data is overlaid for comparison, shown in black with associated statistical and systematic uncertainties. Included in these error bars are the systematic uncertainties on the signal prediction. The data/MC ratios are also shown.}}
\end{figure*}


Seven reconstructed parameters, chosen for their strong selective power, are used to train the BDT model to reject neutrino-induced backgrounds. To remove \numu{} CC interactions, the BDT trains on the \code{Pandora} shower score, track PID score, and number of subclusters (isolated two-dimensional charge segments) making up the reconstructed shower. To remove \pizero{} events, the BDT trains on the average shower Moli\`ere angle, d$E$/d$x$ on the collection plane, and both the three-dimensional and smallest two-dimensional distance on the collection plane between the interaction vertex and the start of the shower. See Supplemental Material for distributions of these selection parameters \cite{SupplementalMaterial}.


\begin{table*}[t]
\centering\footnotesize
\caption{Selection efficiencies and purities (\%) at different cut stages for Run~1 FHC and Run~3 RHC. Efficiencies and purities are cumulative after each successive cut and are evaluated on simulated events with respect to the signal definition described in Sec. \ref{introduction}.}
\label{tab:fhc_rhc_eff_pur}
\begin{ruledtabular}
\begin{tabular}{l l c c c c}
\textbf{Cut Goal} & \textbf{Cut Definition} & \multicolumn{2}{c}{\textbf{Run~1 FHC}} & \multicolumn{2}{c}{\textbf{Run~3 RHC}} \\ [3pt]
 & & \textbf{Eff.\ (\%)} & \textbf{Pur.\ (\%)} & \textbf{Eff.\ (\%)} & \textbf{Pur.\ (\%)} \\
\hline
Neutrino Identification & Number of Neutrinos Identified = 1 & 82.5 & 0.4 & 82.7 & 0.3 \\ [6pt]
\multirow{2}{*}{Containment}
 & Reconstructed Vertex in FV & 77.1 & 0.7 & 77.6 & 0.6 \\
 & Contained Fraction $>$ 0.9 & 61.5 & 1.6 & 61.6 & 1.4 \\ [6pt]
\multirow{3}{*}{Signal Definition Constraints}
 & No. Showers Contained = 1 & 45.7 & 3.7 & 45.3 & 3.2 \\
 & No. Tracks Contained = 1 & 40.3 & 4.5 & 39.8 & 3.9 \\
 & Track Kinetic Energy $>$ 40\,MeV & 39.3 & 4.6 & 38.7 & 4.0 \\ [6pt]
\multirow{2}{*}{Loose CC $\nu_\mu$ Rejection}
 & Track PID (Proton/Muon Log-Likelihood) $<$ 0.35 & 31.7 & 10.0 & 31.5 & 8.5 \\
 & \texttt{Pandora} Shower Score $<$ 0.3 & 30.3 & 12.5 & 30.4 & 10.8 \\ [6pt]
\multirow{3}{*}{Loose $\pi^0$ Rejection}
 & Distance Between Track \& Shower $<$ 12\,cm & 25.9 & 16.5 & 25.7 & 14.3 \\
 & Collection Plane $\mathrm{d}E/\mathrm{d}x$ $<$ 7\,MeV/cm & 24.0 & 16.4 & 23.8 & 14.1 \\
 & Average Shower Molière Angle $<$ 15$\degree$ & 22.9 & 23.7 & 22.7 & 21.4 \\ [6pt]
 BDT Score & BDT Score $<$ 0.55 (FHC), 0.575 (RHC) & 13.7 & 78.5 & 12.3 & 74.0 \\
 
\end{tabular}
\end{ruledtabular}
\end{table*}


The trained model assigns each input event a score based on how closely it resembles signal. The final selection stage applies a cut on this BDT score, with the threshold value chosen based on a scan of the full selection's performance on the testing sample using different BDT score criteria. Two metrics are used to evaluate the performance of the selection. The efficiency, which quantifies how well signal is retained, is defined as the ratio between the number of true selected and true generated signal events. The purity quantifies how well the selection performs at removing backgrounds and is defined as the ratio between the number of selected signal events and the total selected events. Both metrics are estimated from the simulated event rates. The optimal BDT cut is chosen by evaluating the efficiency and purity as functions of the BDT score and selecting a point that improves background rejection relative to the baseline linear selection with minimal sacrifice in efficiency. The linear selection, whose cut definitions are summarized in the Supplemental Material \cite{SupplementalMaterial}, is adapted for NuMI from the MicroBooNE low-energy excess search selection for $1e + Np$ interactions in the BNB \cite{linear_selection, linear_selection_prd}.

Results of the BDT on data, compared with the full MC+EXT sample, are shown in Fig. \ref{fig:bdtscore}. In these plots, the predicted event rate is broken down into the signal channel $\left(\nu_{e}\,\mathrm{CC}0\pi Np\right)$, CC \numu{} or \numubar{} backgrounds with $\left(\nu_{\mu}\,\mathrm{CC}\pi^{0}\right)$ and without $\left(\nu_{\mu}\,\mathrm{CC}\right)$ a \pizero{} present in the final topology, neutral current (NC)  \numu{} or \numubar{} backgrounds with $\left(\nu_{\mu}\,\mathrm{NC}\pi^{0}\right)$ and without $\left(\nu_{\mu}\,\mathrm{NC}\right)$ a \pizero{} present in the final topology, and \nue{} or \nuebar{} backgrounds ($\nu_{e}\,\mathrm{NC}$, $\nu_{e}\,\mathrm{CC}$ other). Irreducible \nuebar{} contamination $\left(\bar{\nu}_{e}+\,^{40}{\rm Ar} \rightarrow 1e+Np+0\,\pi\right)$, which is indistinguishable from signal events in the LArTPC, is shown separately $\left(\bar{\nu}_{e}\,\mathrm{CC}0\pi Np\right)$. Data (black) is overlaid for comparison. For the Run~1 FHC sample, events with a BDT score $> 0.55$ are accepted. This yields an estimated efficiency of 13.7\% and a purity of 78.5\%. For the Run~3 RHC sample, events with a BDT score $> 0.575$ are accepted. This yields an estimated efficiency of 12.3\% and purity of 74.0\%. The data/MC agreement is well covered by statistical and systematic uncertainties for both the signal and background dominated regions. FHC and RHC event distributions after the BDT score cuts are available in the Supplemental Material \cite{SupplementalMaterial}. See Table~\ref{tab:fhc_rhc_eff_pur} for a summary of the cumulative selection efficiencies and purities at each cut stage for Run~1 FHC and Run~3 RHC. 

The FHC+RHC selected events have a combined efficiency of 12.7\% and are displayed in Fig. \ref{fig:selected_full} separated into simulated signal and backgrounds. The largest background contribution is $\bar{\nu}_{e}\,\mathrm{CC}0\pi Np$ interactions, which comprises an estimated 7.4\% of the final selected event count. Figure \ref{fig:selected_full}(a) shows the selected events as a function of reconstructed electron energy, defined as the sum of the calorimetric energy of the electromagnetic shower and the rest mass energy of the final-state electron. The rightmost bin is overflow and includes all events with electron energy $> 1.22$\,GeV. Figure \ref{fig:selected_full}(b) displays the selected events as a function of reconstructed visible energy $\left(E_{\mathrm{visible}}\right)$, defined as the kinetic energy sum of the electromagnetic shower $\left(E_{\mathrm{shower}}\right)$ and all tracks in the event with greater than 40\,MeV kinetic energy $\left(E_{\mathrm{track}}\right)$, 

\begin{equation}
    E_{\mathrm{visible}} = E_{\mathrm{shower}} + \sum E_{\mathrm{track}}. 
\end{equation}
\newline
\noindent The rightmost bin is overflow and includes all events with visible energy $> 1.85\,\mathrm{GeV}$. The selected events as a function of the cosine of the reconstructed opening angle between the momenta of the electromagnetic shower $\left(\vec{p}_{\mathrm{e}}\right)$ and the longest track above the $40\,\mathrm{MeV}$ kinetic energy threshold $\left(\vec{p}_{\mathrm{p}}\right)$,

\begin{equation}
    \cos{\theta_{ep}} = \frac{\vec{p}_{e} \cdot \vec{p}_{p}}{||\vec{p}_{e}|| \: ||\vec{p}_{p}||},
\end{equation}
\newline
\noindent are displayed in Fig. \ref{fig:selected_full}(c). Finally, the reconstructed proton multiplicity is shown in Fig. \ref{fig:selected_full}(d), defined as the number of tracks above the $40\,\mathrm{MeV}$ kinetic energy threshold that are fully contained in the fiducial volume. The rightmost bin is overflow and includes all events with three or more tracks in the final topology.

In total, 203 NuMI data events pass the selection. The metric $\chi^{2}$ per degree of freedom ($n$), where $n$ is the number of bins in the distribution, quantifies the agreement between data and prediction by accounting for statistical and systematic uncertainty (including correlations) as described in Sec. \ref{uncertainties}. The $\chi^2/n$ is equal to 4.3/5, 10.0/7, 4.4/5, and 0.6/3 for the total selected events as a function of reconstructed electron energy, visible energy, cosine of the opening angle, and proton multiplicity, respectively.

The measured signal is estimated by subtracting the predicted background distribution from data in each bin reconstructed. A total of 155 NuMI data events remain after the background subtraction, and background-subtracted data distributions in several variables are shown in Fig. \ref{fig:selected}. The $\chi^2/n$ is equal to 5.6/5, 11.5/7, 5.7/5, and 0.8/3 for the background-subtracted selected events as a function of reconstructed electron energy, reconstructed visible energy, the cosine of the reconstructed opening angle, and reconstructed proton multiplicity, respectively.

\section{Sources of Uncertainty}
\label{uncertainties}

Several sources of uncertainty affect MicroBooNE measurements, some of which are common to all analyses, while others are specific to those utilizing the NuMI beam. Sources of uncertainty common to all analyses include systematic effects from neutrino cross section and secondary particle re-interaction models used to simulate event rates, uncertainties arising from limitations in understanding the detector response to interacting particles, and poorly constrained predictions of out-of-cryostat (dirt) events. Uncertainties in NuMI-specific analyses include systematic effects from the hadron production and geometric beamline models used in the NuMI flux prediction, and uncertainty in the estimated POT delivered to the NuMI beamline. Finally, statistical fluctuations, due to the finite number of events in beam-on data and simulated events in the MC+EXT simulation, affect all measurements. 

Systematic uncertainties in the MC+EXT simulation can change the event count, the detector response, and the estimated efficiency of the selection algorithm---all of which impact the cross-section measurement. To quantify this impact, parameters associated with each uncertainty are identified and varied to generate alternate universe (UV) event rates distinct from the central value (CV) distribution. Variations are created by re-simulating entire event samples with alternate parameters or by employing a reweighting scheme in which interactions in the CV distribution are weighted to produce the UV event rate. Reweighting factors are generated using either a multisim or unisim approach. Multisim reweights take into account correlations between dependent parameters by randomly sampling them simultaneously \cite{multisim}. Unisim reweights, on the other hand, treat the parameters as independent and are constructed by varying each one individually by $\pm\,1\sigma$. 

For each source of systematic uncertainty, a representative set of FHC+RHC systematic variations is produced. The effect on the reconstructed event rate in bin $i$ for each universe is evaluated as 

\begin{equation}
\label{eqn:event_rate}
x_i \equiv N_{\mathrm{reco}\:i}^{\mathrm{UV}},
\end{equation}
\newline
where $N_{\mathrm{reco}\:i}^{\mathrm{UV}}$ is the selected total (signal + background) prediction in universe $i$. From these variations, the covariance matrix is constructed as 

\begin{equation}
\label{eqn:covariance}
    \mathrm{cov}(i, j) = \frac{1}{N_{\mathrm{univ}}}\sum_{k=1}^{N_{\mathrm{univ}}} 
    \left(x^{k}_i - x^{\mathrm{CV}}_i\right)\left(x^{k}_j - x^{\mathrm{CV}}_j\right),
\end{equation}
\newline
where $x^k_i$ and $x^{\mathrm{CV}}_i$ are the reconstructed event rates in universe $k$ and the central-value (CV) sample, respectively, and $N_{\mathrm{univ}}$ is the total number of systematic variations. For cross-section--related uncertainties, only the effects on the reconstructed--to--true migration, signal selection efficiency, and background rate are propagated when constructing the covariance matrix. The total covariance matrix is then obtained by summing the covariance matrices from each source of uncertainty. The combined covariance matrix is used as input for the cross-section extraction and is provided in the Supplemental Material \cite{SupplementalMaterial}. 

In this work, model uncertainties for hadron production \cite{numi_flux}, neutrino cross-section models \cite{genie, genie-untuned, genie_tune}, and secondary particle re-interactions \cite{geant4-reweight} are evaluated using event reweights generated via the multisim approach. 

\noindent The impact of the $\nu_{e}+\,^{40}{\rm Ar} \rightarrow 1e + Np + 0\,\pi$ cross-section model on the reconstruction of kinematic quantities and the estimated efficiency is taken into account. Unisim reweights are used to quantify the impact of uncertainties on the beamline geometry \cite{numi_beam}, dirt interaction models, POT counting \cite{leo-thesis}, number of targets in the detector fiducial volume, and a subset of neutrino cross-section parameters not encapsulated in the multisim reweights \cite{genie_tune}. Detector response uncertainty is assessed by simulating a set of UV event samples, each representing a particular detector response variation. These include variations in the light yield, the free electron recombination model, space charge effects, and TPC wire response \cite{detsys}. Given the limited statistics available to determine bin-to-bin uncertainties and correlations, we evaluate detector variations as a function of the reconstructed position of the neutrino vertex in the drift direction and assign a 10\% uncertainty due to detector effects in each bin. This uncertainty is treated as flat and uncorrelated between bins. 

The bin-by-bin fractional systematic uncertainties on the background-subtracted prediction and the largest single category of background, \nuebar{} \npsel{}, shown as a function of reconstructed electron energy, visible energy, $\cos{\theta_{ep}}$, and proton multiplicity, are shown in the Supplemental Material \cite{SupplementalMaterial}.


\begin{table}
\caption{\label{tab:uncertainty_breakdown} Average uncertainty contributions (\%) to the background-subtracted event rates for the data cross-section measurement.}
\begin{ruledtabular}
\begin{tabular}{lllll}
\textbf{Source of Uncertainty} & $\boldsymbol{E_{e}}$ & $\boldsymbol{E_{\mathrm{visible}}}$ & $\boldsymbol{\cos{\theta_{ep}}}$ & $\boldsymbol{N_{\mathrm{proton}}}$\\
\hline
Flux & 26.0 & 26.9  & 26.2 & 27.8 \\ 
Cross Section & 7.3 & 8.4 & 8.4 & 10.0 \\ 
Particle Re-Interactions & 2.1 & 2.7 & 2.1 & 3.0 \\ 
Detector Response & 10.0 & 10.0 & 10.0 & 10.0 \\ 
POT Counting & 2.0 & 2.0 & 2.0 & 2.0 \\
Target Counting & 1.0 & 1.0 & 1.0 & 1.0 \\ 
Dirt & 0.1 & 0.1 & 0.1 & 0.1 \\ 
Data Statistics & 21.6 & 26.4 & 22.4 & 24.0 \\
MC+EXT Statistics & 5.4 & 6.7 & 5.6 & 6.2 \\
\hline
Total (syst.) & 29.5  & 31.0 & 30.2 & 32.3 \\
Total (stat. + syst.)  & 36.7 & 40.9 & 37.7 & 40.8 \\ 
\end{tabular}
\end{ruledtabular}

\end{table} 


Fractional uncertainty contributions averaged across all bins are listed in Table \ref{tab:uncertainty_breakdown} for electron energy, visible energy, the cosine of the opening angle, and proton multiplicity. These uncertainties are derived from the covariance matrix, which is first normalized by the product of the event rates in the corresponding bins to produce the fractional covariance matrix. Each diagonal element represents the variance in that bin, and the square root gives the fractional uncertainty. These fractional uncertainties are then averaged across all bins to provide the total fractional uncertainty. The dominant systematic uncertainty in the background-subtracted event rates used for the data cross-section measurement of $E_{e}$, $E_{\mathrm{visible}}$, $\cos{\theta_{ep}}$, and $N_{\mathrm{proton}}$ arises from the flux, with an average fractional uncertainty of 26.7\%. At low electron and visible energies (0.02 to 0.42\,GeV and 0.05 to 0.65\,GeV, respectively), the largest source of uncertainty is from the flux, specifically due to limitations in hadron production modeling. Statistical uncertainty dominates at higher energies above 0.42\,GeV for electron energy and 0.65\,GeV for visible energy due to limited events in the beam-on data samples. For non-energy variables, these contributions are comparable.  

\section{Cross Section Extraction}
\label{extraction} 

Flux-averaged differential cross sections are reported as a function of the outgoing electron energy, the total visible energy, and the cosine of the opening angle between the electron and the leading proton. In addition, the total cross section is reported. In the present article, we do not report a differential cross section as a function of proton multiplicity. While such a result would provide useful information for testing neutrino interaction models, the unfolding procedure needed to correct for imperfect multiplicity reconstruction is particularly sensitive to detection thresholds and modeling details. Fully addressing these challenges for a robust proton multiplicity measurement is therefore reserved for future work.

The total cross section, $\sigma$, is measured by evaluating the expression

\begin{equation}
\label{eqn:total_xsec_formula}
    \sigma = \frac{N - B}{\epsilon \times N_{\rm target} \times \Phi_{\nu_{e}}},
\end{equation}
\noindent where $N$ and $B$ are the number of selected events in data (203.0) and the estimated number of selected background events (47.6), $\epsilon$ is the selection efficiency (12.8\%), $N_{\rm target}$ is the number of collision targets in the fiducial volume and $\Phi_{\nu_{e}}$ is the integrated \nue{} flux prediction. No regularization is used for the total cross section. 

The number of collision targets in the detector $\left(N_{\rm target}\right)$ is equal to the number of nucleons within the fiducial volume that serve as potential targets for neutrino interactions. This is calculated using the assumed density of liquid argon ($1.38\,\mathrm{g}\,/\mathrm{cm}^{3}$), the MicroBooNE fiducial volume ($5.08\times10^7$\,cm$^3$), and the average mass per nucleon in argon $\left(0.999\,\mathrm{g}\,/\,\mathrm{mol}\right)$. The integrated \nue{} flux prediction $\Phi_{\nu_{e}}$ is estimated using a POT-weighted sum of the integrated FHC and RHC \nue{} flux predictions shown in Fig. \ref{fig:flux}. Integrated flux values and POT contributions from FHC and RHC are shown in Table \ref{tab:pot_flux}. Only the flux above 60\,MeV is considered; muon decay-at-rest events occurring at neutrino energies too low to contribute to the background-subtracted event rate are excluded. The resulting flux is included as part of the data release corresponding to this paper. 

The flux-integrated differential cross section, $\left\langle\frac{\mathrm{d}\sigma}{\mathrm{d}x}\right\rangle_{i}$, is given by

\begin{equation}
\label{eqn:xsec_formula}
    \left\langle\frac{\mathrm{d}\sigma}{\mathrm{d}x}\right\rangle_{i} = \frac{\Sigma_{j}U_{ij}\left(N_{j} - B_{j}\right)}{N_{\rm target} \times \Phi_{\nu_{e}} \times \Delta x_{i}},
\end{equation}
\newline
\noindent where $N_{j}$ and $B_{j}$ are the number of selected events in data and the estimated number of selected background events in the measured bin $j$, $N_{\rm target}$ is the number of collision targets in the fiducial volume, $\Phi_{\nu_{e}}$ is the integrated \nue{} flux prediction, and $\Delta x_{i}$ is the width of bin $i$ in the observed variable $x$. The term $U_{ij}$ is the unfolding matrix, and corrects for both bin-migration effects and selection efficiencies. The unfolding matrix provides an estimate of the true signal counts for each bin when applied to the background-subtracted data.


\begin{table}
\caption{\label{tab:pot_flux} Integrated flux values and POT contributions from Run~1 FHC and Run~3 RHC portions of the NuMI dataset. The integrated \nue{} flux includes a requirement that $E_{\nu} > 60$\,MeV to remove contributions from low-energy electron neutrinos produced by muon decays at rest in the absorber.}
\begin{ruledtabular}
\begin{tabular}{lll}
\textbf{Horn Current} & \textbf{Integrated \nue{} flux} & \textbf{NuMI POT}\\
& $\boldsymbol{\left[/\,\mathrm{cm}^{2}/\,\mathrm{POT}\right]}$ & \\
\hline \rule{0pt}{2.5ex}
FHC (Run~1) & $1.22\times 10^{-11}$ & $2.0\times10^{20}$ \\ \rule{0pt}{2.5ex}
RHC (Run~3) & $9.02\times 10^{-12}$ & $5.0\times10^{20}$

\end{tabular}
\end{ruledtabular}

\end{table}


The background-subtracted event rate $M$ is measured using reconstructed quantities. These reconstructed quantities are affected by selection efficiency, which impacts the event count in each bin, and detector response, which can shift reconstructed observables away from their true values (smearing). To meaningfully compare and constrain theoretical models with data, such effects need to be unfolded from the event rate via a response matrix $\bm{R}$ encoding the strength of the inefficiency and smearing. In theory, if $\bm{R}$ is invertible, this can be done via direct inversion.

In practice, directly inverting $\bm{R}$ can amplify small noise fluctuations in the data, leading to difficulty in interpreting the results. To mitigate this issue, it is common to regularize by introducing a controlled amount of bias in the final result to suppress amplified variance.

In this work, differential cross sections are unfolded with the Wiener singular value decomposition (Wiener-SVD) method \cite{wsvd}, which takes advantage of deconvolution techniques traditionally used in digital signal processing and has been employed for previous results within the MicroBooNE Collaboration \cite{numi-inclusive, wsvd-numu, TKI1, TKI2}. In Wiener SVD, the degree of regularization is determined by minimizing the sum of variance and the introduced bias of the result. An advantage of the Wiener-SVD method is that $\chi^{2}$ is independent of the regularization, and thus remains consistent before and after unfolding. This is important to accurately assess the validity of model predictions used in neutrino studies. 

As input, the Wiener-SVD method requires a background-subtracted event rate, an associated covariance matrix (combining the statistical uncertainty from data with systematic uncertainties), a response matrix quantifying the selection efficiency and post-selection smearing, and a signal model for constructing the Wiener filter. The systematic uncertainties and the response matrix are both derived from the MicroBooNE tune prediction. See Supplemental Material for the covariance and response matrices \cite{SupplementalMaterial}.

The method returns an unfolded event rate $\hat{s}$ and covariance matrix, both of which are converted to cross-section units via normalization by flux, bin width, and $N_{\rm target}$. An additional smearing matrix that encodes the introduced bias, $\bm{A_c}$, is also returned, which can be applied to theoretical predictions to enable a direct comparison with the unfolded result. A series of closure tests and fake data studies were performed to validate the Wiener-SVD method before applying it to NuMI data.

\section{Results}
\label{results}

\begin{figure}
\centering

\begin{tikzpicture} \draw (0, 0) node[inner sep=0] {
\includegraphics[width=.44\textwidth]{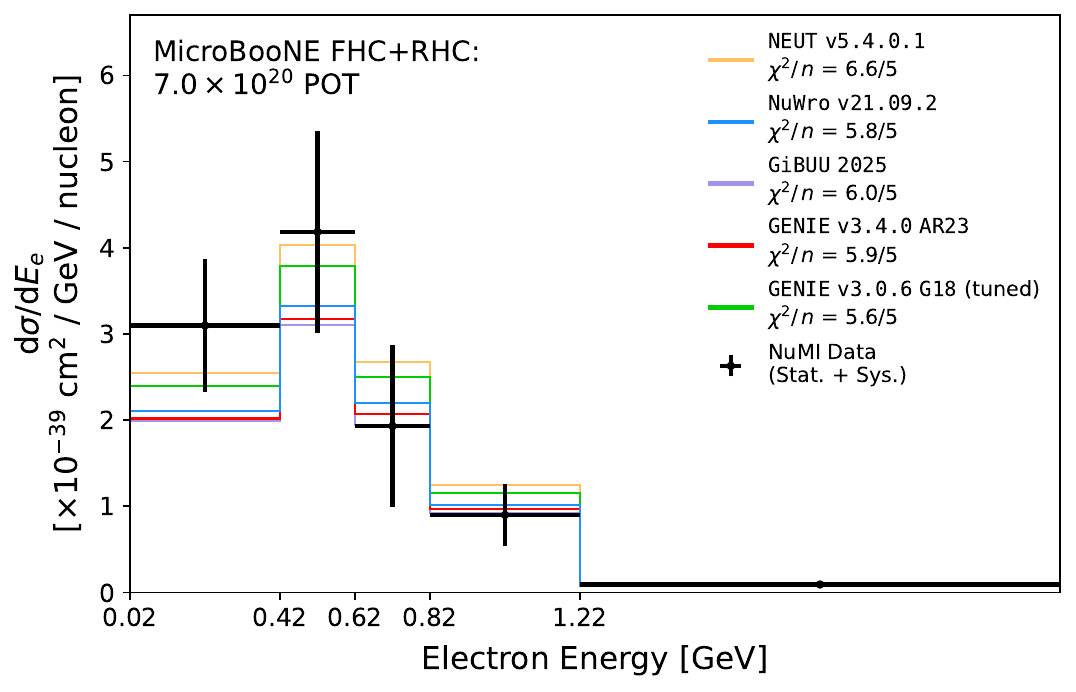}};
  \draw (-2.7, -1.55) node {\textbf{(a)}};
\end{tikzpicture}

\begin{tikzpicture} \draw (0, 0) node[inner sep=0] {
\includegraphics[width=.44\textwidth]{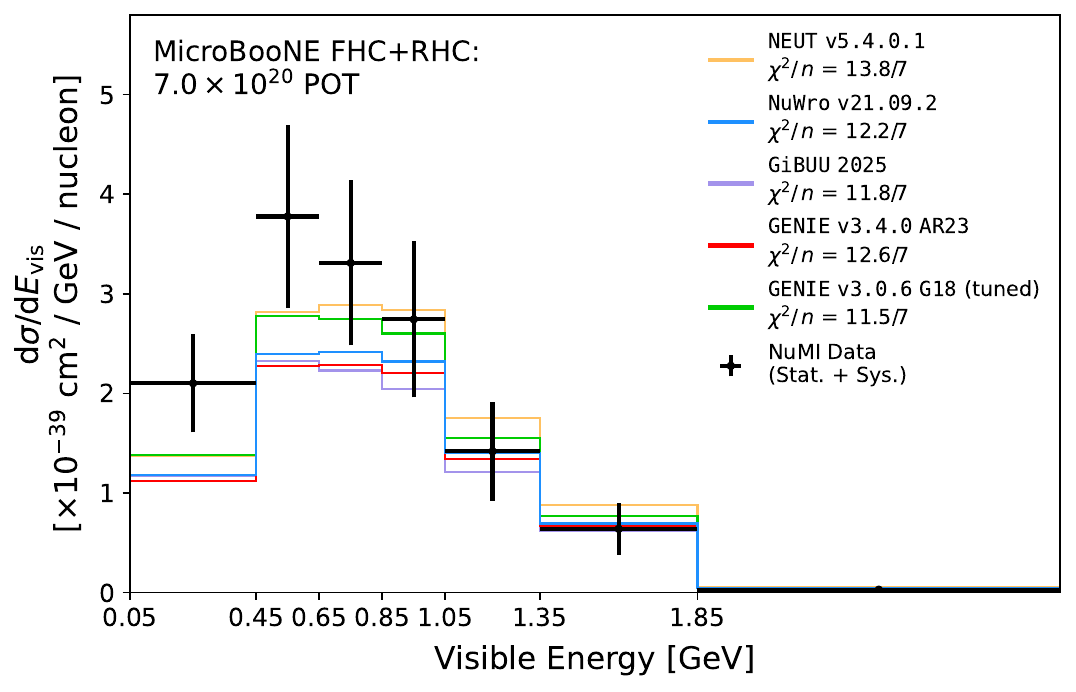}};
  \draw (-2.7, -1.5) node {\textbf{(b)}};
\end{tikzpicture}

\begin{tikzpicture} \draw (0, 0) node[inner sep=0] {
\includegraphics[width=.46\textwidth]{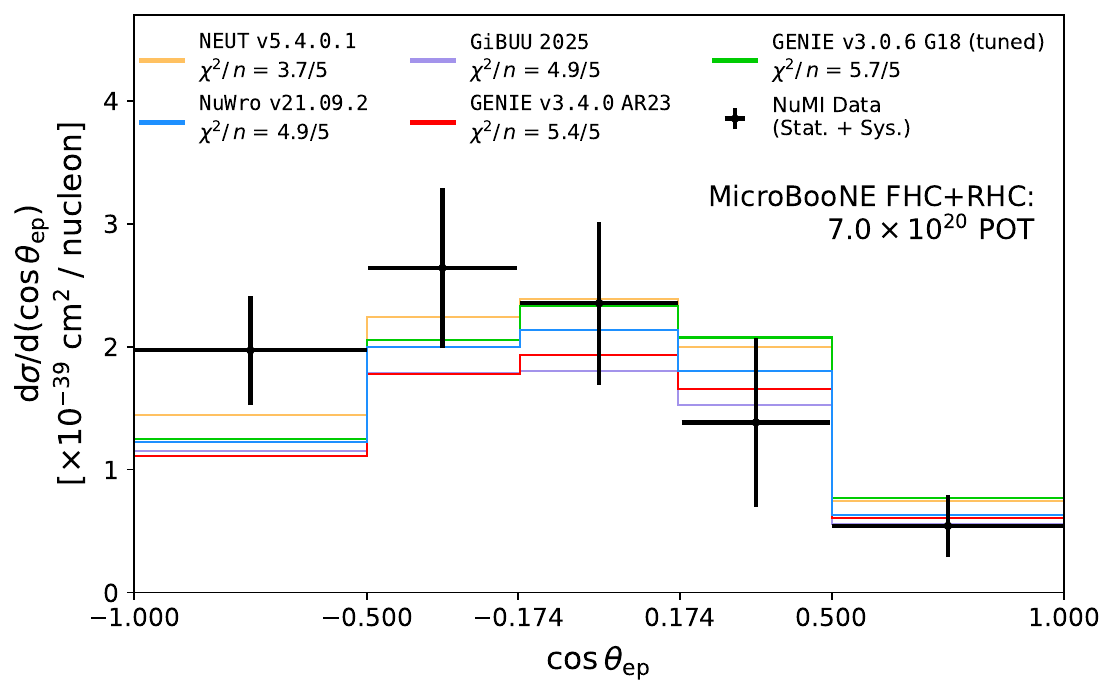}};
  \draw (-2.7, -1.45) node {\textbf{(c)}};
\end{tikzpicture}

\caption{\label{fig:results} {The unfolded NuMI data result (black) as a function of electron energy (a), visible energy (b), and $\cos{\theta_{ep}}$ (c), displayed with unfolded statistical and systematic uncertainties. Comparisons are made to predictions from \code{NEUT v5.4.0.1} (orange), \code{NuWro v.21.09.2} (blue), \code{GiBUU 2025} (purple), \code{GENIE v03.4.2 AR23} (red), and \code{GENIE v03.0.6 G18} tuned (green).}}
\end{figure}
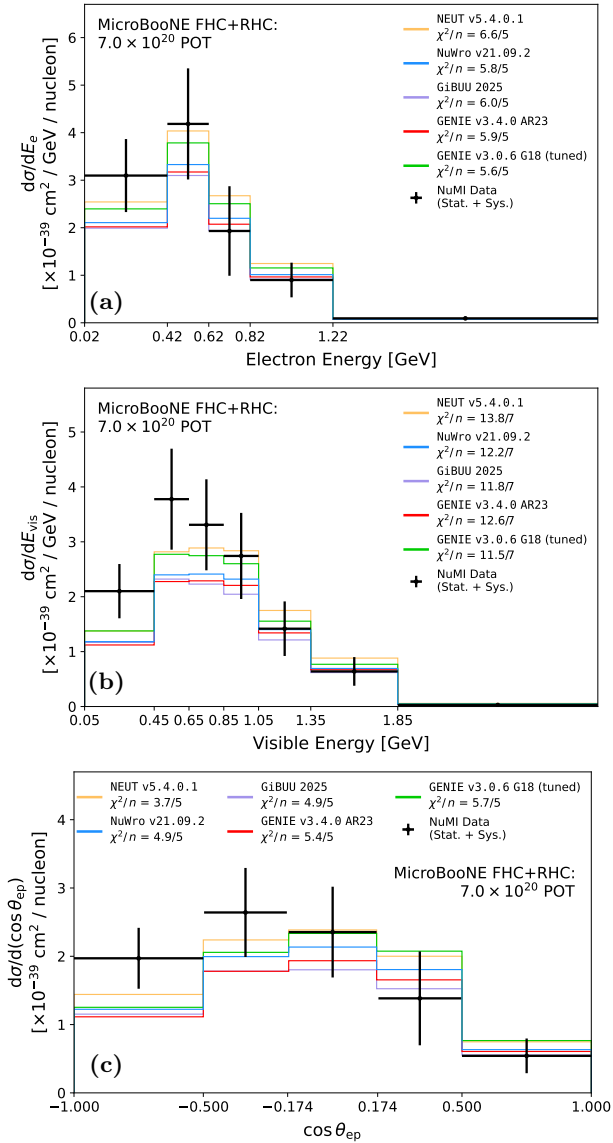

The total cross section is measured to be $\left[4.1 \pm 0.3 \, \mathrm{(stat.) \pm 1.1 \, \mathrm{(syst.)}}\right] \times 10^{-39}\;\mathrm{cm}^2/\,\mathrm{{nucleon}}$. The unfolded differential cross sections in electron energy $\left(E_{e}\right)$, visible energy $\left(E_{\mathrm{visible}}\right)$, and the cosine of the opening angle $\left(\cos\:\theta_{ep}\right)$ are shown in in Fig. \ref{fig:results}. %

The unfolded results are compared to predictions from several commonly used neutrino event generators: \code{NEUT v5.4.0.1} \cite{neut1, neut2}, \code{NuWro v.21.09.2} \cite{nuwro1, nuwro2}, \code{GiBUU 2025} \cite{gibuu}, \code{GENIE v03.4.2 AR23\_20i\_00\_000} (labeled \code{GENIE v03.4.2 AR23}) \cite{genie-untuned}, and the MicroBooNE \code{GENIE} tune (labeled \code{GENIE v03.0.6 G18} (tuned)) \cite{genie_tune}. These predictions were generated using the signal definition described at the beginning of Sec. \ref{selection}. They were processed using the $\texttt{NUISANCE}$ comparison framework \cite{nuisance}. All predictions have been smeared by $\bm{A_c}$ to enable a direct comparison with the unfolded cross section. Agreement with different generators is assessed with a $\chi^2/n$ metric using the covariance of the unfolded data result, available in the Supplemental Material \cite{SupplementalMaterial} along with the additional smearing matrices.


Table \ref{tab:total_xsec} shows a comparison of the measured total cross section to each neutrino event generator. All generator predictions are in good agreement with the measured result within statistical and systematic uncertainties. The data slightly prefers the higher cross section predicted by \code{NEUT v5.4.0.1} compared to the lower cross sections predicted by \code{GENIE v03.0.6 G18} (tuned), \code{NuWro v.21.09.2}, \code{GENIE v03.4.2 AR23}, and \code{GiBUU 2025}. 


\begin{table}
\caption{\label{tab:total_xsec}Comparison of the total unfolded cross section to commonly employed neutrino event generators.}
\begin{ruledtabular}
\begin{tabular}{lcc}
\textbf{Generator} & $\boldsymbol{\sigma\;[10^{-39}\;\mathrm{cm}^2/\,\mathrm{{nucleon}]}}$ & $\boldsymbol{\chi^2/\,n}$ \\
\hline\hline
\textbf{Unfolded Data} & \multicolumn{2}{c}{\small\begin{tabular}[c]{@{}c@{}}4.1 $\pm$ 0.3 (stat.) $\pm$ 1.1 (syst.)\end{tabular}} \\
\hline\hline
\code{NEUT v5.4.0.1}         & 4.2& 0.0/1 \\
\code{NuWro v.21.09.2}          & 3.4& 0.3/1 \\
\code{GiBUU 2025}            & 3.2& 0.6/1 \\
\code{GENIE v03.4.2 AR23}      & 3.3& 0.5/1 \\
\code{GENIE v03.0.6 G18} (tuned) & 3.9& 0.0/1 \\
\end{tabular}
\end{ruledtabular}
\end{table}


Table \ref{tab:chi2} summaries the generator comparisons for the differential cross sections in $E_{\mathrm{e}}$, $E_{\mathrm{visible}}$, and $\cos(\theta_{\mathrm{ep}})$ shown in Fig. \ref{fig:results}. Across all variables, the model predictions agree with the data within $2\sigma$. For the electron energy variable, the $\chi^2/n$ ranges between 5.6/5 (\code{GENIE v3.0.6 G18} tuned) and 6.6/5 (\code{NEUT v5.4.0.1}). For the visible energy variable, the $\chi^2/n$ ranges between 11.5/7 (\code{GENIE v3.0.6 G18} tuned) and 13.8/7 (\code{NEUT v5.4.0.1}). Finally, for the cosine of the opening angle variable, the $\chi^2/n$ ranges between 3.7/5 (\code{NEUT v5.4.0.1}) and 5.7/5 (\code{GENIE v3.0.6 G18} tuned). Slightly poorer agreement is seen in the visible energy compared with the other two variables, with model predictions agreeing with data within $1.9\sigma$  compared to $1.2\sigma$ for electron energy and $1.0\sigma$ for the cosine of the opening angle. Overall, the results suggest that existing generators effectively model the data within the sensitivity of this measurement.


\begin{table}
\caption{\label{tab:chi2} $\chi^2/n$ values comparing the measured differential cross sections to commonly employed neutrino event generators, computed using the covariance of the unfolded result. }
\begin{ruledtabular}
\begin{tabular}{llll}
\textbf{Generator} & $\boldsymbol{E_{e}}$ & $\boldsymbol{E_{\mathrm{visible}}}$ & $\boldsymbol{\cos{\theta_{ep}}}$\\
\hline
\code{NEUT v5.4.0.1} & 6.6/5 & 13.8/7 & 3.7/5  \\
\code{NuWro v.21.09.2} & 5.8/5 & 12.2/7 & 4.9/5 \\
\code{GiBUU 2025} & 6.0/5 & 11.8/7 & 4.9/5 \\
\code{GENIE v03.4.2 AR23} & 6.9/5 & 12.6/7 & 5.4/5 \\
\code{GENIE v03.0.6 G18} (tuned) & 5.6/5 & 11.5/7 & 5.7/5 \\

\end{tabular}
\end{ruledtabular}

\end{table}


Compared with the first exclusive \nue{} \npsel{} cross-section measurement in the BNB \cite{bnb-exclusive}, this analysis observes model underprediction in electron energy, whereas the earlier measurement observed model overprediction in the same observable. However, these differences are not significant, and both results agree within 1$\sigma$. The two measurements differ in several important respects: they are performed using data from two different neutrino beams with different neutrino energy spectra, are subject to different dominant uncertainties, employ different unfolding techniques (D'Agostini versus Wiener-SVD), and compare the data to different versions of neutrino event generators. Despite these differences, both measurements observe overall agreement between data and generator predictions within uncertainties.

\section{Conclusion \& Outlook}

This work presents the extraction of single-differential electron-neutrino charged-current cross sections on argon with at least one proton and no pions in the final topology. Data from the interactions of particles from the NuMI beam at the Fermi National Accelerator Laboratory was collected using the MicroBooNE detector. A robust event selection and full evaluation of uncertainties are developed to extract exclusive differential cross sections as a function of outgoing electron energy, visible energy, and the cosine of the opening angle between the electron and the most energetic proton. Interaction rates and associated uncertainties as a function of proton multiplicity are also reported. 

The total cross section is measured to be $\left[4.1 \pm 0.3 \, \mathrm{(stat.) \pm 1.1 \, \mathrm{(syst.)}}\right] \times 10^{-39}\;\mathrm{cm}^2/\,\mathrm{{nucleon}}$. Predictions from commonly employed neutrino event generators show good compatibility with data for the electron energy, visible energy, and cosine of the opening angle variables. 

One of the dominant sources of uncertainty is the neutrino flux, which is challenging to model due to the MicroBooNE detector being significantly off-axis to the NuMI beam. Additional hadron production data covering the off-axis phase space could help to constrain this uncertainty in the future and substantially reduce the associated uncertainties. As is often the case for measurements with exclusive signal definitions, a leading limitation of this analysis is event statistics. Future studies in MicroBooNE will leverage approximately three times the currently available NuMI dataset and will be able to extract individual electron-neutrino cross sections on argon from the forward horn current and reverse horn current portions of MicroBooNE’s NuMI dataset. Aside from increased statistics, MicroBooNE will see improved reconstruction developments, where a greater efficiency will provide even higher statistics than with the full dataset alone.

\section*{ACKNOWLEDGEMENTS}

This document was prepared by the MicroBooNE collaboration using the resources of the Fermi National Accelerator Laboratory (Fermilab), a U.S. Department of Energy, Office of Science, Office of High Energy Physics HEP User Facility. Fermilab is managed by Fermi Forward Discovery Group, LLC, acting under Contract No. 89243024CSC000002. MicroBooNE is supported by the following: the U.S. Department of Energy, Office of Science, Offices of High Energy Physics and Nuclear Physics; the U.S. National Science Foundation; the Swiss National Science Foundation; the Science and Technology Facilities Council (STFC), part of United Kingdom Research and Innovation (UKRI); the Royal Society (United Kingdom); the UKRI Future Leaders Fellowship; the NSF AI Institute for Artificial Intelligence and Fundamental Interactions; the Ford Foundation Predoctoral Fellowship; and the European Union’s Horizon 2020 research and innovation programme under the Marie Sk\l{}odowska-Curie grant agreement No. 101003460 (PROBES).

Additional support for the laser calibration system and cosmic ray tagger was provided by the Albert Einstein Center for Fundamental Physics, Bern, Switzerland. We also acknowledge the contributions of technical and scientific staff to the design, construction, and operation of the MicroBooNE detector as well as the contributions of past collaborators to the development of MicroBooNE analyses, without whom this work would not have been possible. 

For the purpose of open access, the authors have applied a Creative Commons Attribution (CC BY) public copyright license to any Author Accepted Manuscript version arising from this submission.

\bibliography{references_main}

\end{document}


\title{Measurements of the differential charged current cross section on argon for electron neutrinos with final-state protons in MicroBooNE: Supplemental Material}
   
\date{\today}

\maketitle

\section{\numu{} Charged Current Background Rejection}

\begin{figure}[h]
\centering

\begin{tikzpicture} \draw (0, 0) node[inner sep=0] {
\includegraphics[width=.45\textwidth]{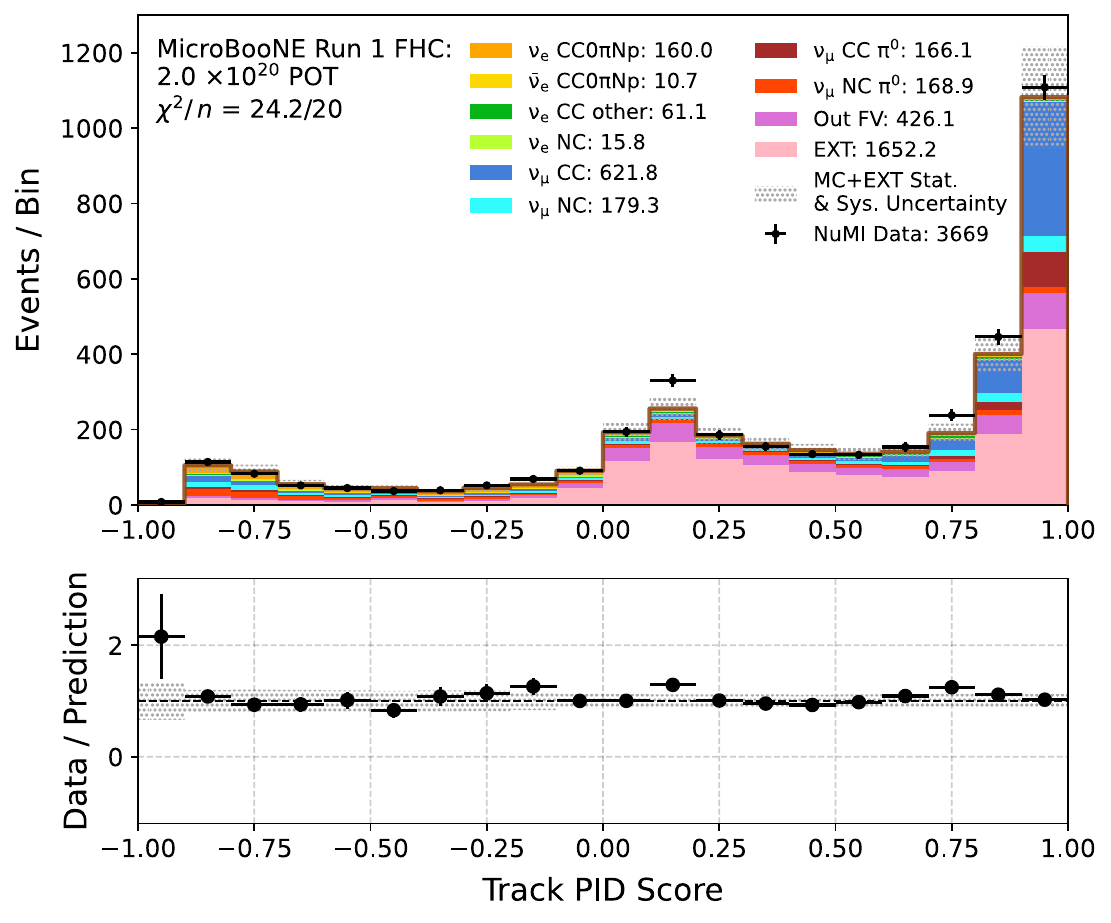}};
  \draw (-2.5, -1.15) node {\textbf{(a)}};
\end{tikzpicture}
\begin{tikzpicture} \draw (0, 0) node[inner sep=0] {
\includegraphics[width=.45\textwidth]{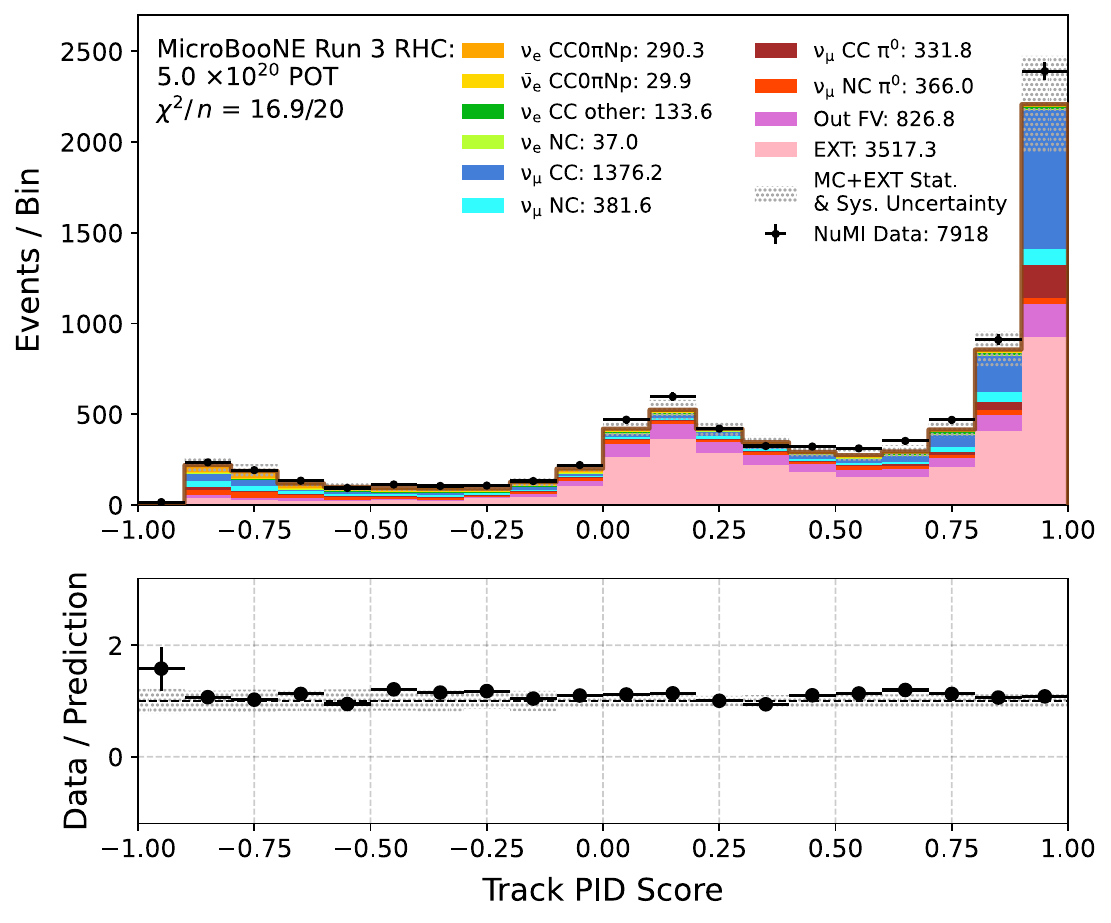}};
  \draw (-2.5, -1.15) node {\textbf{(b)}};
\end{tikzpicture}
\caption{\label{trkpid_presel}{Run~1 FHC (a) and Run~3 RHC (b) event rates as a function of the log-likelihood track PID score for interactions passing the preselection.}}
\end{figure}

\begin{figure}[h]
\centering
\begin{tikzpicture} \draw (0, 0) node[inner sep=0] {
\includegraphics[width=.45\textwidth]{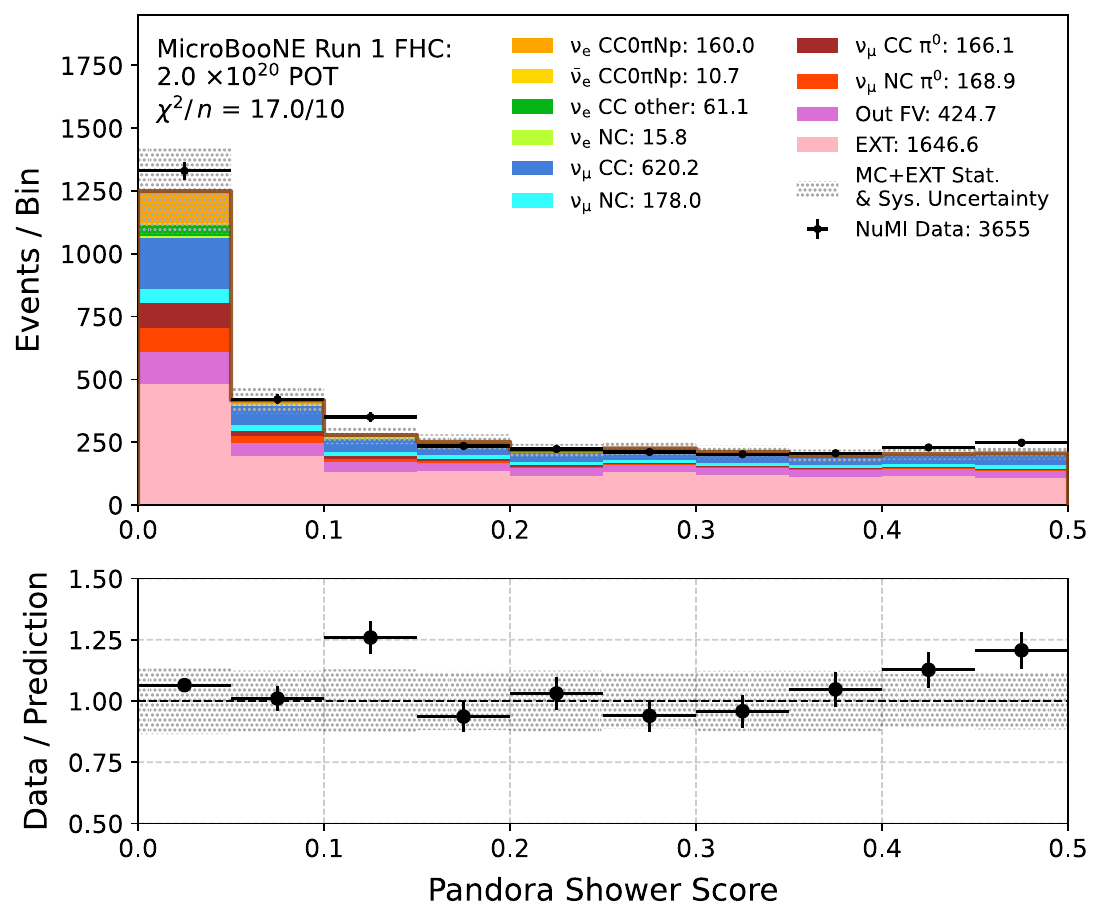}};
  \draw (-2.7, -1.15) node {\textbf{(a)}};
\end{tikzpicture}
\begin{tikzpicture} \draw (0, 0) node[inner sep=0] {
\includegraphics[width=.45\textwidth]{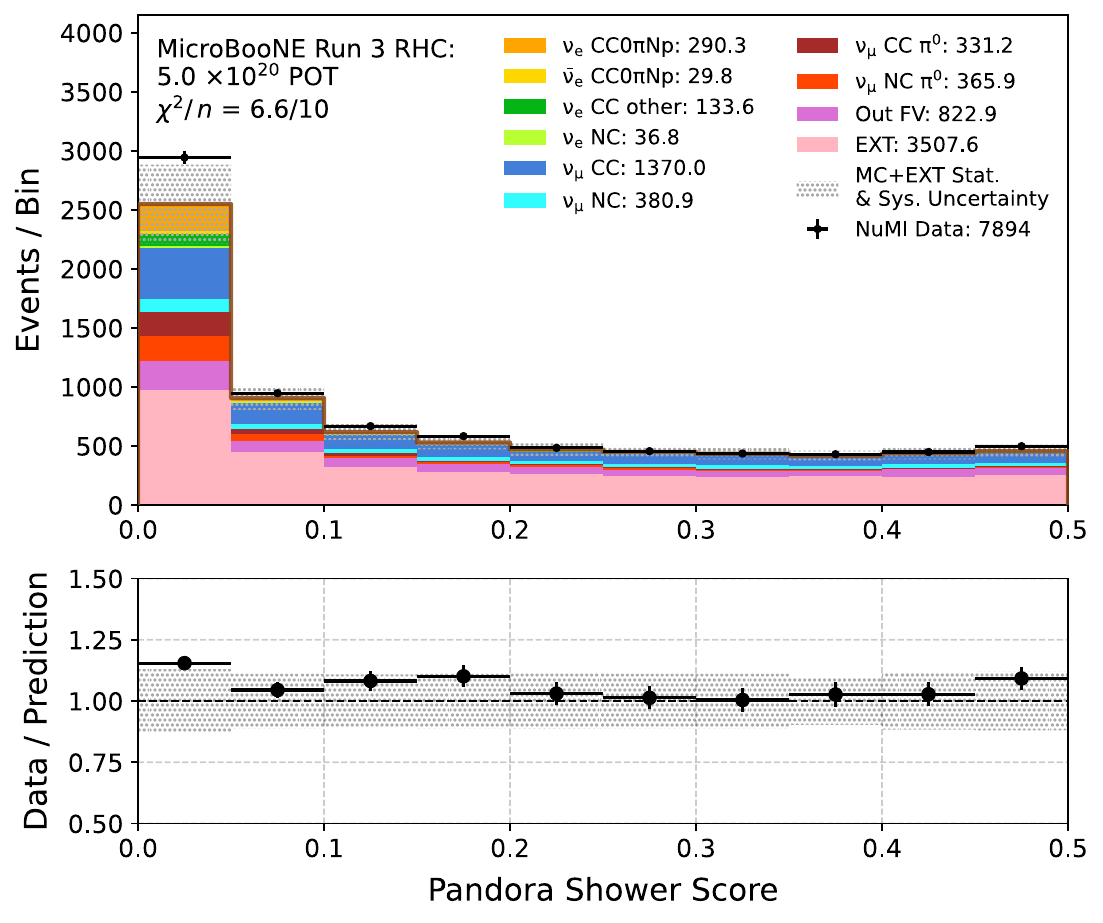}};
  \draw (-2.5, -1.15) node {\textbf{(b)}};
\end{tikzpicture}
\caption{\label{shr_score_presel}{Run~1 FHC (a) and Run~3 RHC (b) event rates as a function of \code{Pandora} shower score for interactions passing the preselection.}}
\end{figure}

Plots in this section illustrate the agreement between NuMI prediction and data for the reconstructed observables used to reject \numu{} CC interactions. Figures \ref{trkpid_presel} and \ref{shr_score_presel} show the Run~1 FHC and Run~3 RHC event rates as a function of the log-likelihood track particle identification (PID) score and \code{Pandora} shower score, respectively.


The predicted event rate is broken down into the signal channel (orange), \numu/\numubar{} backgrounds with (red) and without (blue) a \pizero{} present in the final topology, and \nue/\nuebar{} backgrounds (green). Irreducible \nuebar{} contamination ($\bar{\nu}_{e}+\,^{40}{\rm Ar} \rightarrow 1e+Np+0\pi$), which is indistinguishable from signal events in the LArTPC, is shown separately in yellow. EXT (pink) refers to beam-off data, and Out FV refers to simulated neutrino events that occur outside of the fiducial volume. Beam-on data (black) is overlaid for comparison.

The gray error band represents the statistical and systematic uncertainty of the prediction. Detector systematic uncertainty is not included in this band, nor in the calculation of $\chi^{2}/n$.

\clearpage

\section{\pizero{} Background Rejection}

\begin{figure}[h]
\centering
\begin{tikzpicture} \draw (0, 0) node[inner sep=0] {
\includegraphics[width=.49\textwidth]{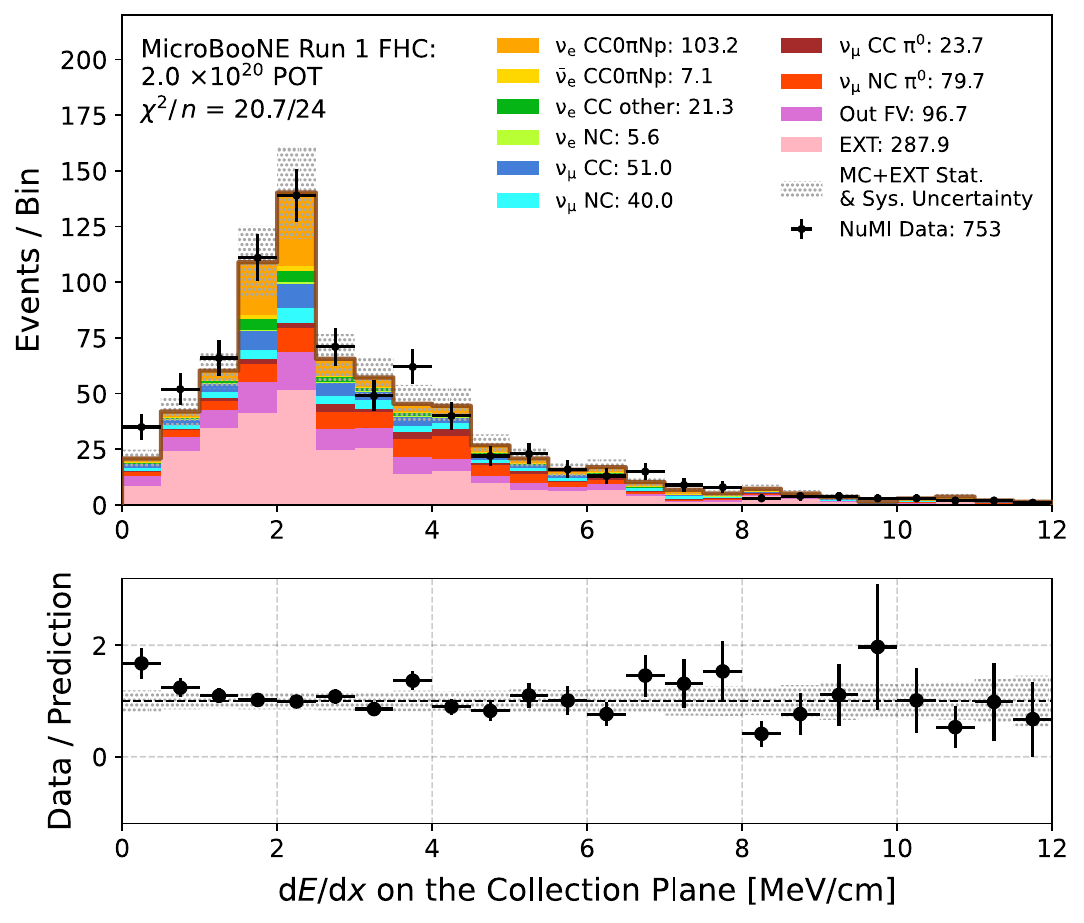}};
  \draw (-3, -1.2) node {\textbf{(a)}};
\end{tikzpicture}
\hfill
\begin{tikzpicture} \draw (0, 0) node[inner sep=0] {
\includegraphics[width=.49\textwidth]{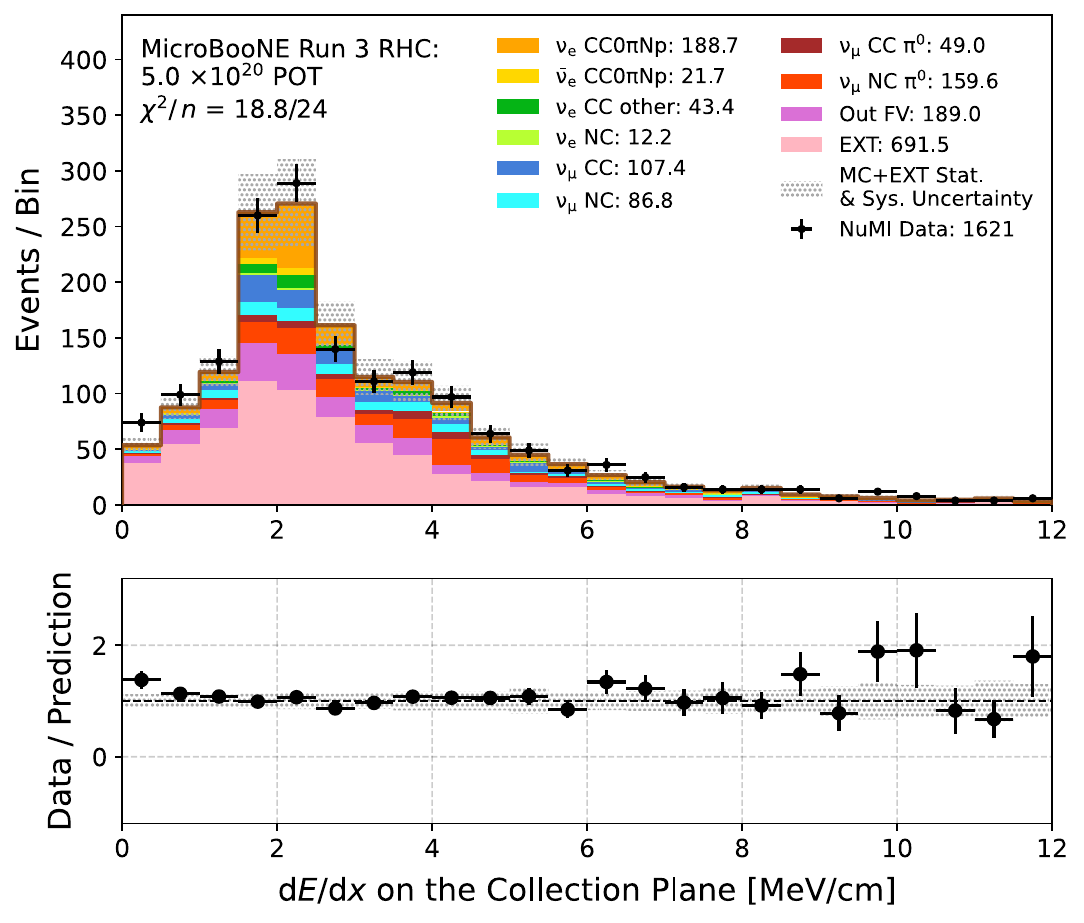}};
  \draw (-3.1, -1.2) node {\textbf{(b)}};
\end{tikzpicture}
\hfill
\caption{\label{dedx}{Run~1 FHC (a) and Run~3 RHC (b) event rates as a function of $\mathrm{d}E/\mathrm{d}x$ on the collection plane for interactions passing the preselection and \numu{} CC background rejection.}}
\end{figure}

\begin{figure}[h]
\centering
\begin{tikzpicture} \draw (0, 0) node[inner sep=0] {
\includegraphics[width=.49\textwidth]{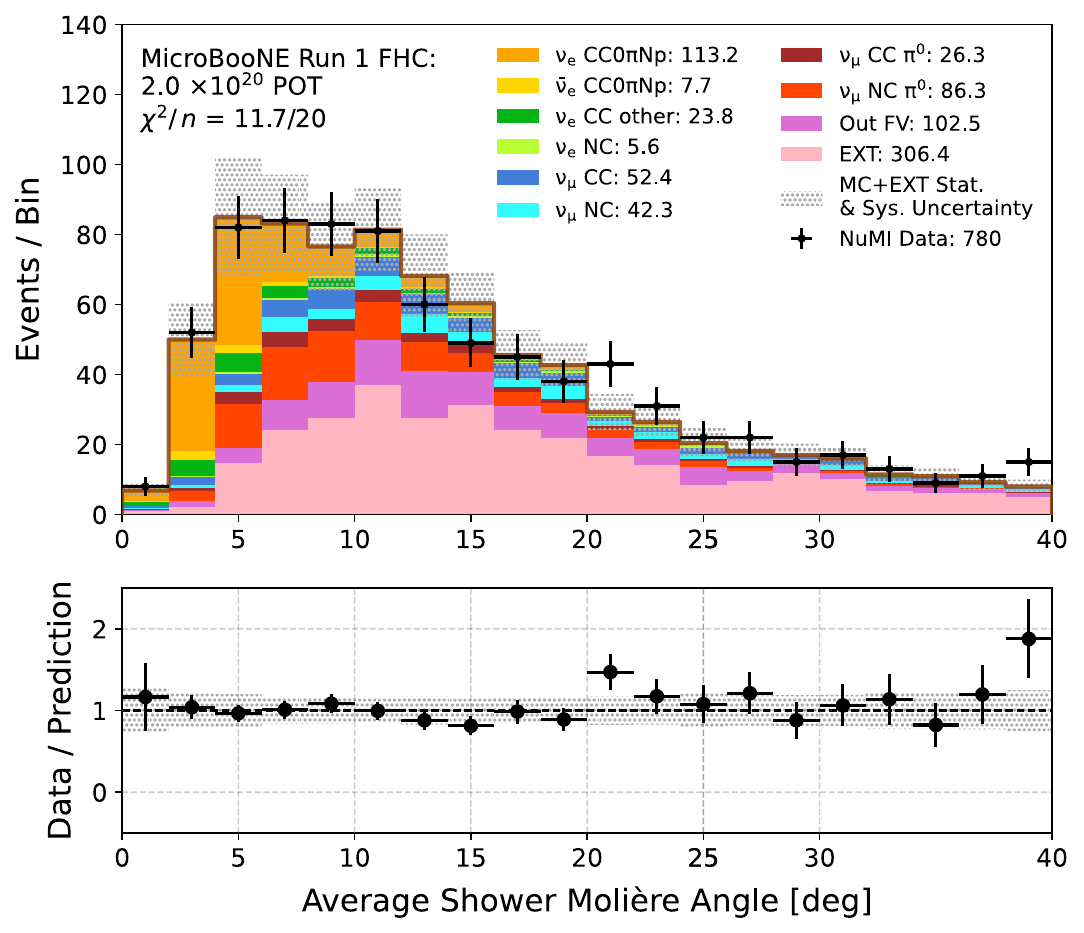}};
  \draw (-3.1, -1.2) node {\textbf{(a)}};
\end{tikzpicture}
\hfill
\begin{tikzpicture} \draw (0, 0) node[inner sep=0] {
\includegraphics[width=.49\textwidth]{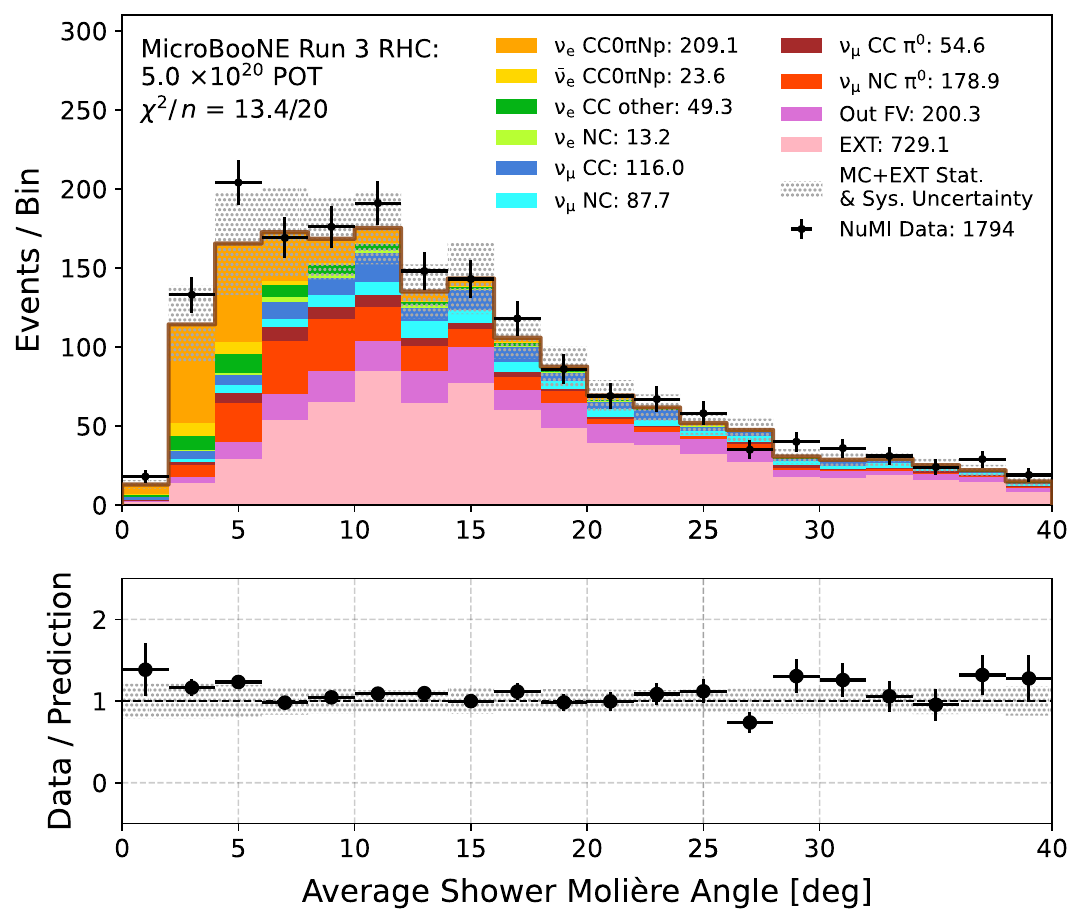}};
  \draw (-3.1, -1.2) node {\textbf{(b)}};
\end{tikzpicture}
\hfill
\caption{\label{moliere}{Run~1 FHC (a) and Run~3 RHC (b) event rates as a function of the average shower Moli\`ere angle for interactions passing the preselection and \numu{} CC background rejection.}}
\end{figure}

\begin{figure}[h]
\centering
\begin{tikzpicture} \draw (0, 0) node[inner sep=0] {
\includegraphics[width=.49\textwidth]{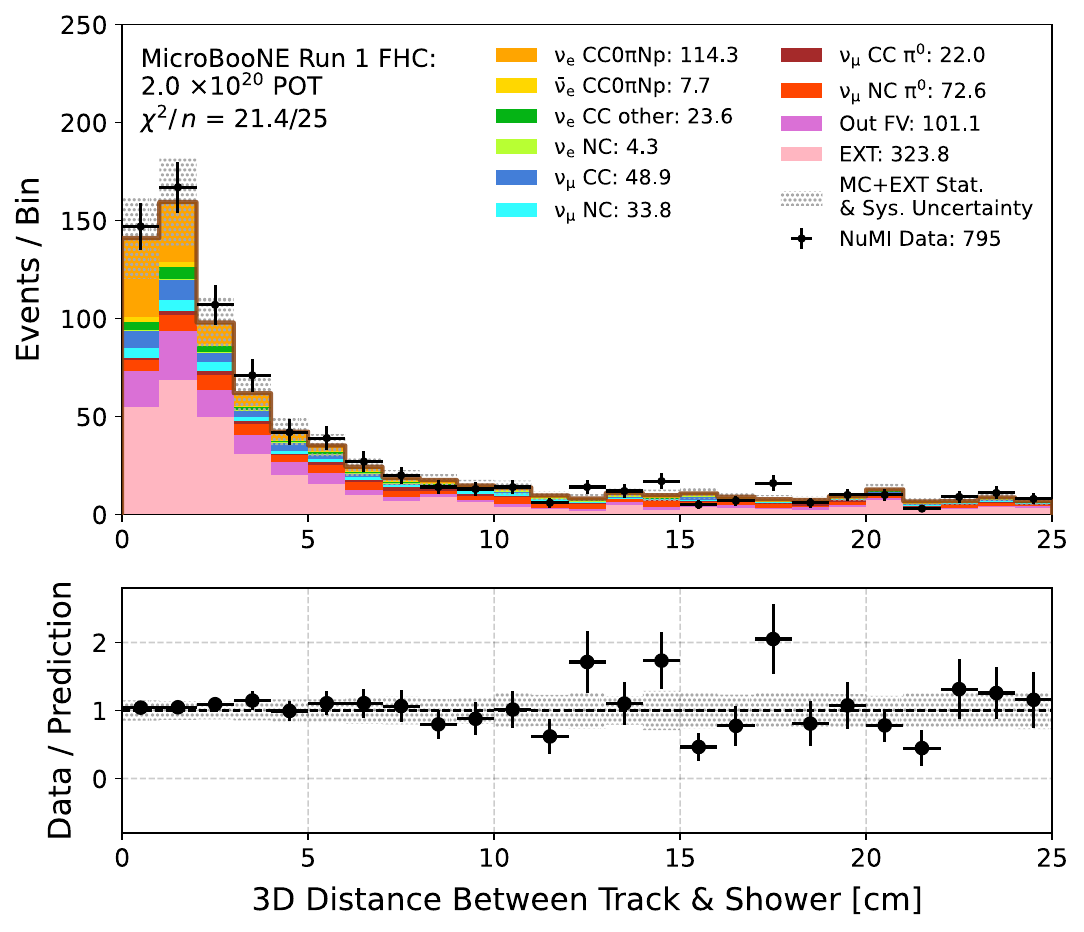}};
  \draw (-3.1, -1.2) node {\textbf{(a)}};
\end{tikzpicture}
\hfill
\begin{tikzpicture} \draw (0, 0) node[inner sep=0] {
\includegraphics[width=.49\textwidth]{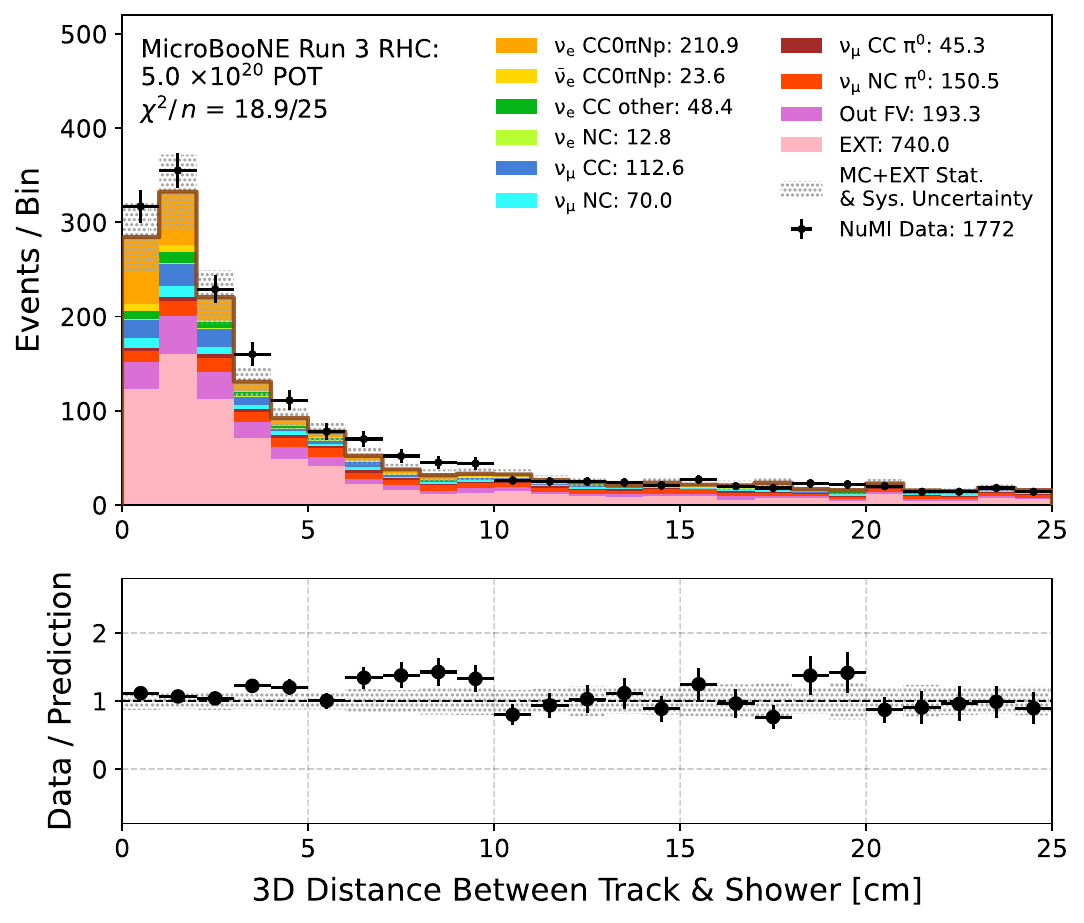}};
  \draw (-3.1, -1.2) node {\textbf{(b)}};
\end{tikzpicture}
\hfill
\caption{\label{tksh_distance}{Run~1 FHC (a) and Run~3 RHC (b) event rates as a function of the three-dimensional distance between the reconstructed shower and leading track objects of the interaction. Shown for events passing the preselection and \numu{} CC background rejection.}}
\end{figure}

Plots in this section illustrate the agreement between NuMI prediction and data for the reconstructed observables used to reject interactions with a \pizero{}. Figures \ref{dedx}, \ref{moliere}, and \ref{tksh_distance} show the Run~1 FHC and Run~3 RHC event rates as a function of $\mathrm{d}E/\mathrm{d}x$ on the collection plane, the three-dimensional distance between the interaction vertex and the start of the shower, and the average shower Moli\`ere angle, respectively. Detector systematic uncertainty is not included in the gray error band nor in the calculation of $\chi^2/n$.


\clearpage

\section{BDT Training Variables}

\begin{figure}[h]
\centering
\begin{tikzpicture} \draw (0, 0) node[inner sep=0] {
\includegraphics[width=.49\textwidth]{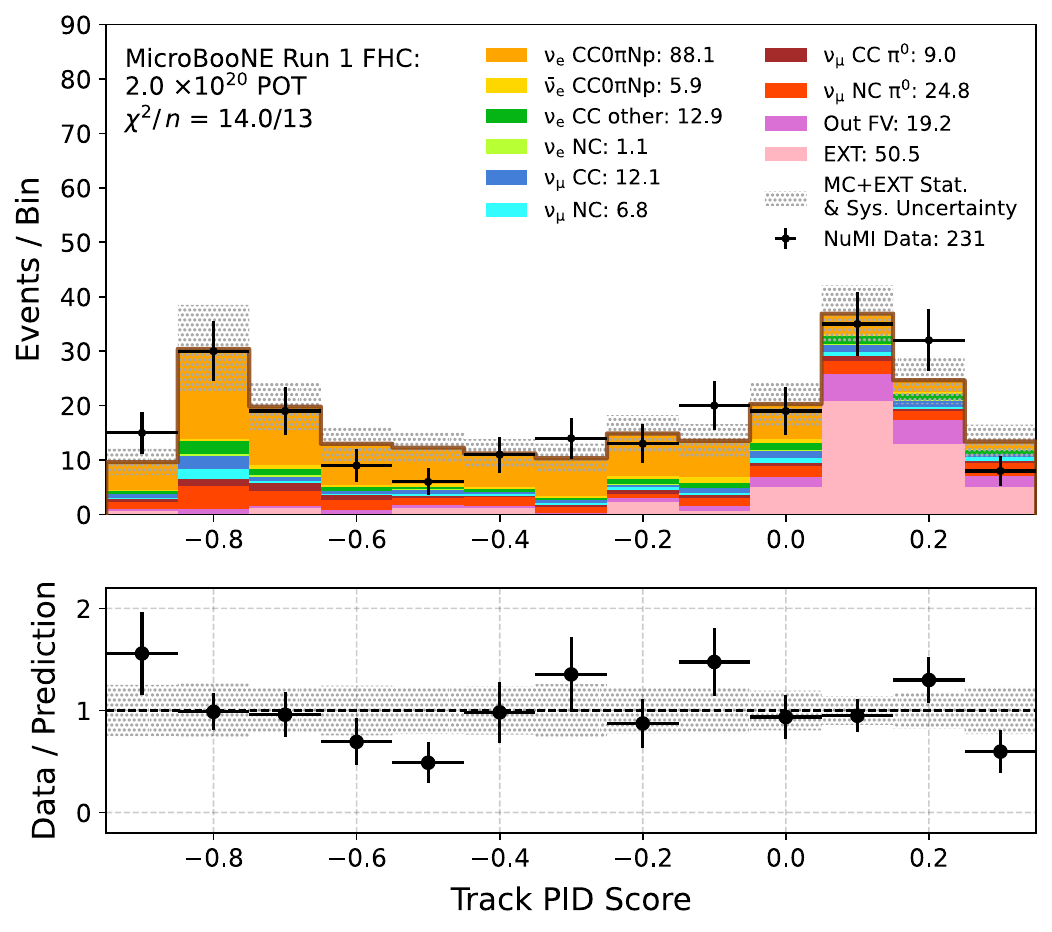}};
  \draw (-2.8, -1.25) node {\textbf{(a)}};
\end{tikzpicture}
\hfill
\begin{tikzpicture} \draw (0, 0) node[inner sep=0] {
\includegraphics[width=.49\textwidth]{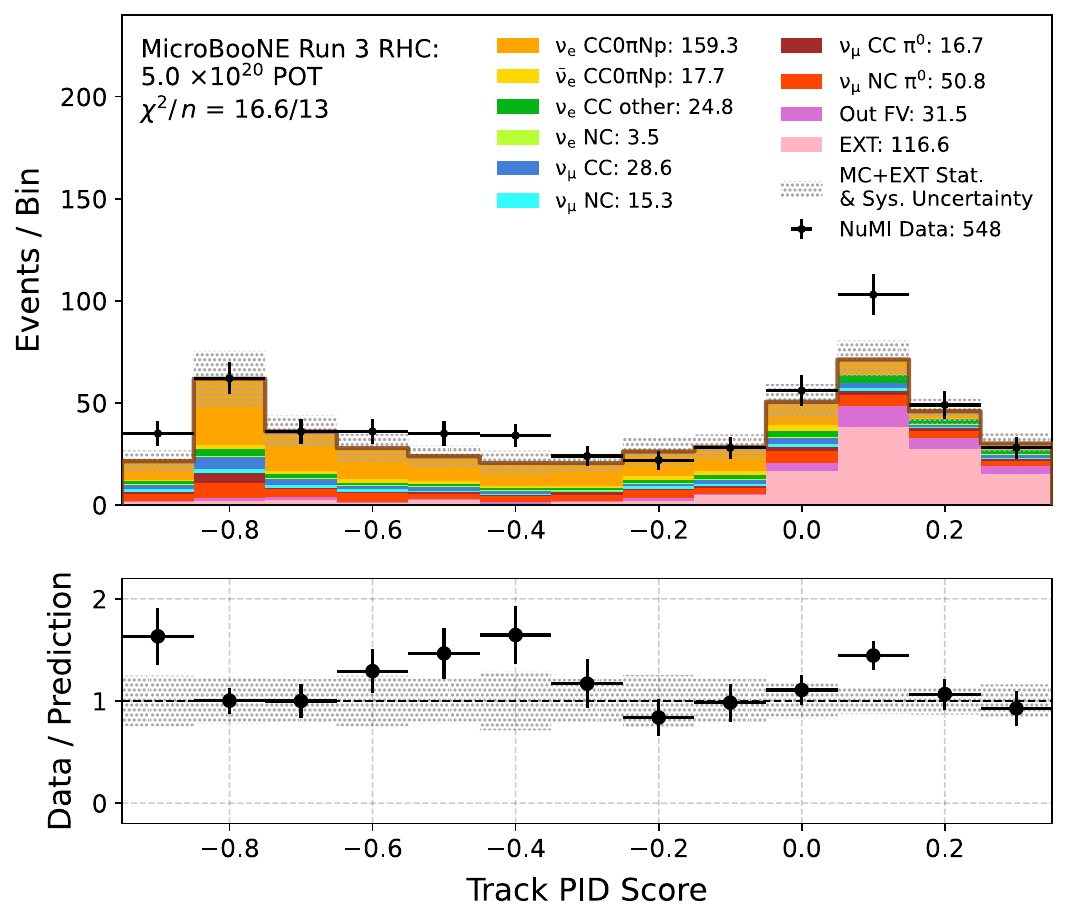}};
  \draw (-2.8, -1.25) node {\textbf{(b)}};
\end{tikzpicture}
\hfill
\caption{\label{trkpid_bdt}{Run~1 FHC (a) and Run~3 RHC (b) event rates as a function of track PID score. Shown for interactions passing the preselection, \numu{} CC rejection, \pizero{} rejection, and additional BDT training constraints.}}
\end{figure}

\begin{figure}[h]
\centering
\begin{tikzpicture} \draw (0, 0) node[inner sep=0] {
\includegraphics[width=.49\textwidth]{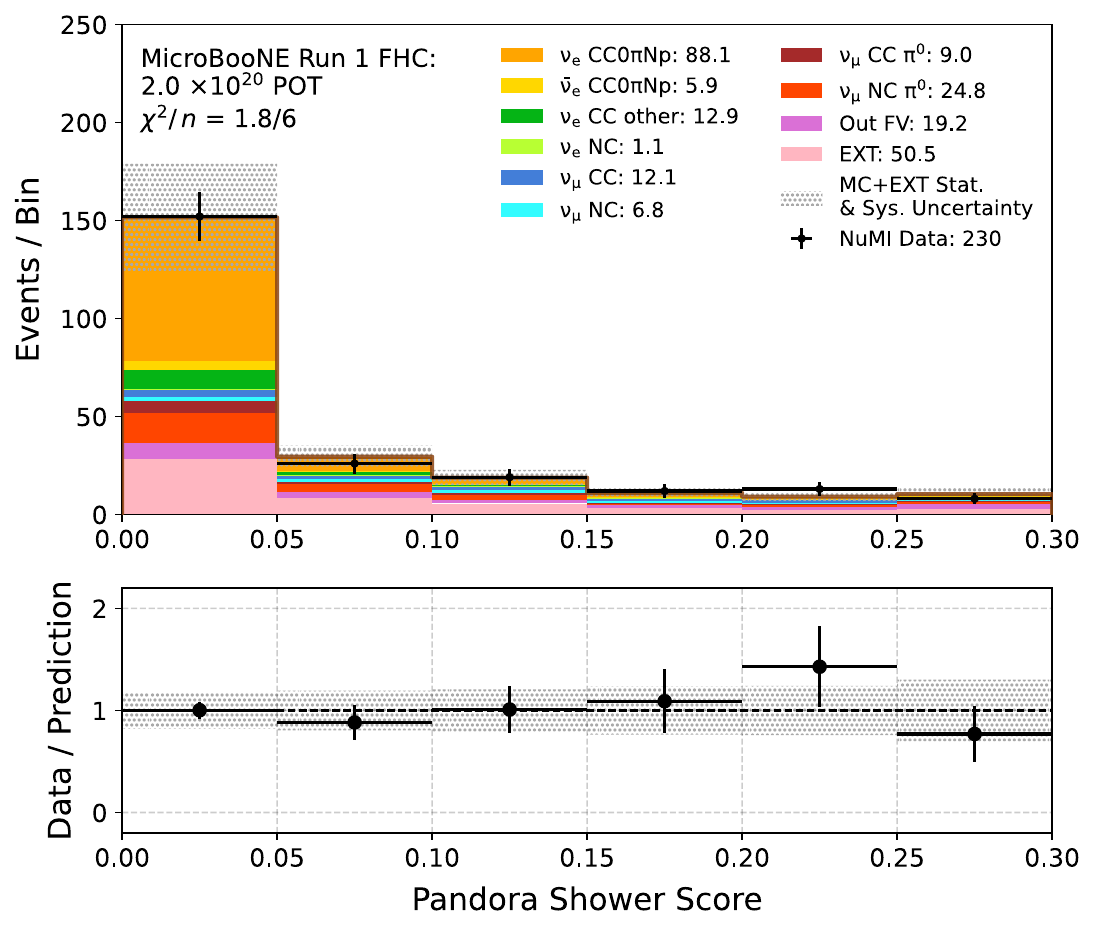}};
  \draw (-3.1, -1.2) node {\textbf{(a)}};
\end{tikzpicture}
\hfill
\begin{tikzpicture} \draw (0, 0) node[inner sep=0] {
\includegraphics[width=.49\textwidth]{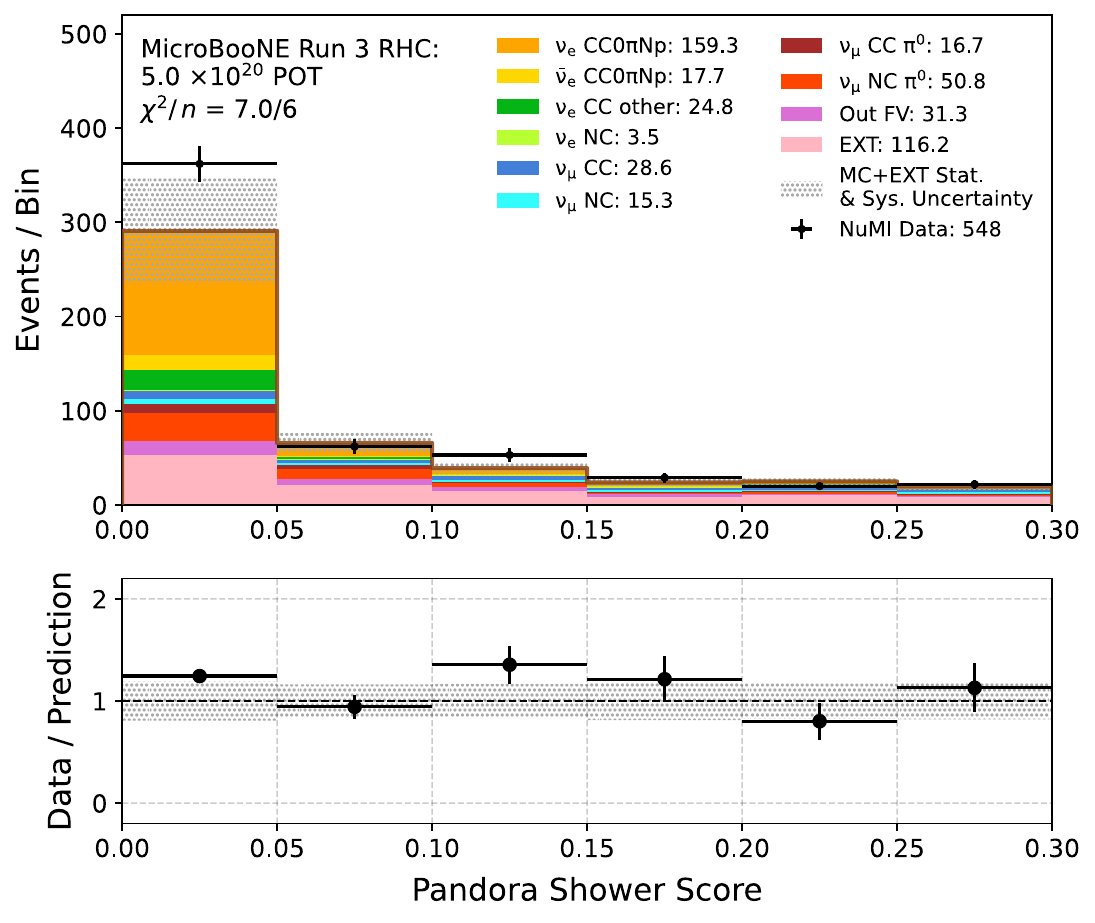}};
  \draw (-3.1, -1.2) node {\textbf{(b)}};
\end{tikzpicture}
\hfill
\caption{\label{shr_score_bdt}{Run~1 FHC (a) and Run~3 RHC (b) event rates as a function of \code{Pandora} shower score. Shown for interactions passing the preselection, \numu{} CC rejection, \pizero{} rejection, and additional BDT training constraints.}}
\end{figure}

\begin{figure}[h]
\centering
\begin{tikzpicture} \draw (0, 0) node[inner sep=0] {
\includegraphics[width=.49\textwidth]{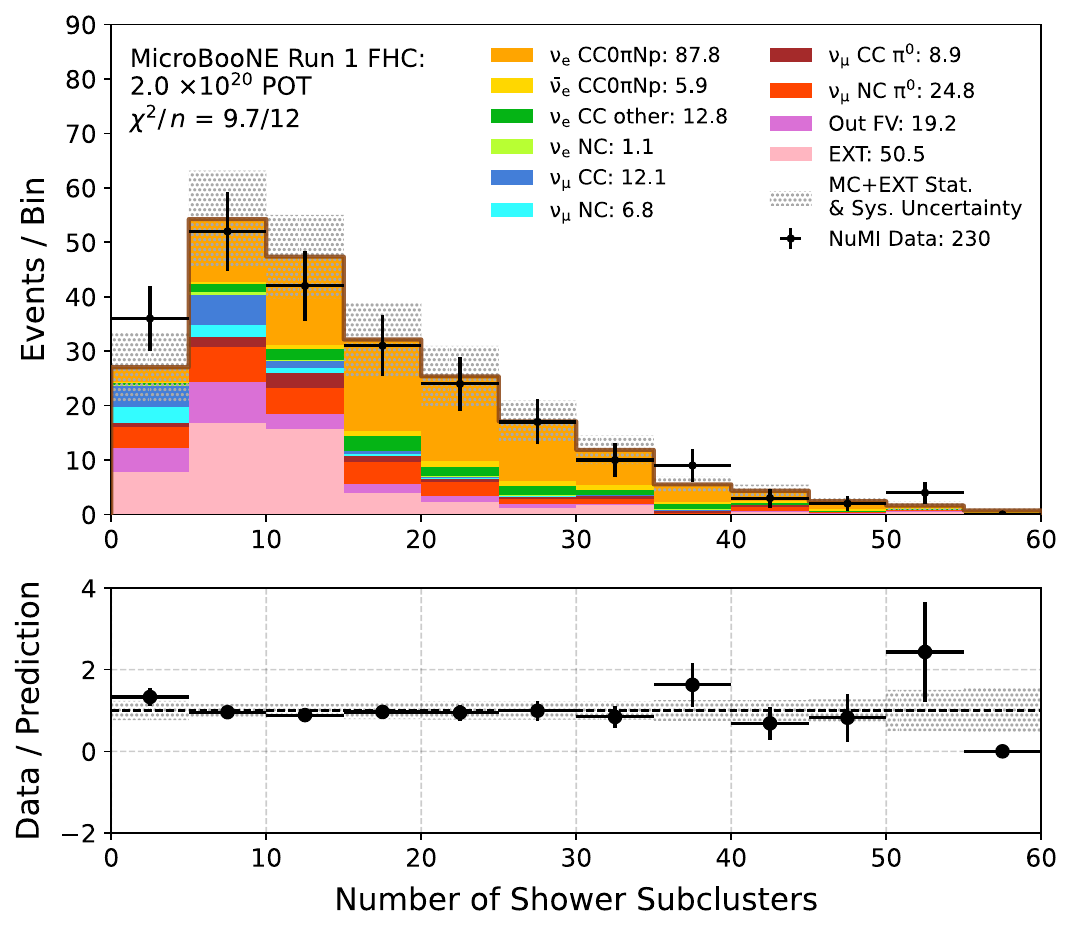}};
  \draw (-3.15, -1.2) node {\textbf{(a)}};
\end{tikzpicture}
\hfill
\begin{tikzpicture} \draw (0, 0) node[inner sep=0] {
\includegraphics[width=.49\textwidth]{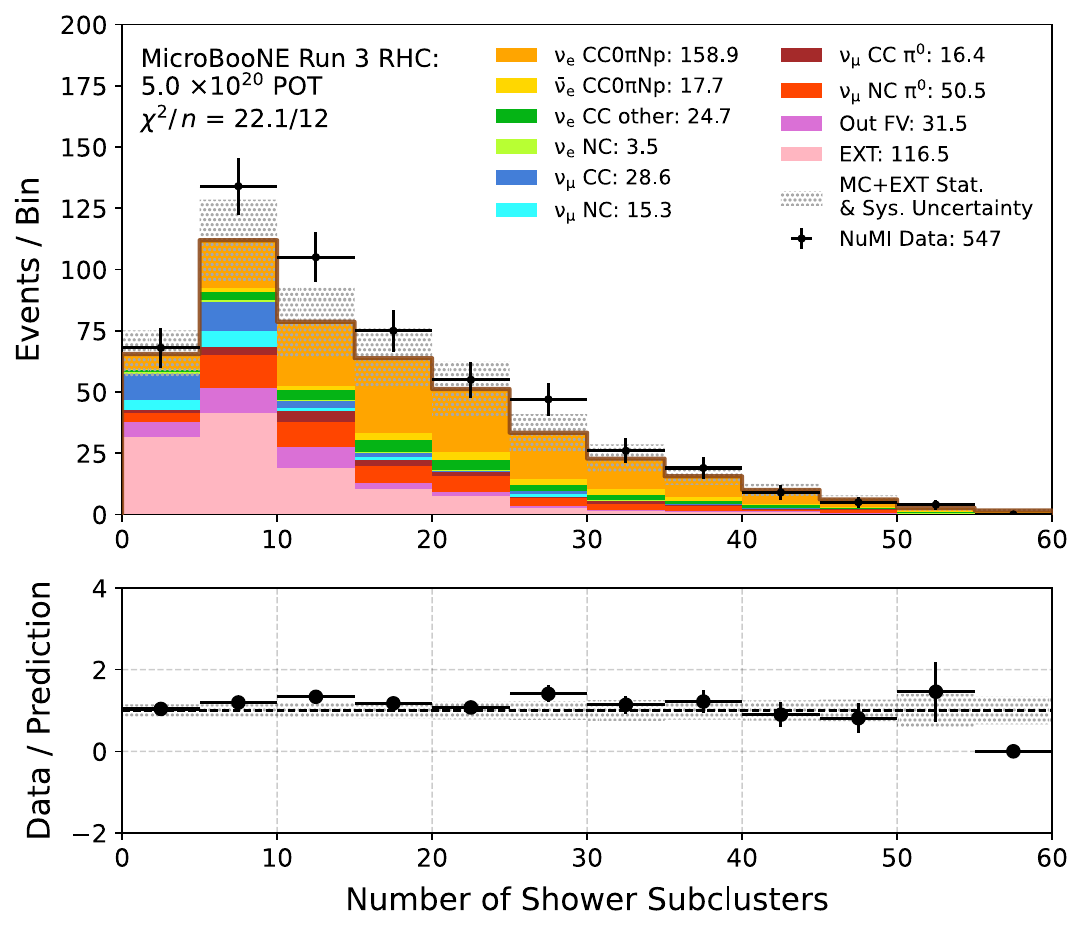}};
  \draw (-3.1, -1.2) node {\textbf{(b)}};
\end{tikzpicture}
\hfill
\caption{\label{subcluster_bdt}{Run~1 FHC (a) and Run~3 RHC (b) event rates as a function of number of subclusters. Shown for interactions passing the preselection, \numu{} CC rejection, \pizero{} rejection, and additional BDT training constraints.}}
\end{figure}

\begin{figure}[h]
\centering
\begin{tikzpicture} \draw (0, 0) node[inner sep=0] {
\includegraphics[width=.48\textwidth]{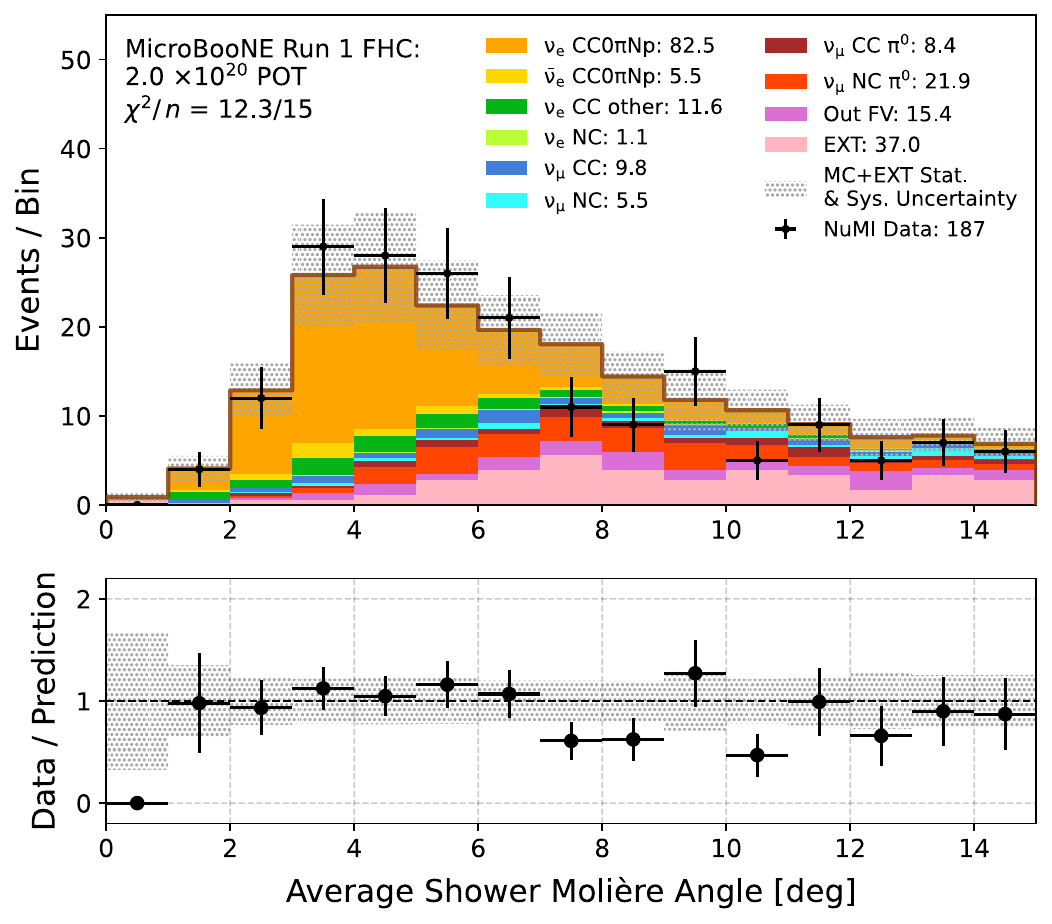}};
  \draw (-3.1, -1.2) node {\textbf{(a)}};
\end{tikzpicture}
\hfill
\begin{tikzpicture} \draw (0, 0) node[inner sep=0] {
\includegraphics[width=.49\textwidth]{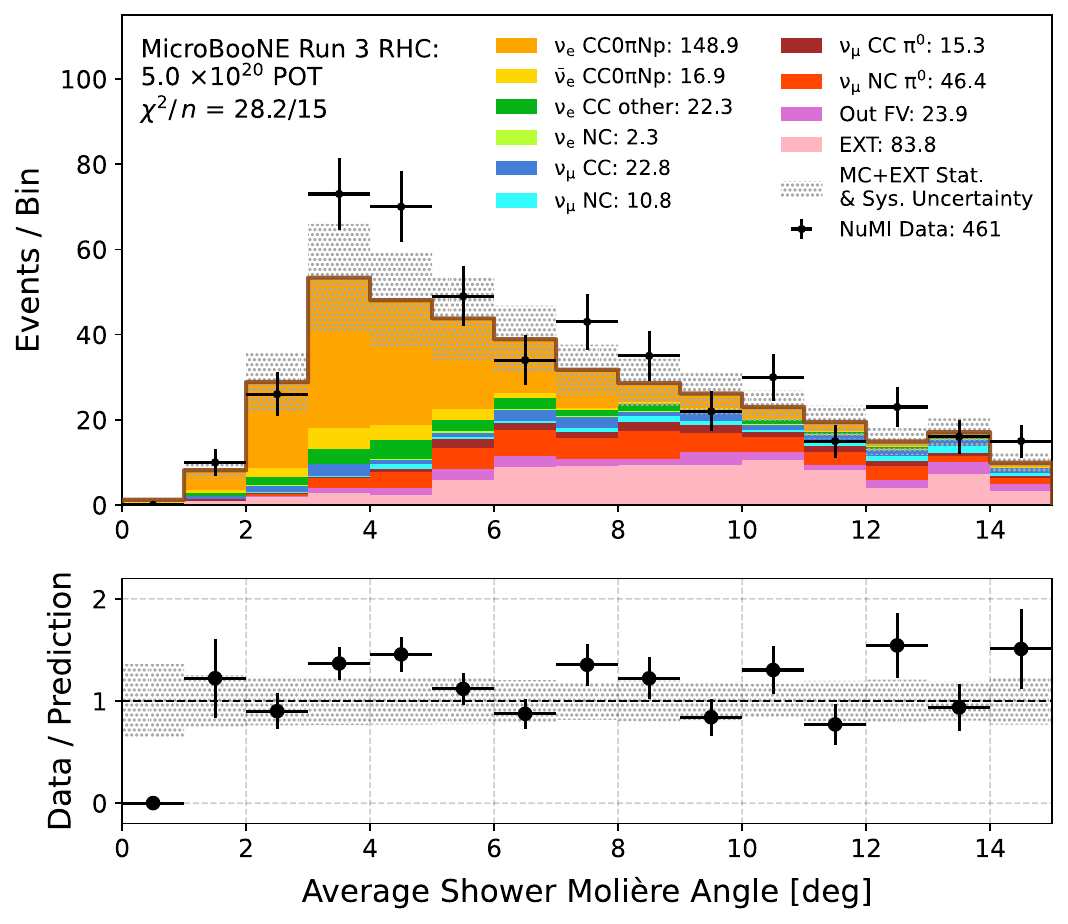}};
  \draw (-3.1, -1.2) node {\textbf{(b)}};
\end{tikzpicture}
\hfill
\caption{\label{moliere_bdt}{Run~1 FHC (a) and Run~3 RHC (b) event rates as a function of average shower Moli\`ere angle. Shown for interactions passing the preselection, \numu{} CC rejection, \pizero{} rejection, and additional BDT training constraints.}}
\end{figure}

\begin{figure}[h]
\centering
\begin{tikzpicture} \draw (0, 0) node[inner sep=0] {
\includegraphics[width=.48\textwidth]{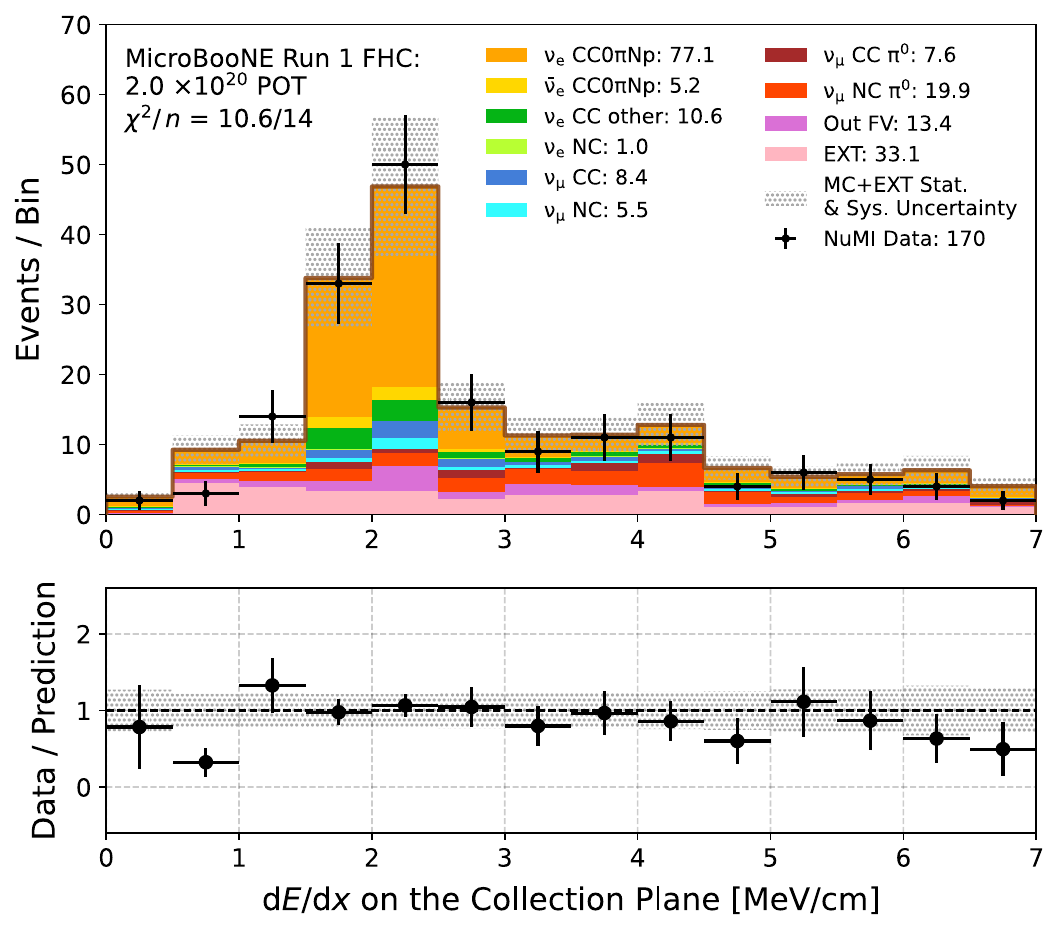}};
  \draw (-3.1, -1.25) node {\textbf{(a)}};
\end{tikzpicture}
\hfill
\begin{tikzpicture} \draw (0, 0) node[inner sep=0] {
\includegraphics[width=.49\textwidth]{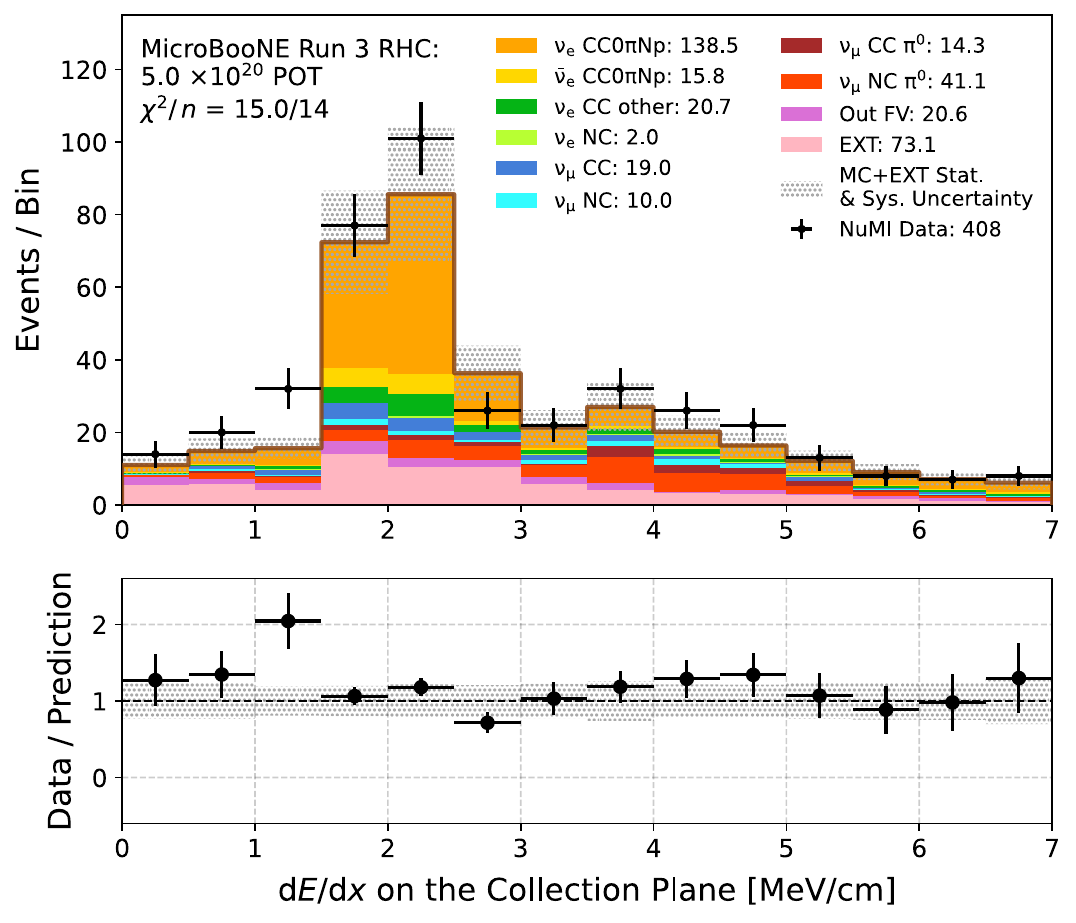}};
  \draw (-3.1, -1.2) node {\textbf{(b)}};
\end{tikzpicture}
\hfill
\caption{\label{dedx_bdt}{Run~1 FHC (a) and Run~3 RHC (b) event rates as a function of $\mathrm{d}E/\mathrm{d}x$ on the collection plane. Shown for interactions passing the preselection, \numu{} CC rejection, \pizero{} rejection, and additional BDT training constraints.}}
\end{figure}

\begin{figure}[h]
\centering
\begin{tikzpicture} \draw (0, 0) node[inner sep=0] {
\includegraphics[width=.49\textwidth]{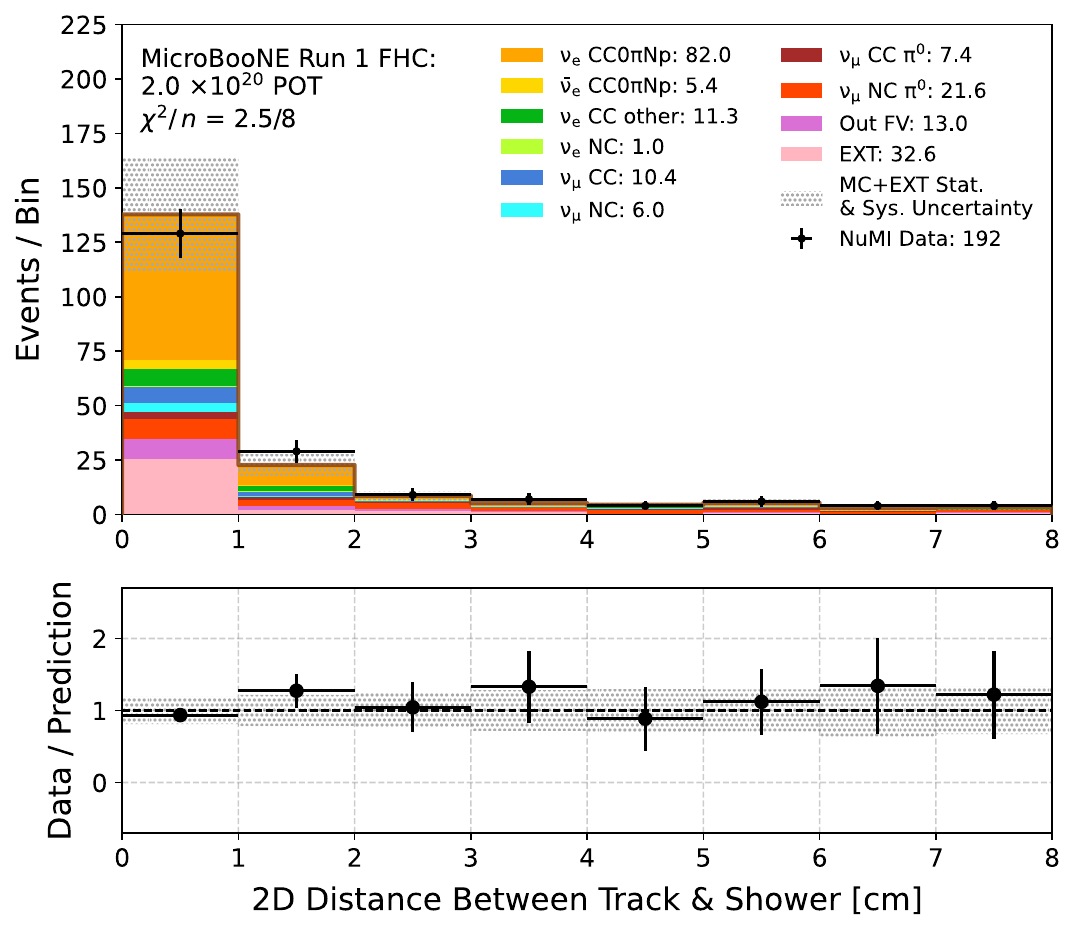}};
  \draw (-3.1, -1.25) node {\textbf{(a)}};
\end{tikzpicture}
\hfill
\begin{tikzpicture} \draw (0, 0) node[inner sep=0] {
\includegraphics[width=.49\textwidth]{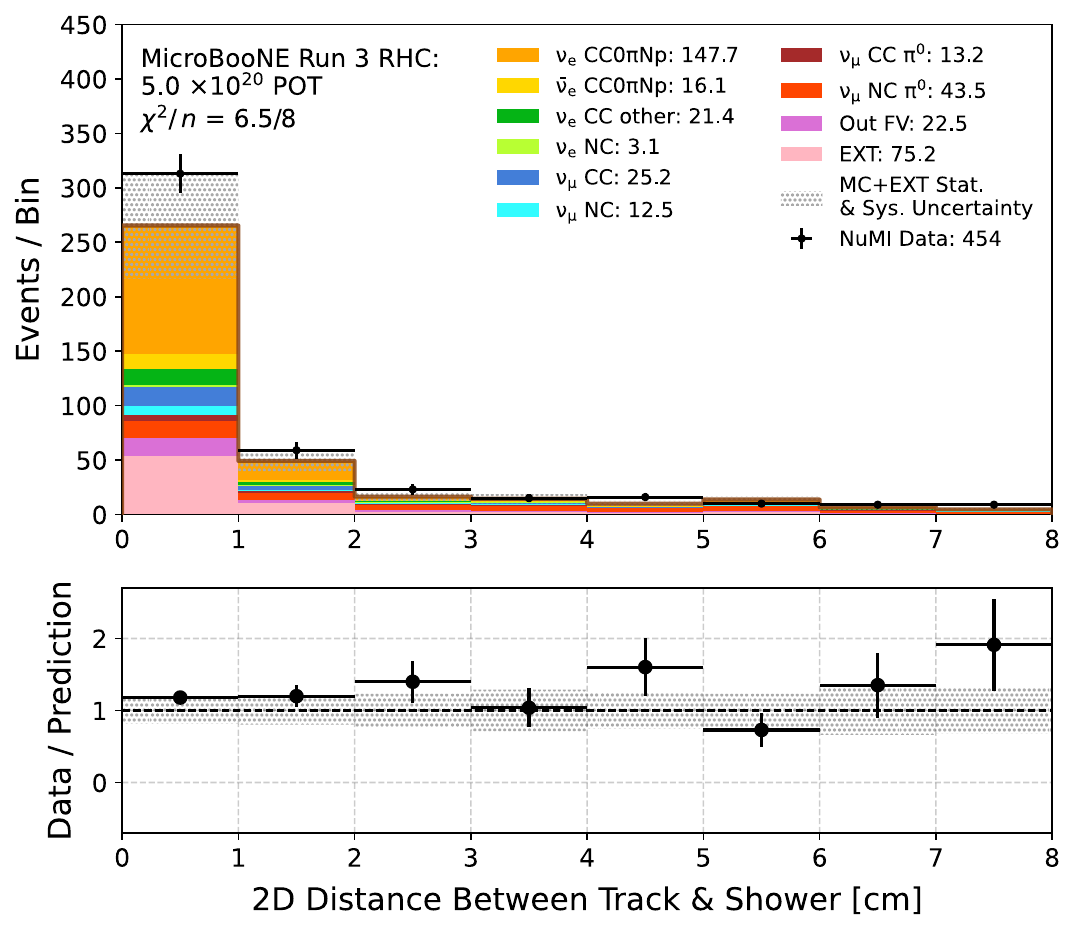}};
  \draw (-3.1, -1.2) node {\textbf{(b)}};
\end{tikzpicture}
\hfill
\caption{\label{tksh_2d_bdt}{Run~1 FHC (a) and Run~3 RHC (b) event rates as a function of the two-dimensional distance between the interaction vertex and the start of the shower. Shown for interactions passing the preselection, \numu{} CC rejection, \pizero{} rejection, and additional BDT training constraints.}}
\end{figure}

\begin{figure}[h]
\centering
\begin{tikzpicture} \draw (0, 0) node[inner sep=0] {
\includegraphics[width=.49\textwidth]{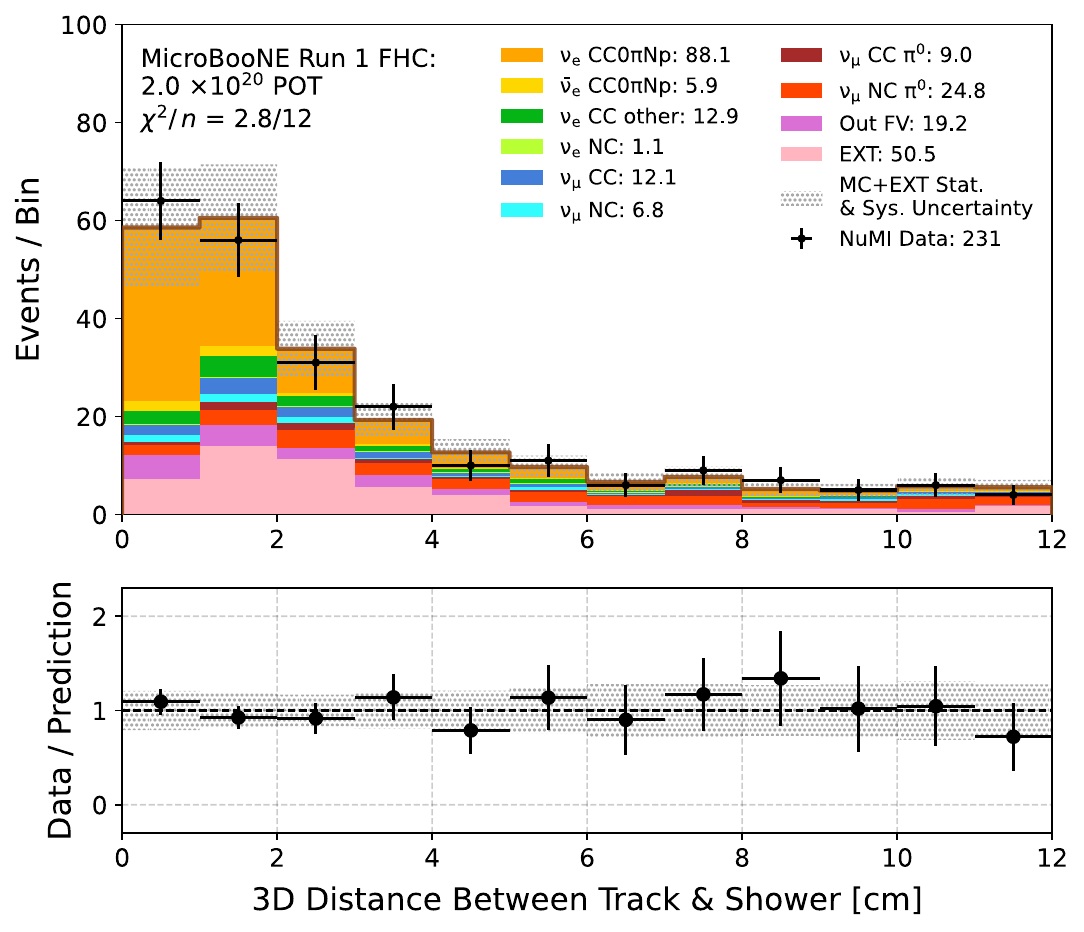}};
  \draw (-3.1, -1.2) node {\textbf{(a)}};
\end{tikzpicture}
\hfill
\begin{tikzpicture} \draw (0, 0) node[inner sep=0] {
\includegraphics[width=.49\textwidth]{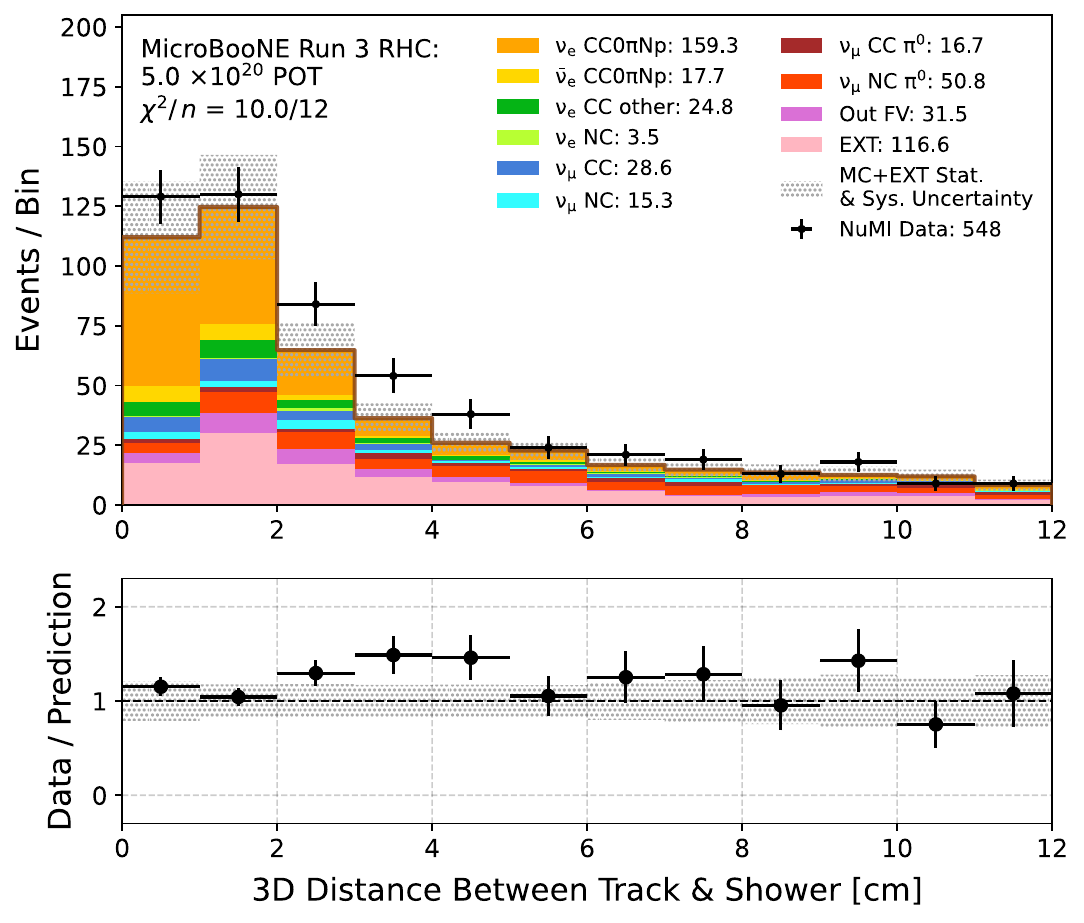}};
  \draw (-3.1, -1.2) node {\textbf{(b)}};
\end{tikzpicture}
\hfill
\caption{\label{tksh_distance_bdt}{Run~1 FHC (a) and Run~3 RHC (b) event rates as a function of the three-dimensional distance between the interaction vertex and the start of the shower. Shown for interactions passing the preselection, \numu{} CC rejection, \pizero{} rejection, and additional BDT training constraints.}}
\end{figure}

Plots in this section illustrate the agreement between NuMI prediction and data for the reconstructed observables used to train the BDT model. To remove \numu{} CC interactions, the BDT trains on the log-likelihood track PID score (Fig. \ref{trkpid_bdt}), \code{Pandora} shower score (Fig. \ref{shr_score_bdt}), and the number of subclusters making up the reconstructed shower (Fig. \ref{subcluster_bdt}). To remove \pizero{} interactions, the BDT trains on the average Moli\`ere angle (Fig. \ref{moliere_bdt}), $\mathrm{d}E/\mathrm{d}x$ on the collection plane (Fig. \ref{dedx_bdt}), and both the two-dimensional (Fig. \ref{tksh_2d_bdt}) and three-dimensional (Fig. \ref{tksh_distance_bdt}) distance between the interaction vertex and the start of the shower. Detector systematic uncertainty is not included in the gray error band nor in the calculation of $\chi^2/n$. 


\clearpage
\newpage
\newpage
\section{Linear Selection}
\label{linear}

The table in this section illustrates the linear selection for Run~1 FHC and Run~3 RHC, adapted from the MicroBooNE low-energy excess search selection for $1e + Np$ interactions in the BNB, which is used to compare with the BDT performance. 

\begin{table*}[h]
\centering\small
\caption{Linear selection (baseline) criteria used as a reference for comparison with the BDT-based selection. The linear selection is adapted for NuMI from the MicroBooNE low-energy excess search selection for $1e + Np$ interactions in the BNB \cite{linear_selection, linear_selection_prd}.}
\label{tab:linear_selection_criteria}
\begin{ruledtabular}
\begin{tabular}{l l}
\textbf{Cut Goal} & \textbf{Cut Definition} \\ [3pt]
\hline
Neutrino Identification & Number of Neutrinos Identified $=1$  \\ [6pt]
\multirow{2}{*}{Containment}
 & Reconstructed Vertex in FV \\
 & Contained Fraction $>$ 0.9 \\ [6pt]
\multirow{3}{*}{Signal Definition Constraints}
 & No. Showers Contained $=1$ \\
 & No. Tracks Contained $> 0$ \\
 & Track Kinetic Energy $>$ 40\,MeV \\ [6pt]
\multirow{2}{*}{CC \(\nu_{\mu}\) rejection}
 & Track PID (Proton/Muon Log-Likelihood) $<$ 0 \\
 & \texttt{Pandora} Shower Score $<$ 0.125 \\ [6pt]
\multirow{3}{*}{\(\pi^{0}\) rejection}
 & Distance Between Track \& Shower $<$ 5\,cm (FHC), 4\,cm (RHC) \\
 & Collection Plane $\mathrm{d}E/\mathrm{d}x$ $<$ 4\,MeV/cm \\
 & Average Shower Molière Angle $<$ 8\degree \\
\end{tabular}
\end{ruledtabular}
\end{table*}

The linear selection yields an estimated efficiency of 13.4\% and a purity of 69.5\% for Run~1 FHC, and 13.0\% and 66.8\% for Run~3 RHC.

\clearpage
\section{Electron Energy}

The material in this section provides additional information for the cross-section measurement as a function of electron energy: the total selected event rates and associated selection efficiencies, fractional systematic uncertainty breakdowns for the signal and $\bar{\nu}_e$ CC$0\pi Np$ background, response matrix, unfolded central-value cross-section values and covariance, and the additional smearing matrix $\bm{A_c}$ returned by the Wiener-SVD method. Unlike in prior sections, detector systematic uncertainty is included in the gray error bands and the calculation of $\chi^2/n$.


\begin{figure}[h]
\centering
\begin{tabular}{cc}
\begin{tikzpicture}
  \node[inner sep=0] {
    \includegraphics[width=0.48\textwidth]{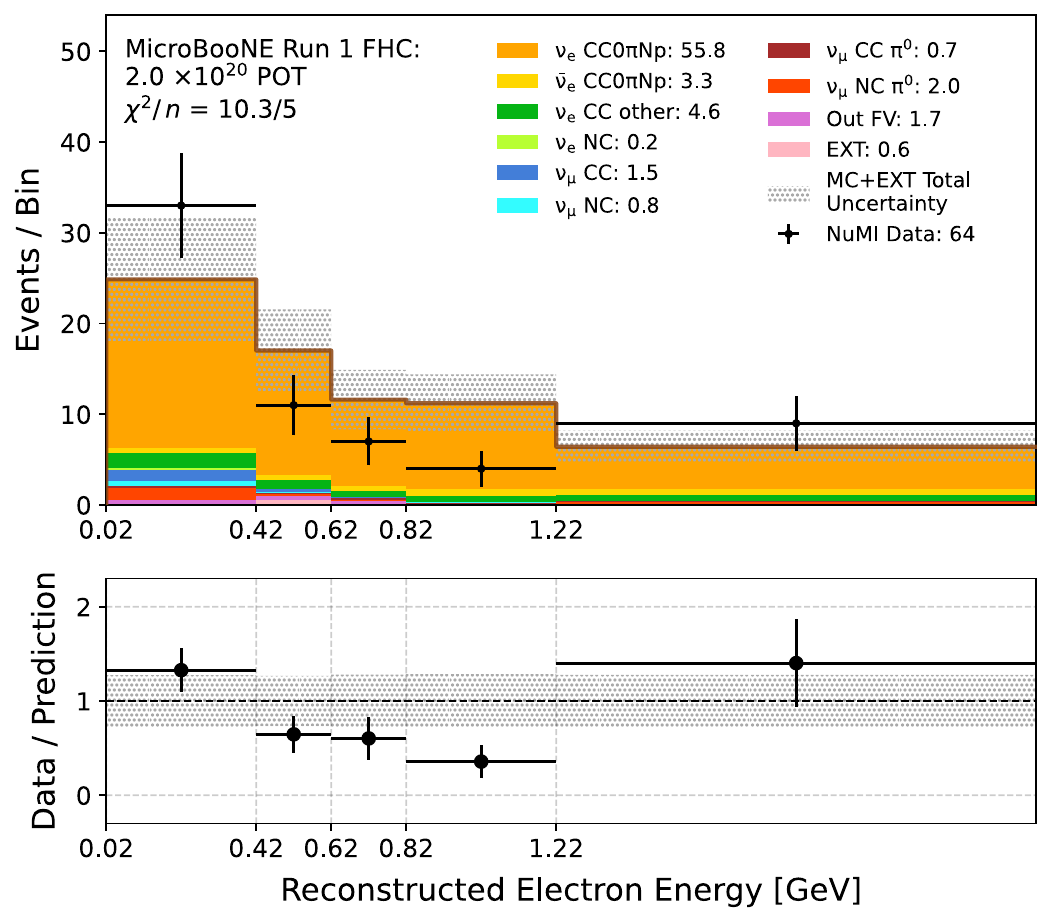}};
  \node at (3.6,-1.3) {\textbf{(a)}};
\end{tikzpicture}
& 
\begin{tikzpicture}
  \node[inner sep=0] {
    \includegraphics[width=0.48\textwidth]{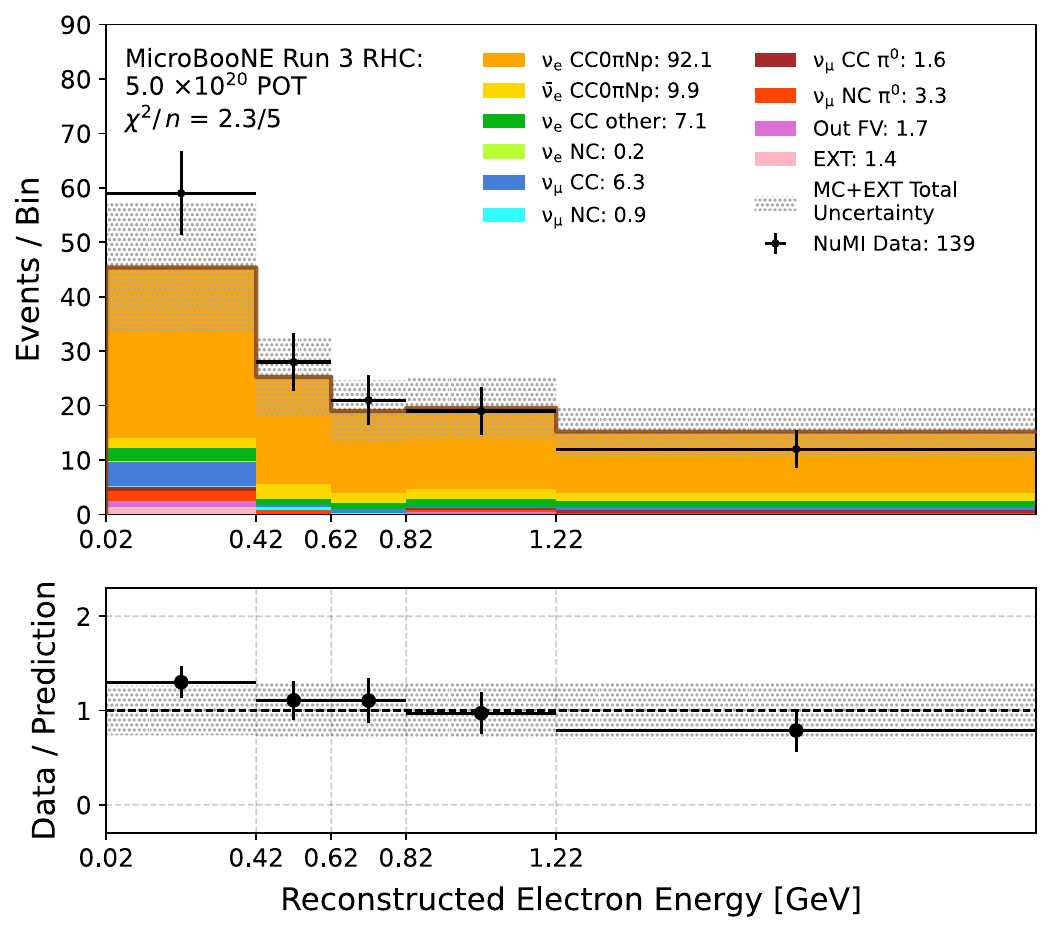}};
  \node at (3.6,-1.3) {\textbf{(b)}};
\end{tikzpicture}
\\[1.2em]
\multicolumn{2}{c}{
\begin{tikzpicture}
  \node[inner sep=0] {
    \includegraphics[width=0.48\textwidth]{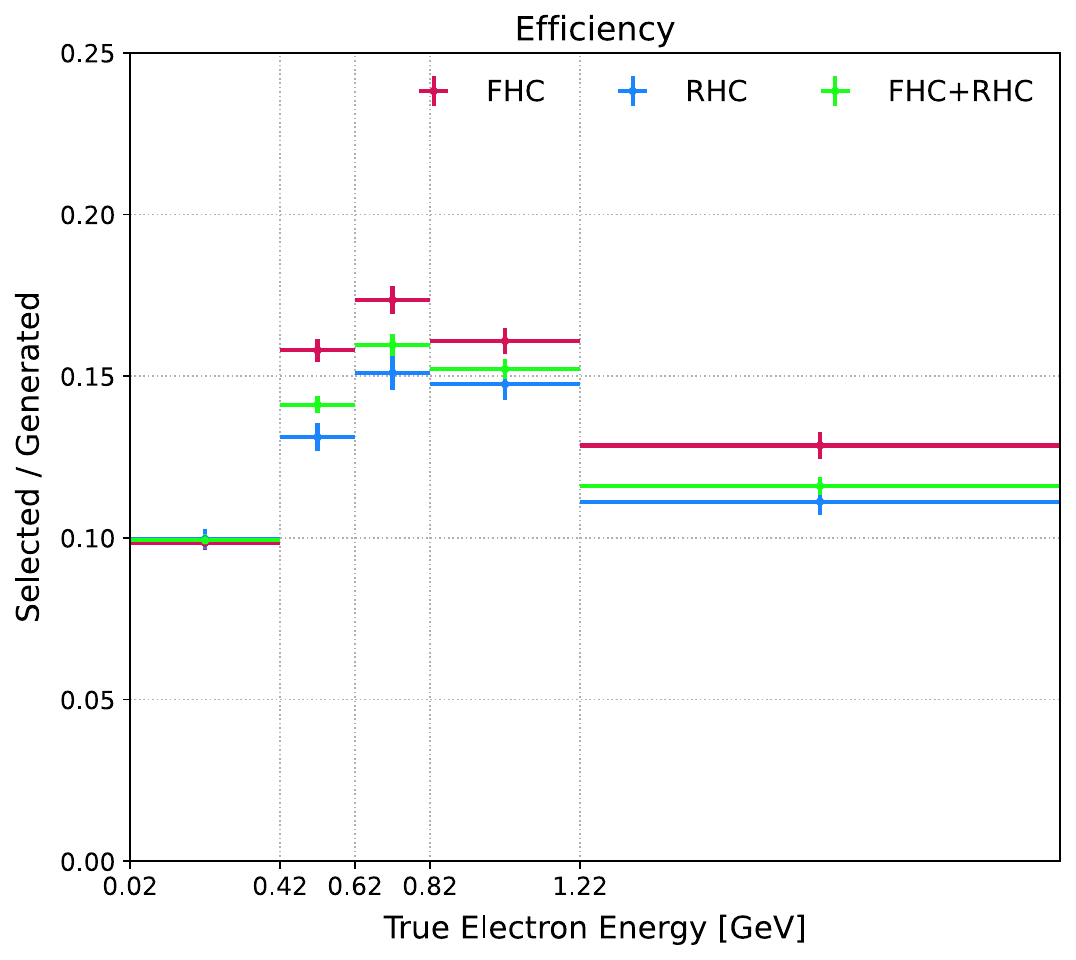}};
  \node at (3.6,-2.6) {\textbf{(c)}};
\end{tikzpicture}
}
\end{tabular}

\caption{Run~1 FHC (a) and Run~3 RHC (b) selected event rates as a function of reconstructed electron energy, and FHC-only, RHC-only, and FHC+RHC selected event rate efficiencies (c) after subtracting background predicted by the tuned \code{GENIE v3.0.6 G18\_10a\_02\_11a} model as a function of true electron energy.}
\label{fig:electron_energy_abc}
\end{figure}


\begin{figure}[h]
\centering
\begin{tabular}{cc}
\begin{tikzpicture}
  \node[inner sep=0] {
    \includegraphics[width=0.48\textwidth]{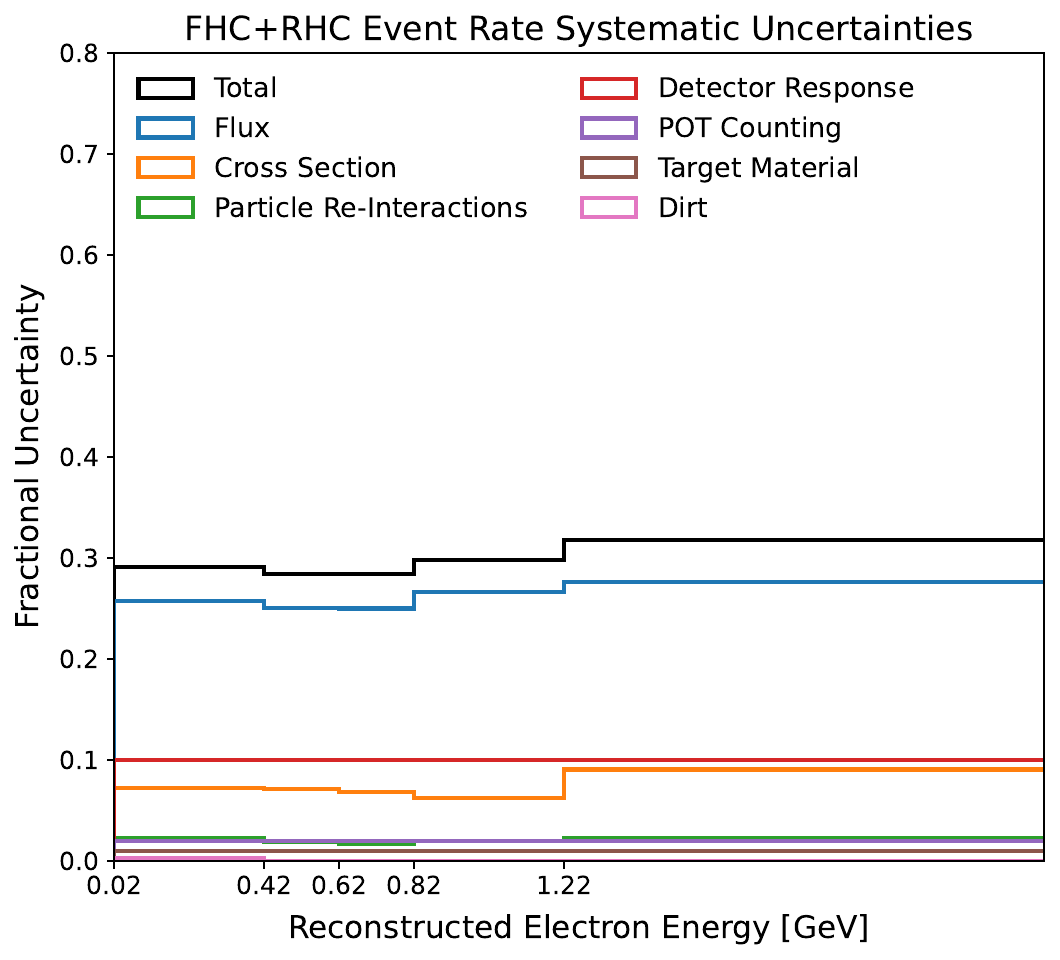}};
  \node at (3.75,3.1) {\textbf{(a)}};
\end{tikzpicture}
&
\begin{tikzpicture}
  \node[inner sep=0] {
    \includegraphics[width=0.48\textwidth]{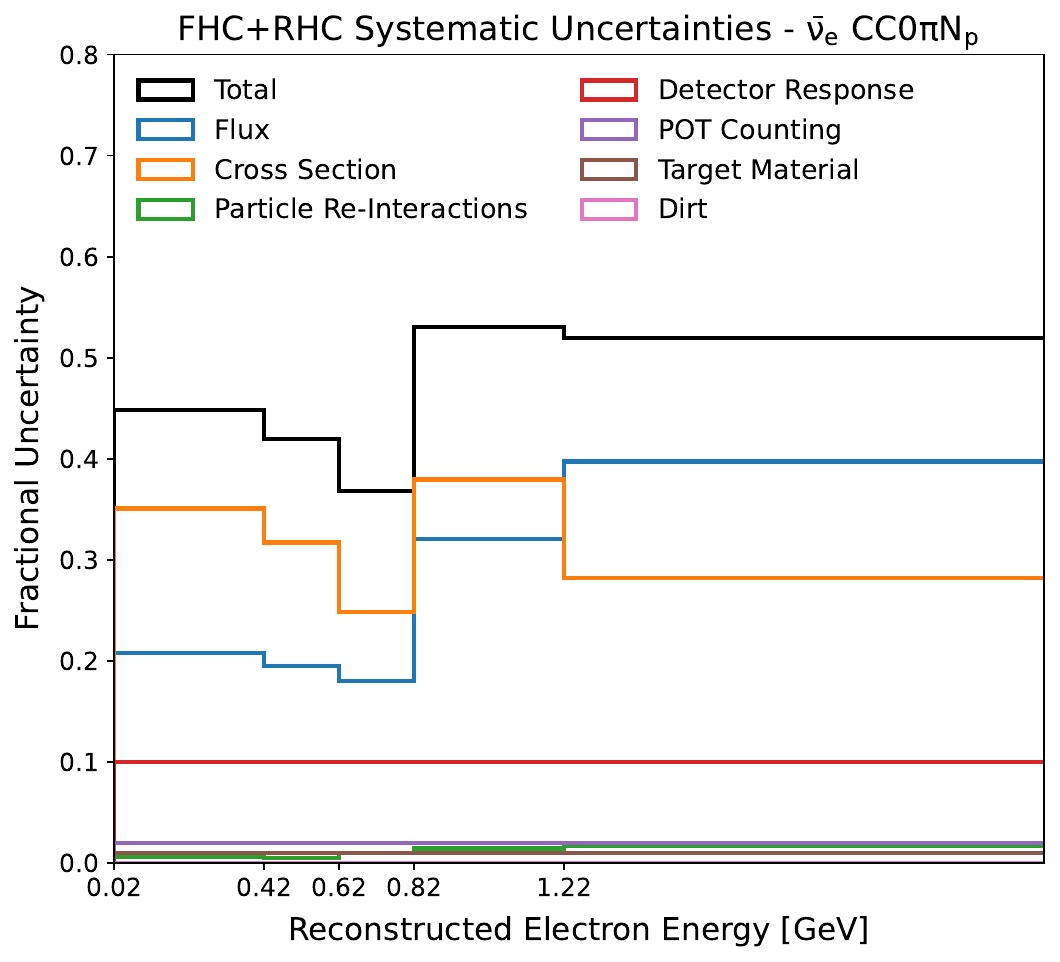}};
  \node at (3.75,3.1) {\textbf{(b)}};
\end{tikzpicture}
\end{tabular}

\caption{Fractional systematic uncertainties after subtracting background predicted by the tuned \code{GENIE v3.0.6 G18\_10a\_02\_11a} model (a) and on background $\bar{\nu}_{e}$ CC$0\pi Np$ events only (b) as a function of reconstructed electron energy.}
\label{fig:electron_energy_unc}
\end{figure}


\begin{figure}[h]
\centering

\includegraphics[width=.7\textwidth]{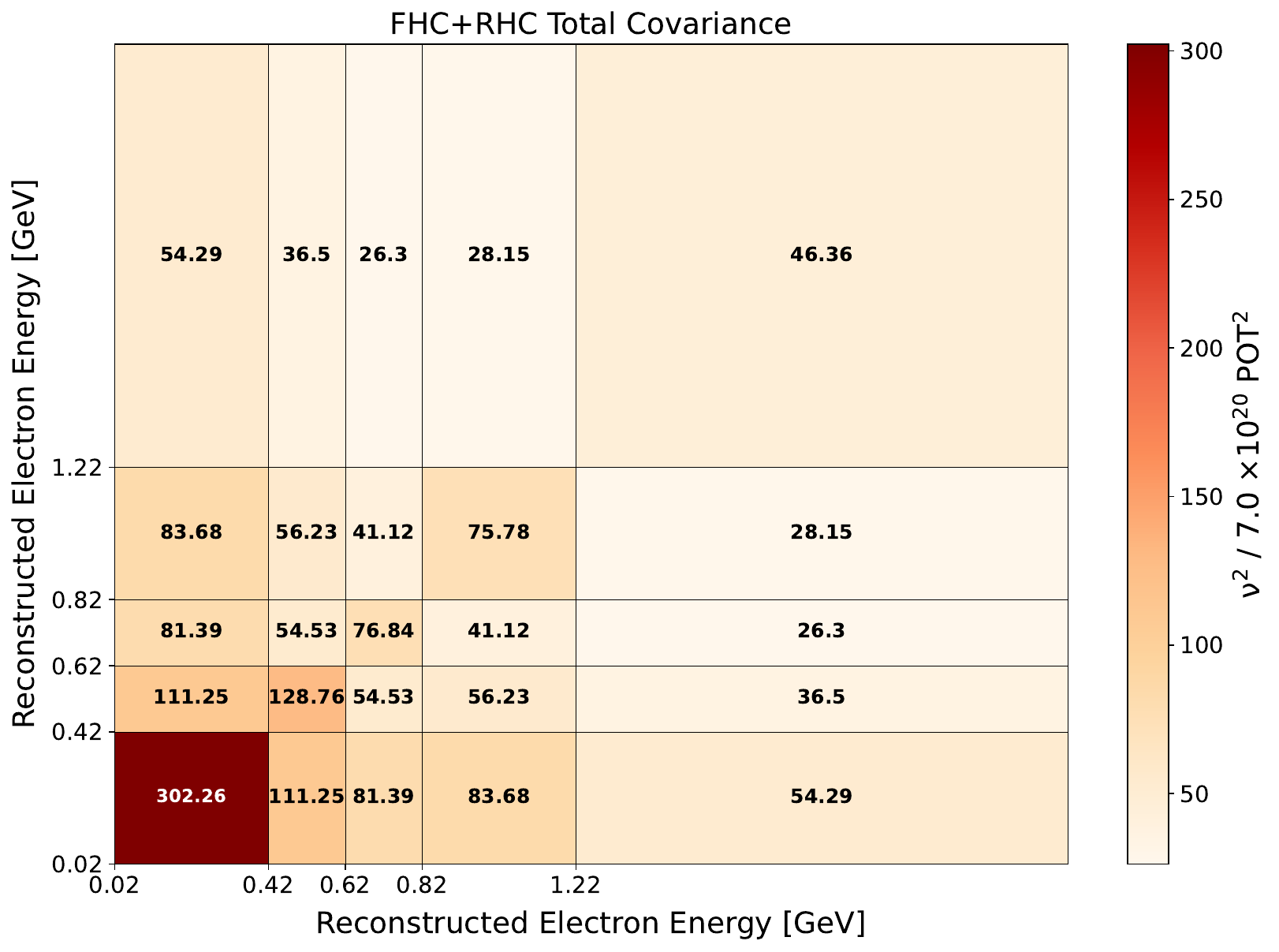}

\caption{{Total covariance (statistical + systematic) matrix for the FHC+RHC background-subtracted event rate as a function of reconstructed electron energy.}}
\end{figure}


\begin{figure}[h]
\centering
\begin{tabular}{cc}
\begin{tikzpicture}
  \node[inner sep=0] {
    \includegraphics[width=0.48\textwidth]{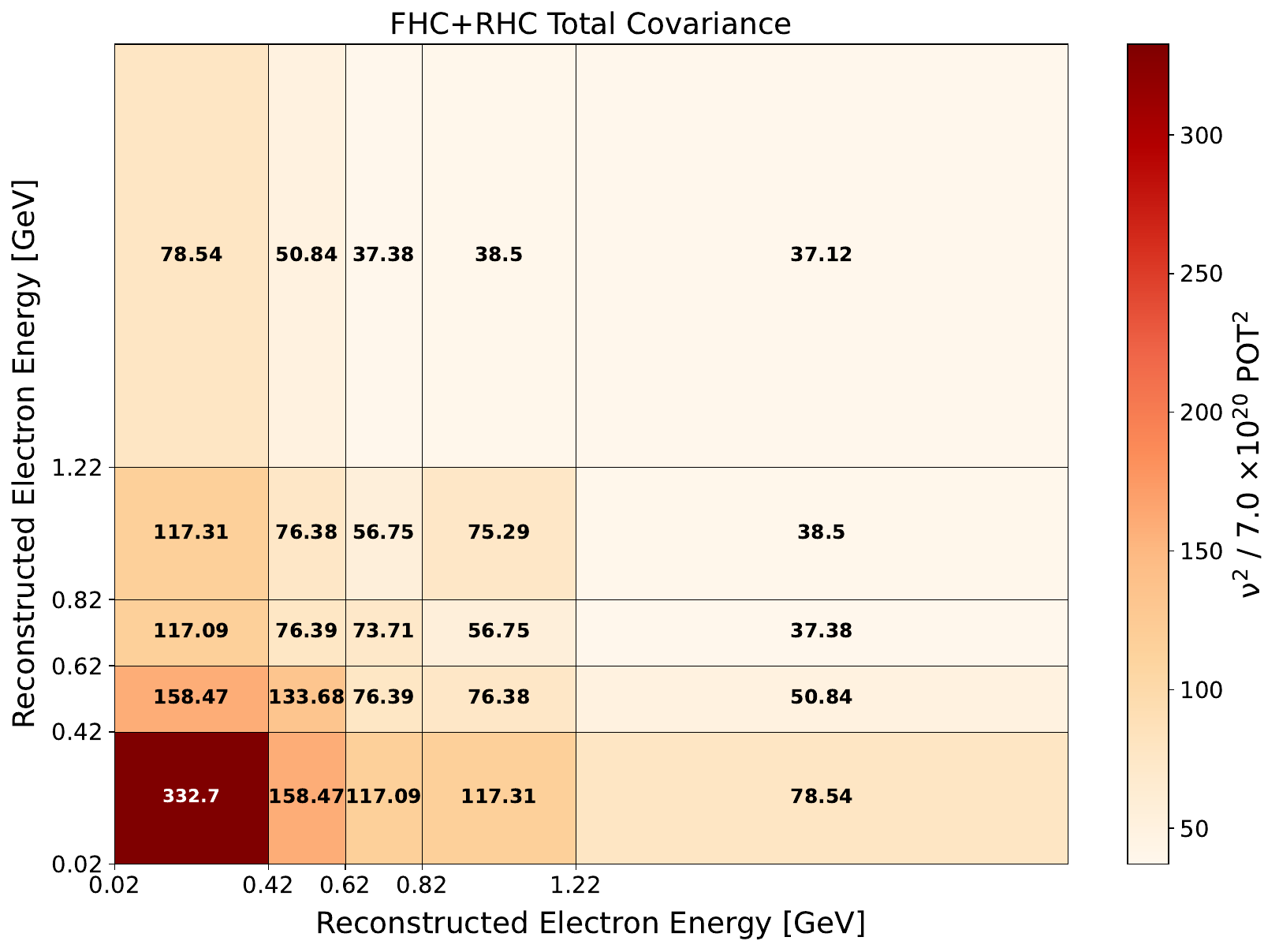}};
  \node at (2.5,2.5) {\textbf{(a)}};
\end{tikzpicture}
&
\begin{tikzpicture}
  \node[inner sep=0] {
    \includegraphics[width=0.48\textwidth]{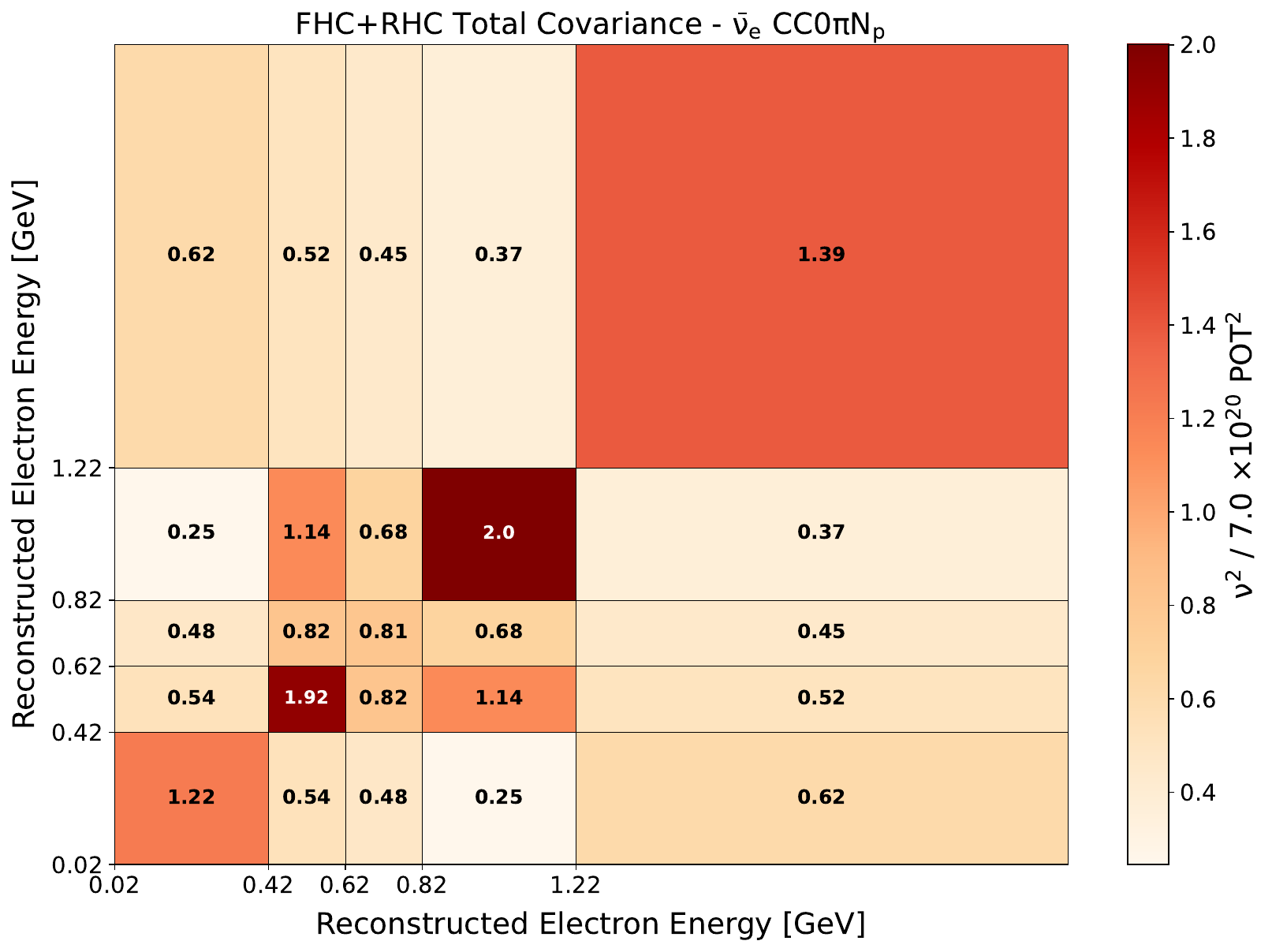}};
  \node at (2.5,2.5) {\textbf{(b)}};
\end{tikzpicture}
\end{tabular}

\caption{Total covariance (statistical + systematic) matrix for the FHC+RHC selected event rate (a) and $\bar{\nu}_{e}$ CC$0\pi Np$ background events only (b) as a function of reconstructed electron energy.}
\end{figure}


\begin{figure}[h]
\centering

\includegraphics[width=.7\textwidth]{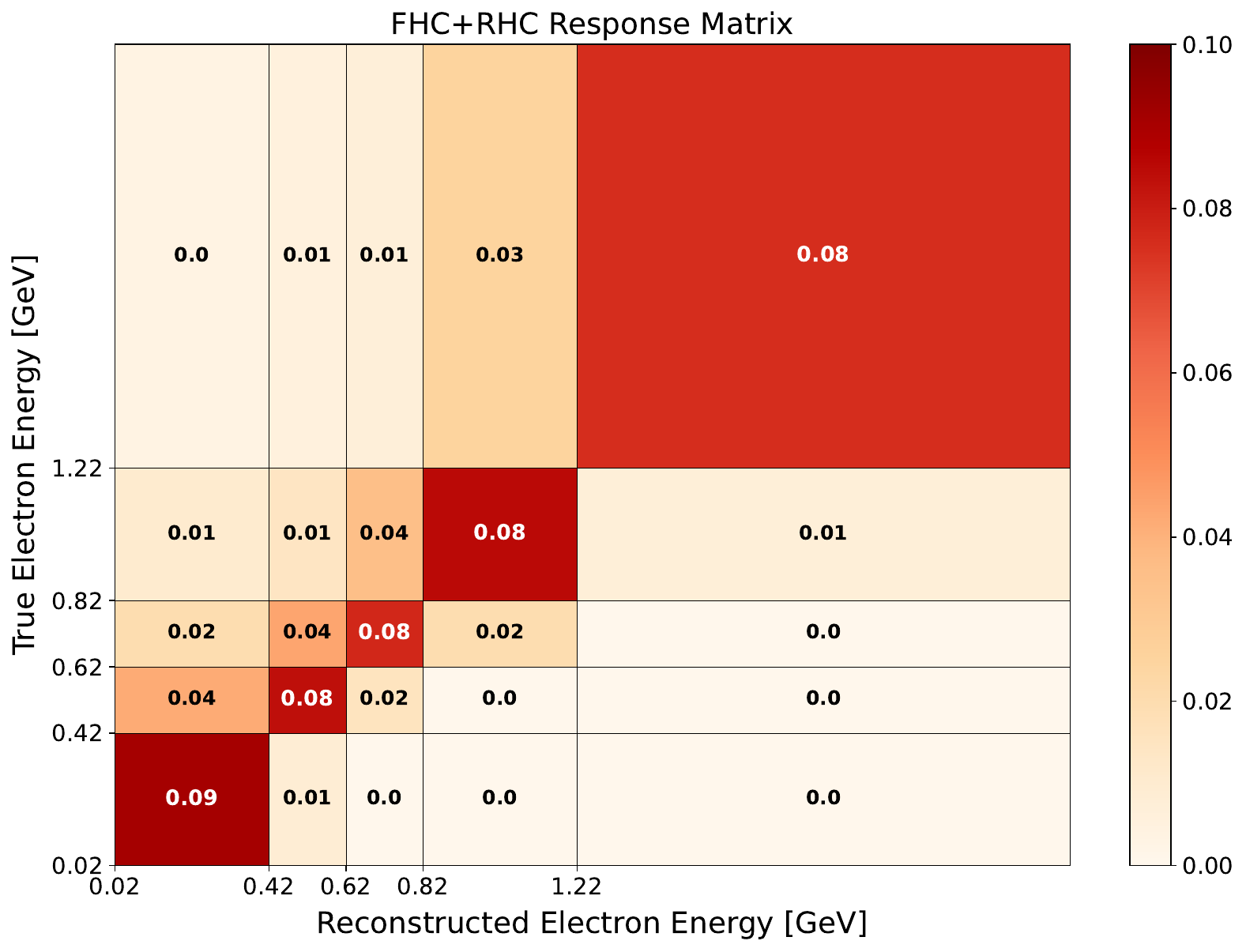}

\caption{{The response matrix, constructed using the \code{GENIE v3.0.6 G18\_10a\_02\_11a} selected signal prediction, as a function of true and reconstructed electron energy. Row $j$ of the matrix is normalized such that $\sum_i r_{ij} = \epsilon_j$, where $i$ is the column and $\epsilon_j$ is the estimated selection efficiency of true bin $j$. This histogram is used as input for the Wiener-SVD unfolding algorithm.}}
\end{figure}


\begin{figure}[h]
\centering

\includegraphics[width=.7\textwidth]{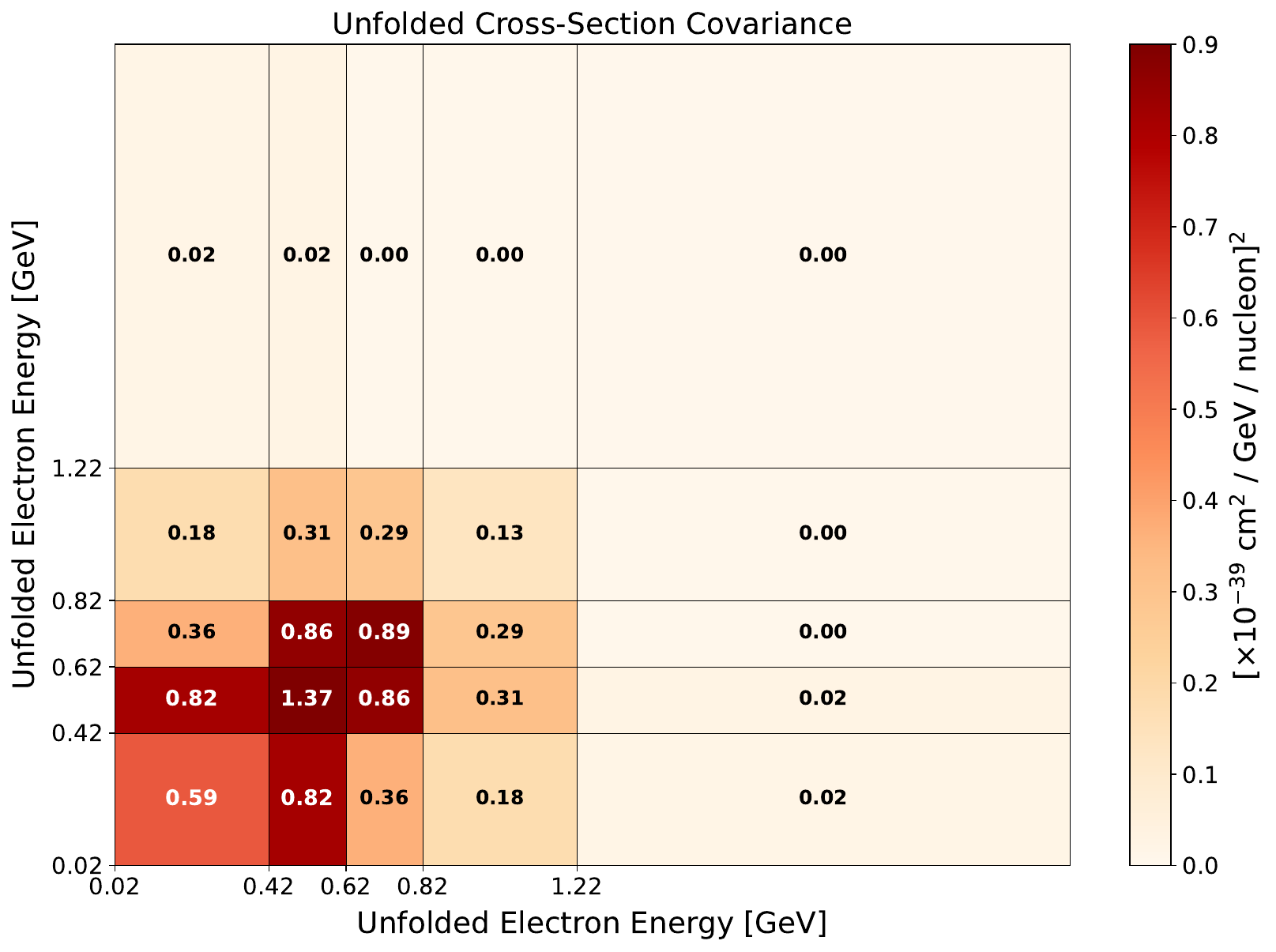}

\caption{{Unfolded covariance of the cross-section result as a function of electron energy, in units of $\left(10^{-39}\,\mathrm{cm}\,/\,\mathrm{GeV}\,/\,\mathrm{nucleon}\right)^{2}$.}}
\end{figure}


\begin{figure}[h]
\centering

\includegraphics[width=.7\textwidth]{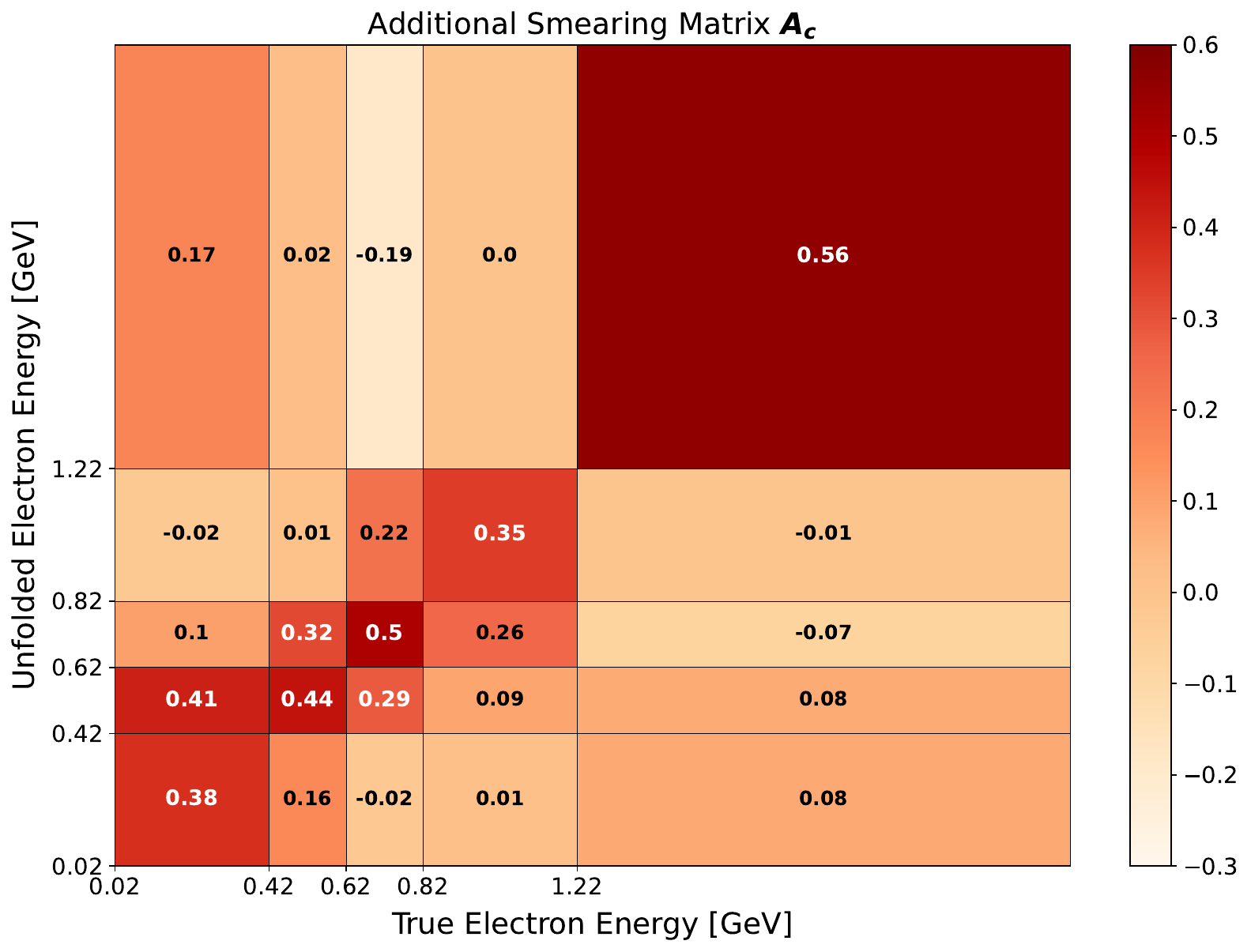}

\caption{{The additional smearing matrix $\bm{A_c}$ returned by the Wiener-SVD method for unfolded electron energy. This is used to smear event rate generator predictions of the event rate into unfolded space.}}
\end{figure}


\clearpage

\begin{table}[h]
\caption{The background-subtracted event count and unfolded cross section as a function of electron energy. The unfolded cross section is in units of $10^{-39}\,\mathrm{cm}\,/\,\mathrm{GeV}\,/\,\mathrm{nucleon}$.}
\begin{ruledtabular}
\begin{tabular}{llllll}
\textbf{Electron energy [GeV]} & \textbf{[0.02, 0.42]} & \textbf{[0.42, 0.62]} & \textbf{[0.62, 0.82]} & \textbf{[0.82, 1.22]} & \textbf{[1.22, $\infty$]} \\
\hline
Background-subtracted event count & 71.58 & 30.01 & 21.95 & 16.60 & 15.21 \\ 
Unfolded cross section & 3.10 & 4.18 & 1.93 & 0.90 & 0.09 \\

\end{tabular}
\end{ruledtabular}

\end{table}
\normalsize

\clearpage 
\clearpage
\section{Visible Energy}

The material in this section provides additional information for the cross-section measurement as a function of visible energy: the total selected event rates and associated selection efficiencies, fractional systematic uncertainty breakdowns for the signal and $\bar{\nu}_e$ CC$0\pi Np$ background, response matrix, unfolded central-value cross-section values and covariance, and the additional smearing matrix $\bm{A_c}$ returned by the Wiener-SVD method. Here, detector systematic uncertainty is included in the gray error bands and the calculation of $\chi^2/n$.


\begin{figure}[h]
\centering
\begin{tabular}{cc}
\begin{tikzpicture}
  \node[inner sep=0] {
    \includegraphics[width=0.48\textwidth]{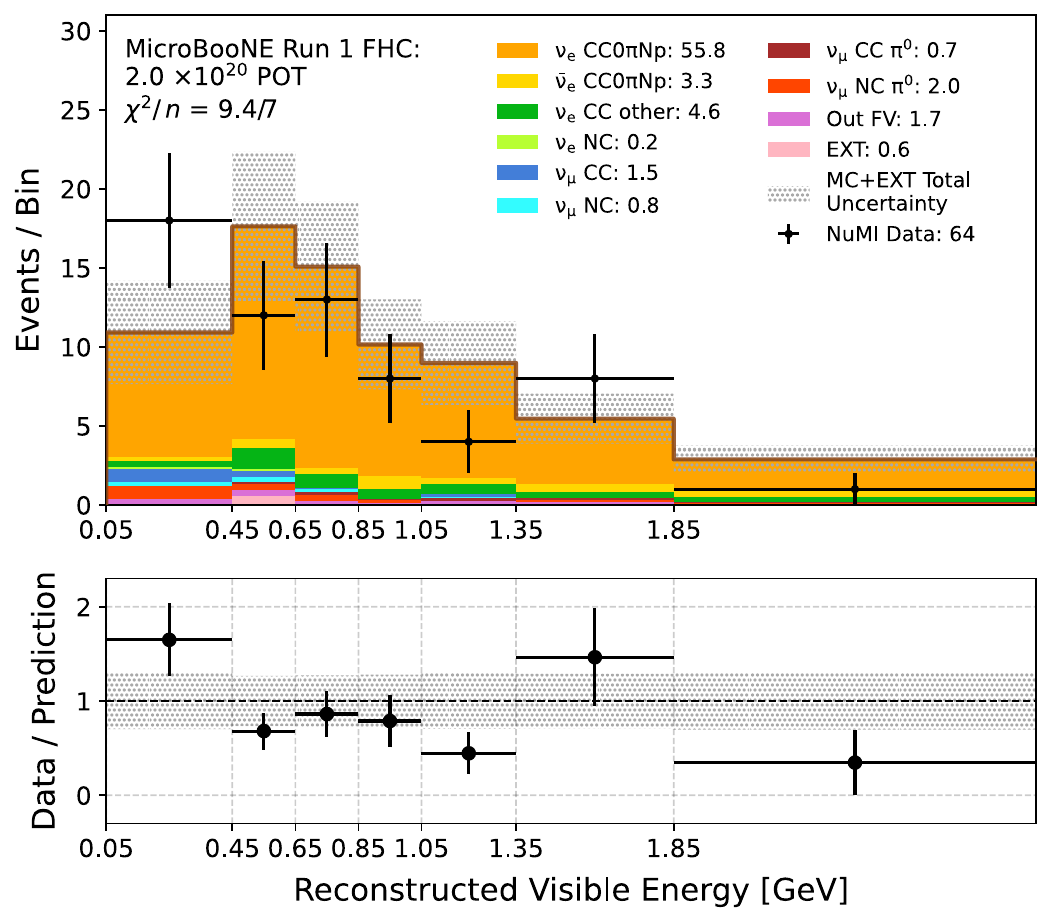}};
  \node at (3.6,-1.3) {\textbf{(a)}};
\end{tikzpicture}
& 
\begin{tikzpicture}
  \node[inner sep=0] {
    \includegraphics[width=0.48\textwidth]{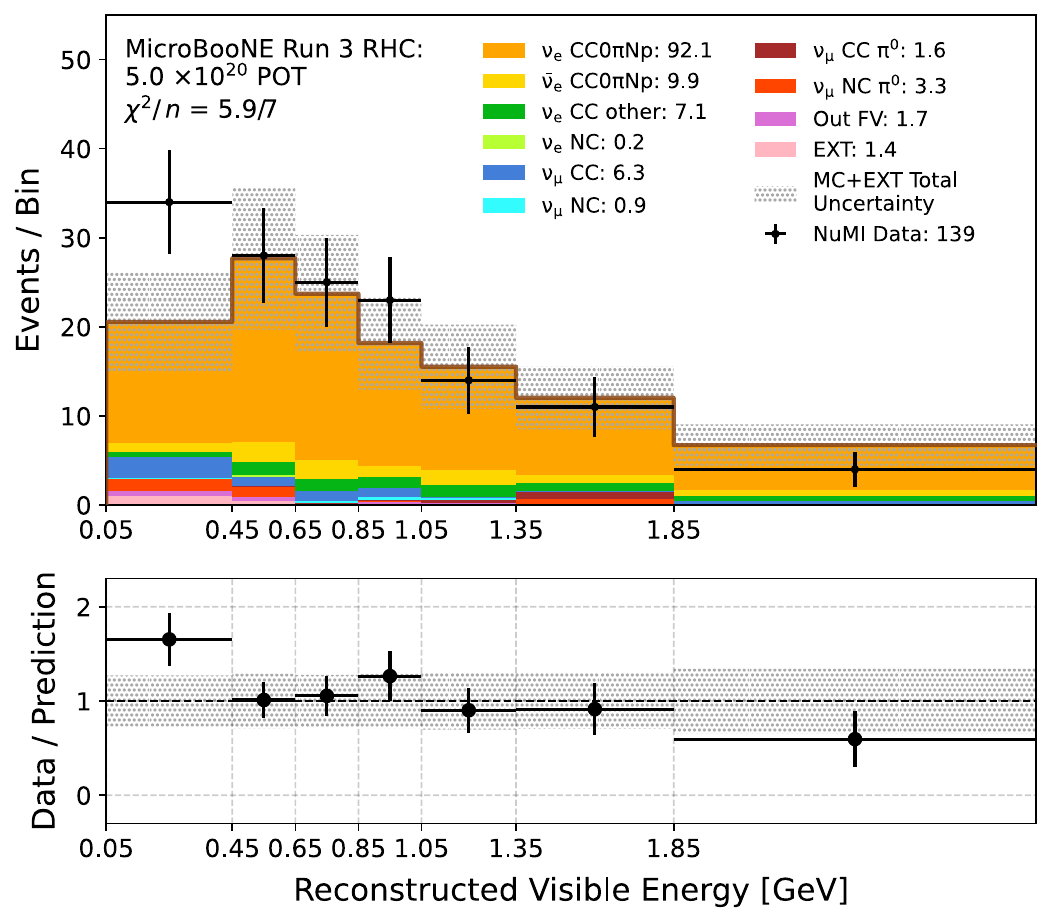}};
  \node at (3.6,-1.3) {\textbf{(b)}};
\end{tikzpicture}
\\[1.2em]
\multicolumn{2}{c}{
\begin{tikzpicture}
  \node[inner sep=0] {
    \includegraphics[width=0.48\textwidth]{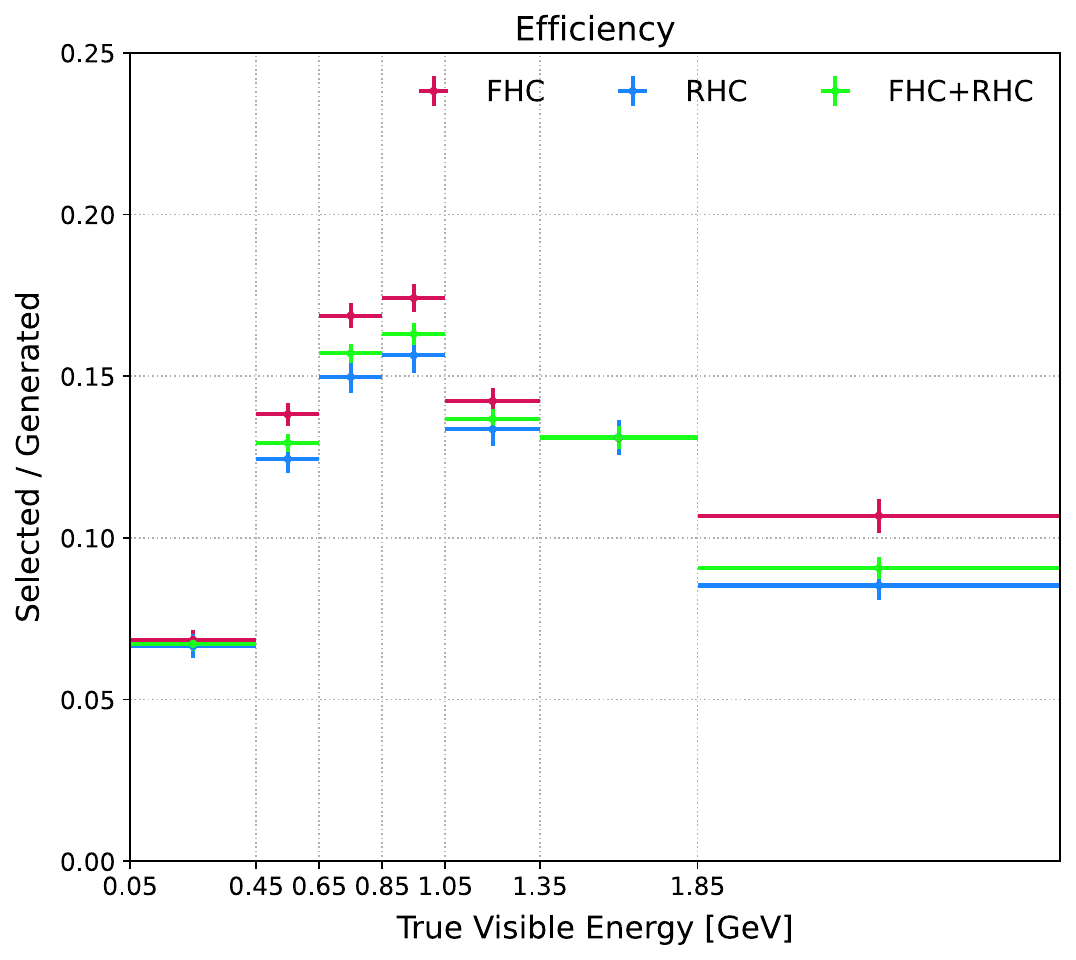}};
  \node at (3.6,-2.6) {\textbf{(c)}};
\end{tikzpicture}
}
\end{tabular}

\caption{Run~1 FHC (a) and Run~3 RHC (b) selected event rates as a function of reconstructed visible energy, and FHC-only, RHC-only, and FHC+RHC selected event rate efficiencies (c) after subtracting background predicted by the tuned \code{GENIE v3.0.6 G18\_10a\_02\_11a} model as a function of true visible energy.}
\label{fig:visible_energy_abc}
\end{figure}


\begin{figure}[h]
\centering
\begin{tabular}{cc}
\begin{tikzpicture}
  \node[inner sep=0] {
    \includegraphics[width=0.48\textwidth]{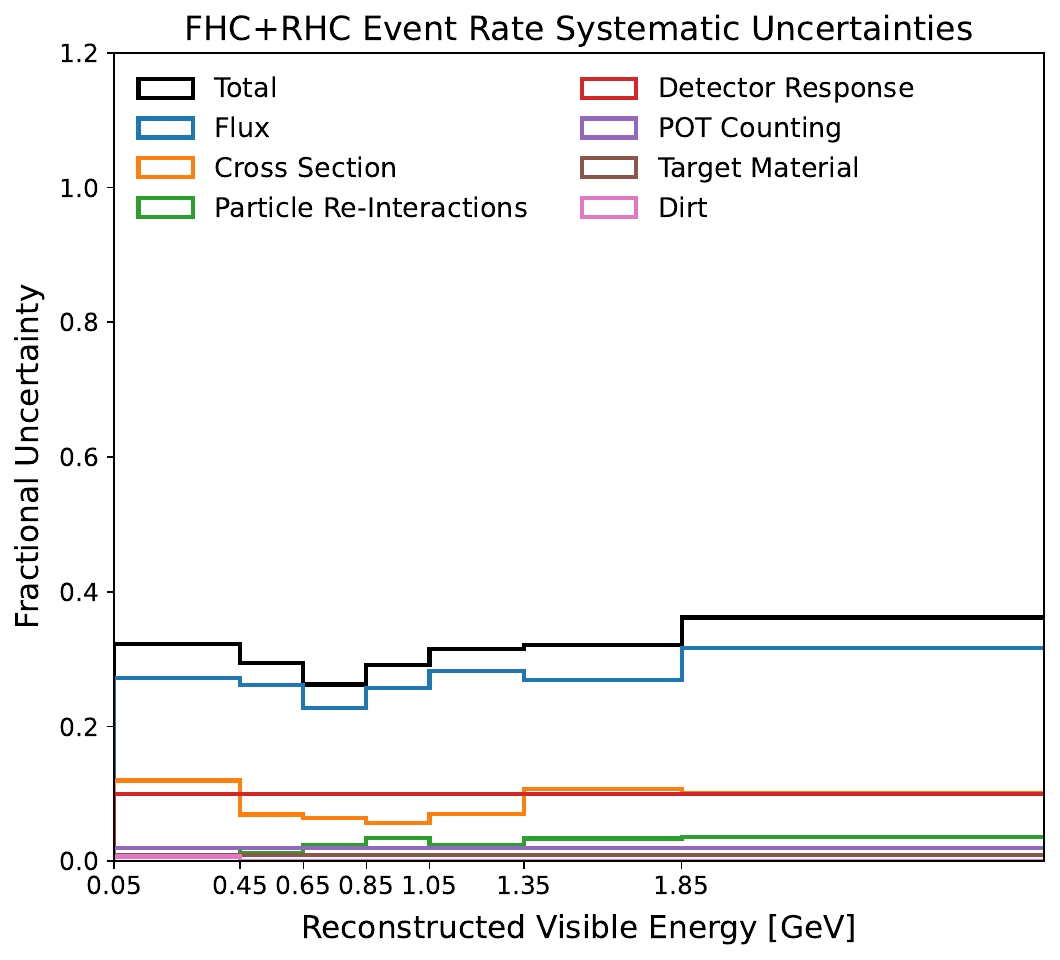}};
  \node at (3.75,3.1) {\textbf{(a)}};
\end{tikzpicture}
&
\begin{tikzpicture}
  \node[inner sep=0] {
    \includegraphics[width=0.48\textwidth]{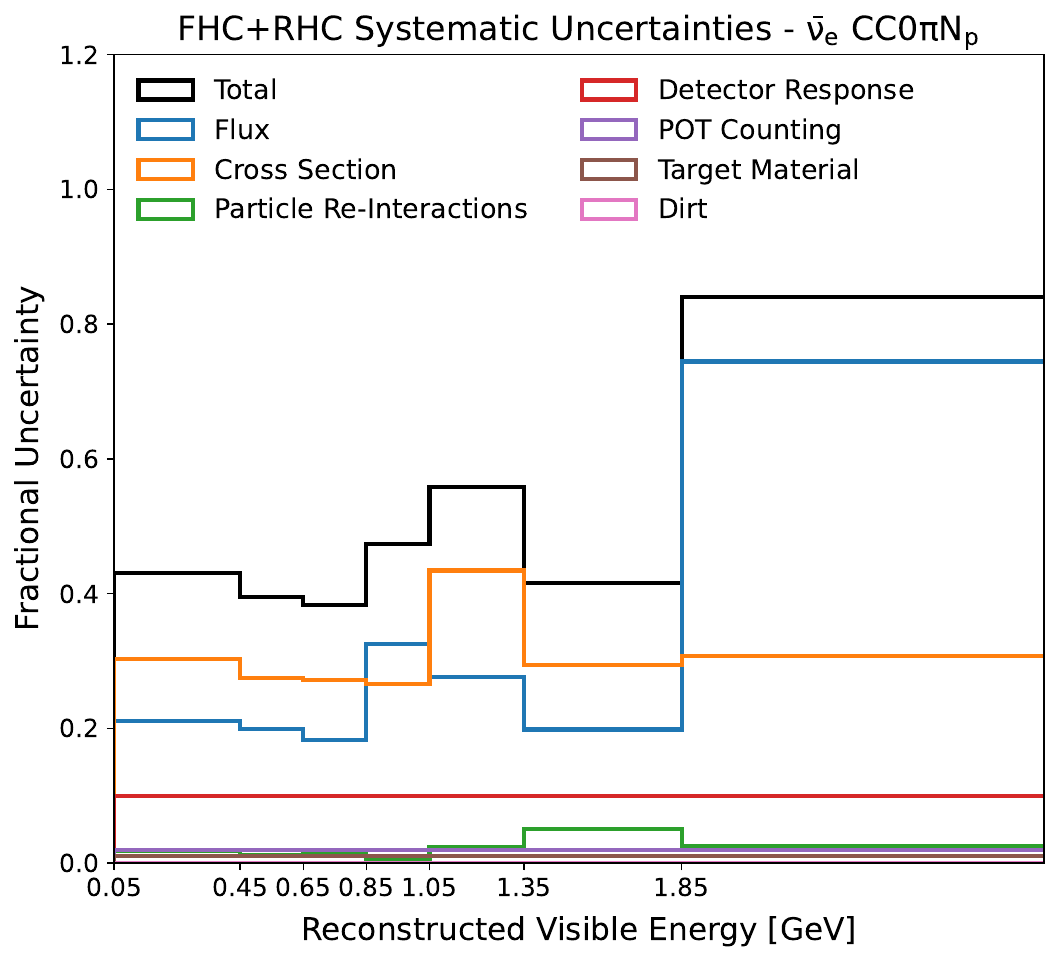}};
  \node at (3.75,3.1) {\textbf{(b)}};
\end{tikzpicture}
\end{tabular}

\caption{Fractional systematic uncertainties after subtracting background predicted by the tuned \code{GENIE v3.0.6 G18\_10a\_02\_11a} model (a) and on background $\bar{\nu}_{e}$ CC$0\pi Np$ events only (b) as a function of reconstructed visible energy.}
\label{fig:visible_angle_unc}
\end{figure}


\begin{figure}[h]
\centering

\includegraphics[width=.7\textwidth]{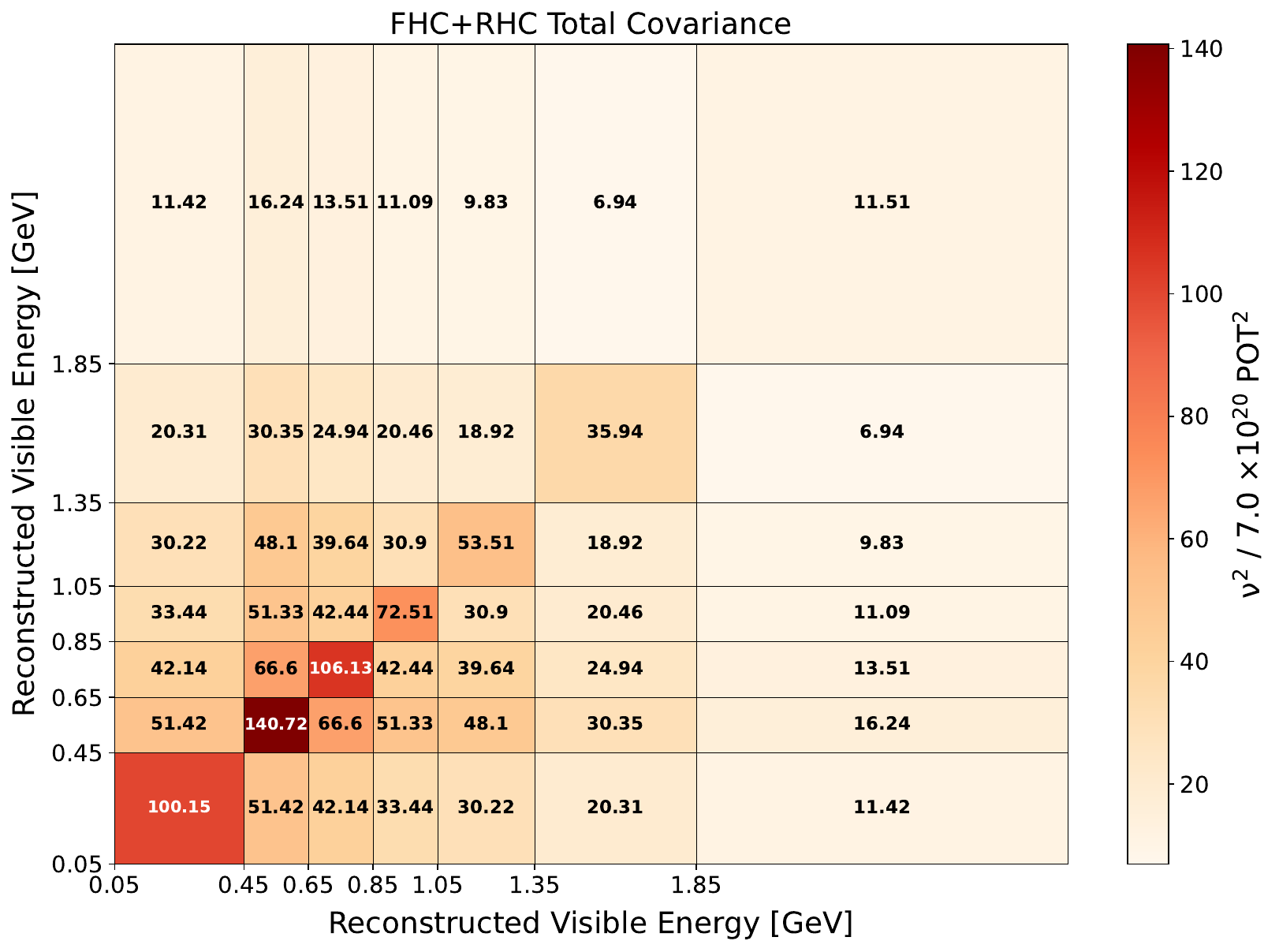}

\caption{{Total covariance (statistical + systematic) matrix for the FHC+RHC background-subtracted event rate as a function of reconstructed visible energy.}}
\end{figure}


\begin{figure}[h]
\centering
\begin{tabular}{cc}
\begin{tikzpicture}
  \node[inner sep=0] {
    \includegraphics[width=0.48\textwidth]{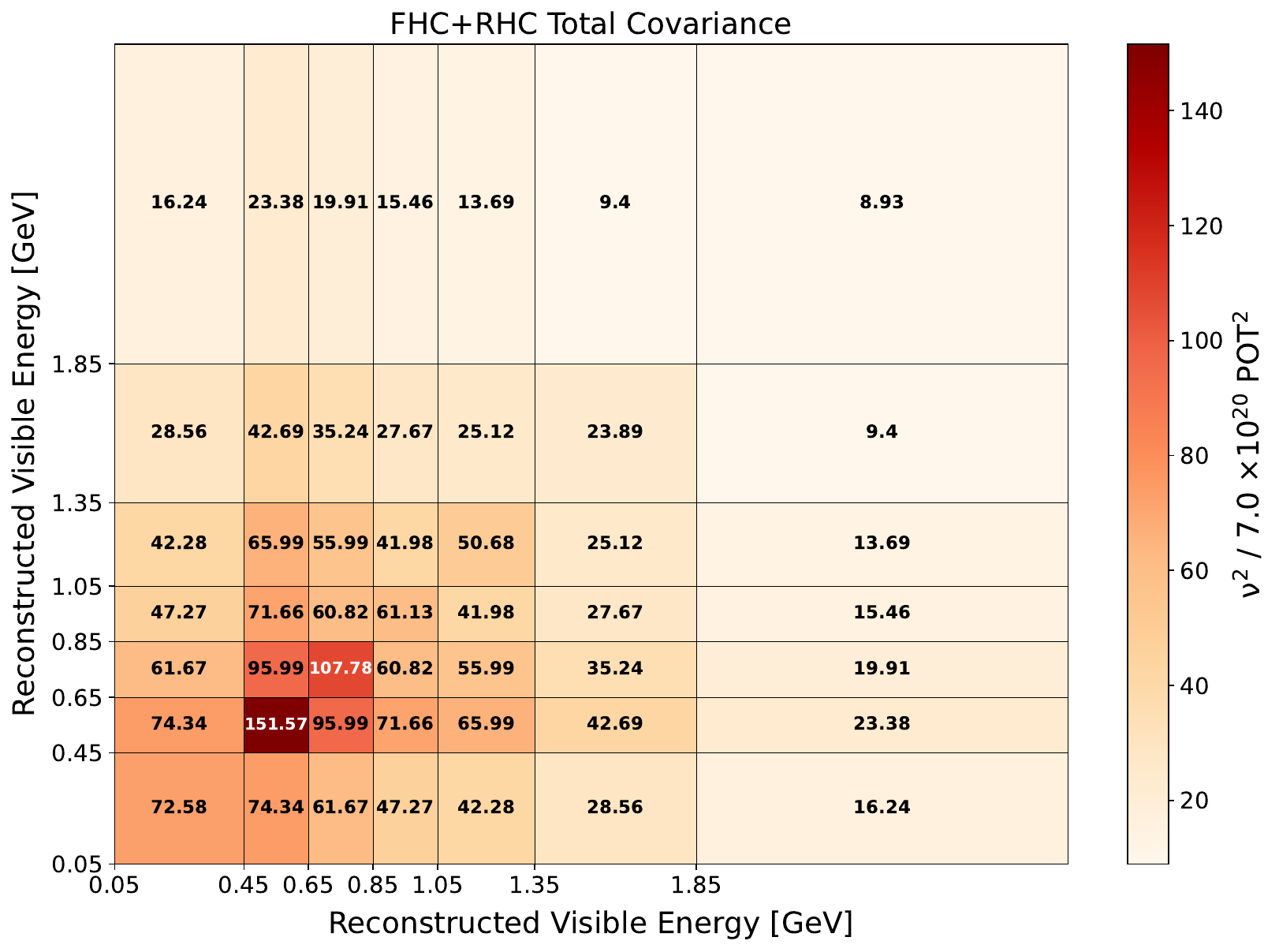}};
  \node at (2.5,2.5) {\textbf{(a)}};
\end{tikzpicture}
&
\begin{tikzpicture}
  \node[inner sep=0] {
    \includegraphics[width=0.48\textwidth]{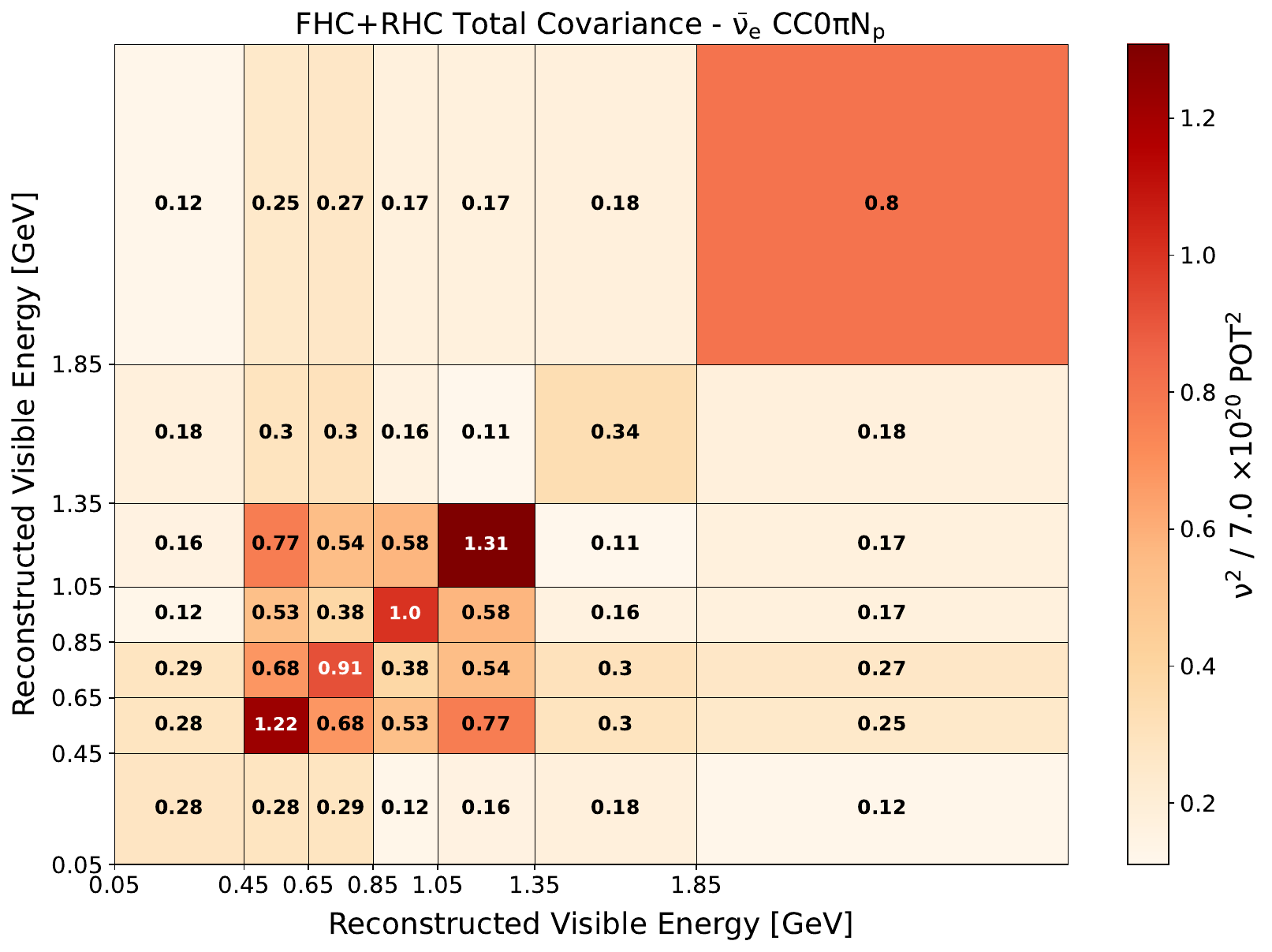}};
  \node at (2.5,2.5) {\textbf{(b)}};
\end{tikzpicture}
\end{tabular}

\caption{Total covariance (statistical + systematic) matrix for the FHC+RHC selected event rate (a) and $\bar{\nu}_{e}$ CC$0\pi Np$ background events only (b) as a function of reconstructed visible energy.}
\end{figure}


\begin{figure}[h]
\centering

\includegraphics[width=.7\textwidth]{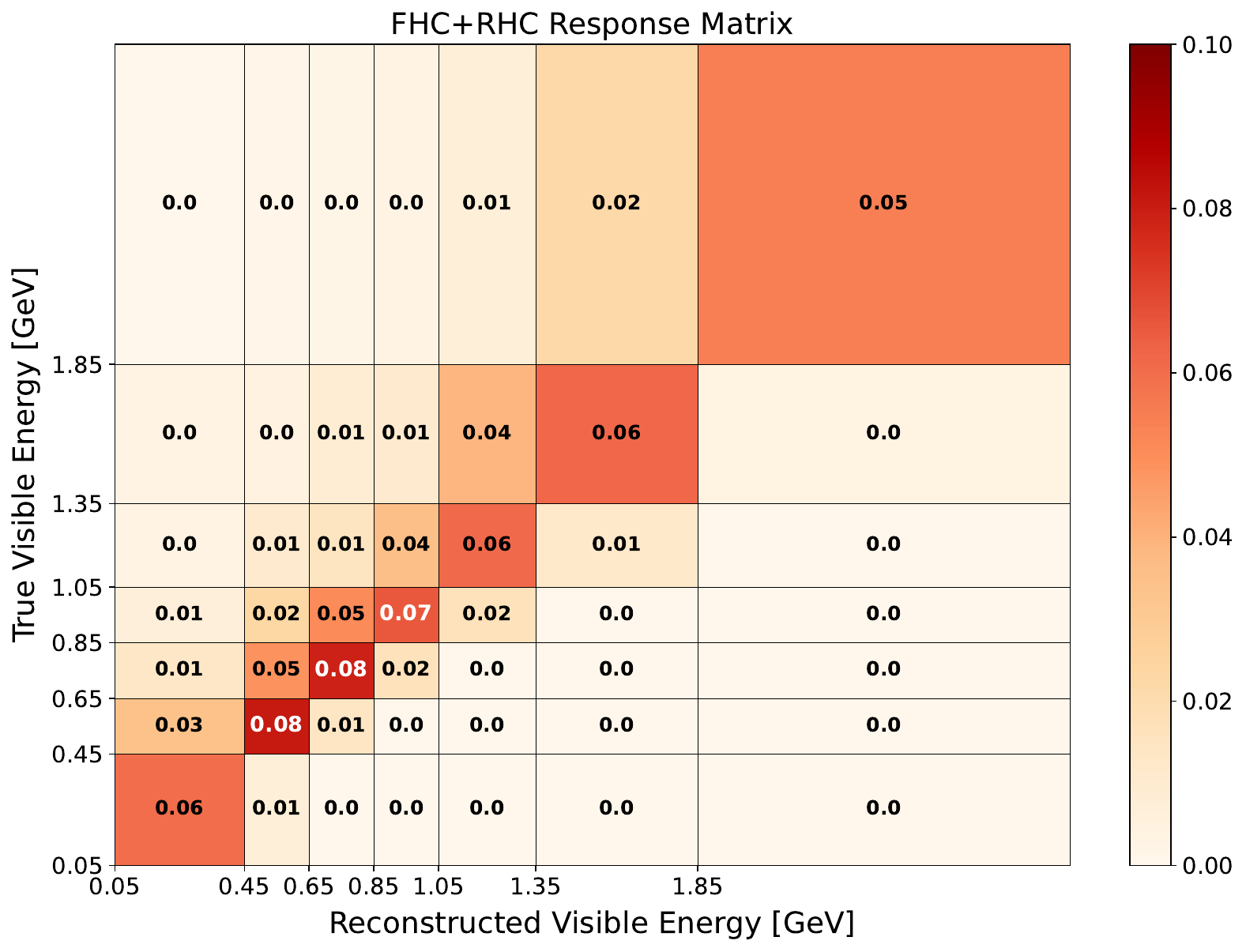}

\caption{{The response matrix, constructed using the \code{GENIE v3.0.6 G18\_10a\_02\_11a} selected signal prediction, as a function of reconstructed and true visible energy. Row $j$ of the matrix is normalized such that $\sum_i r_{ij} = \epsilon_j$, where $i$ is the column and $\epsilon_j$ is the estimated selection efficiency of true bin $j$. This histogram is used as input for the Wiener-SVD unfolding algorithm.}}
\end{figure}


\begin{figure}[h]
\centering

\includegraphics[width=.7\textwidth]{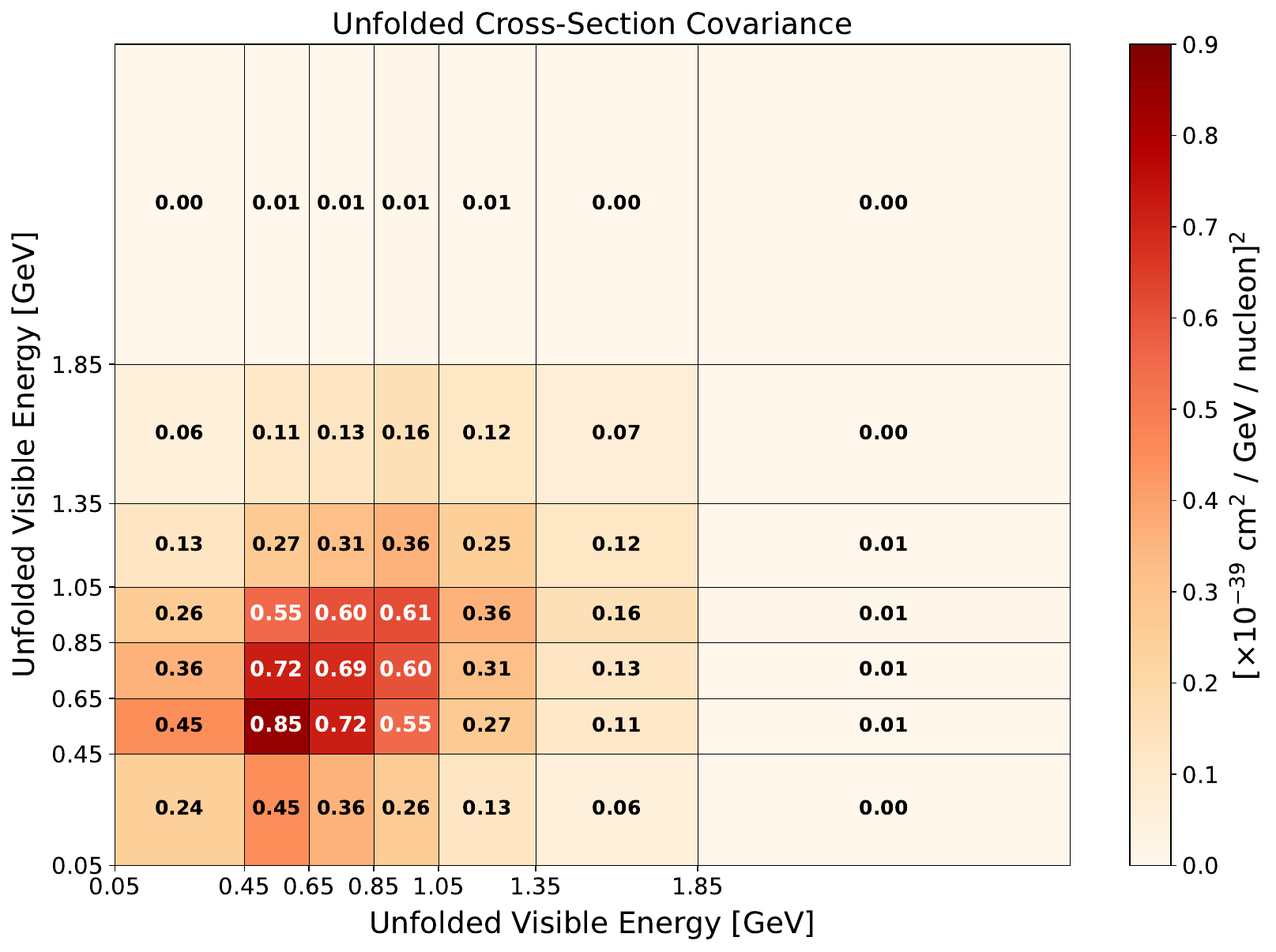}

\caption{{Unfolded covariance of the cross-section result as a function of visible energy, in units of $\left(10^{-39}\,\mathrm{cm}\,/\,\mathrm{GeV}\,/\,\mathrm{nucleon}\right)^{2}$.}}
\end{figure}


\begin{figure}[h]
\centering

\includegraphics[width=.7\textwidth]{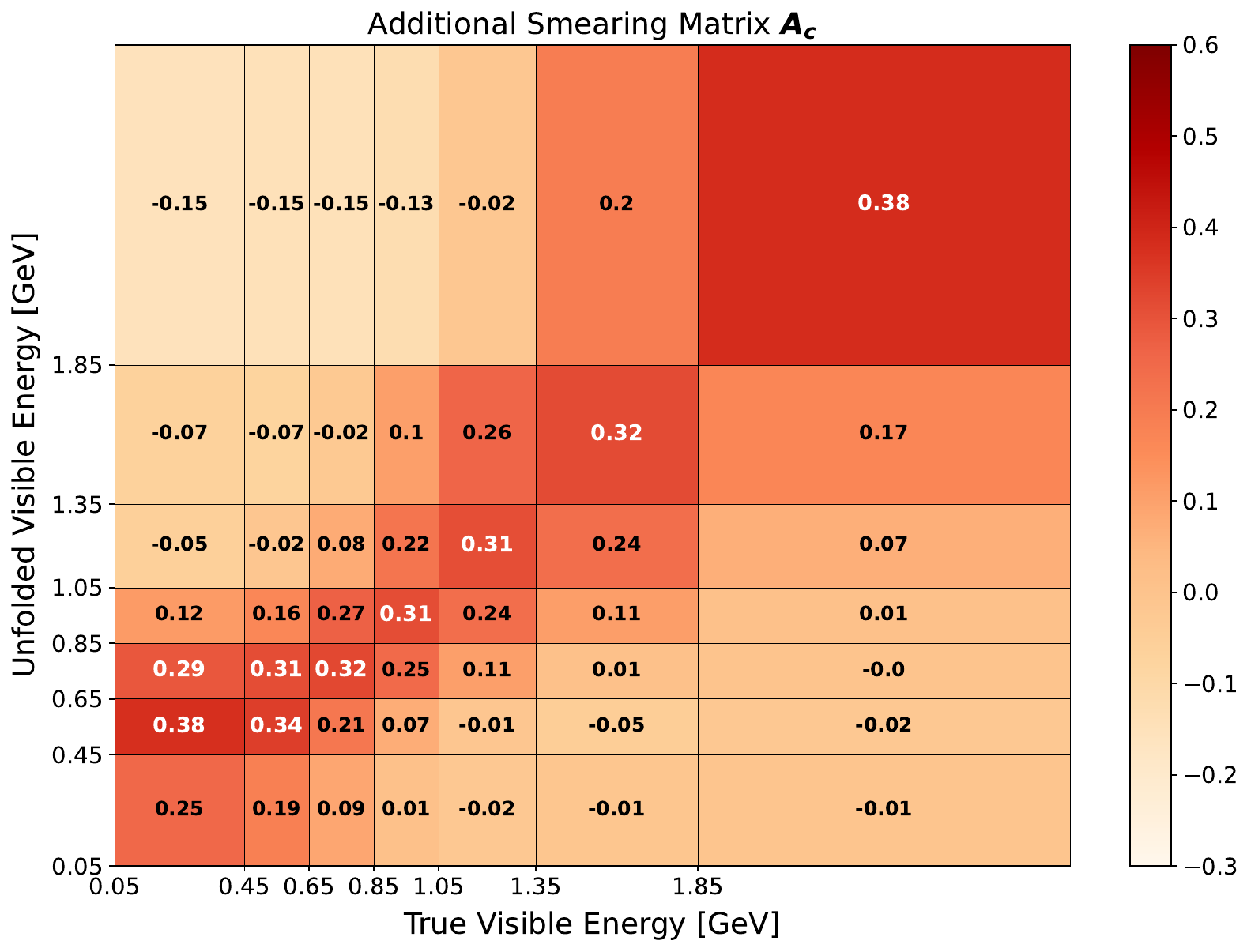}

\caption{{The additional smearing matrix $\bm{A_c}$ returned by the Wiener-SVD method for unfolded visible energy. This is used to smear event rate generator predictions of the event rate into unfolded space.}}
\end{figure}


\clearpage
\clearpage

\begin{table}
\caption{The background-subtracted event count and unfolded cross section as a function of visible energy. The unfolded cross section is in units of $10^{-39}\,\mathrm{cm}/\,\mathrm{GeV}/\,\mathrm{nucleon}$.}
\begin{ruledtabular}
\begin{tabular}{llllllll}
\textbf{Visible energy [GeV]} & \textbf{[0.05, 0.45]} & \textbf{[0.45, 0.65]} & \textbf{[0.65, 0.85]} & \textbf{[0.85, 1.05]} & \textbf{[1.05, 1.35]} & \textbf{[1.35, 1.85]} & \textbf{[1.85, $\infty$]} \\
\hline
Background-subtracted event count & 42.04 & 28.77 & 30.64 & 24.78 & 12.38 & 14.32 & 2.44  \\ 
Unfolded cross section & 2.10 & 3.78 & 3.31 & 2.74 & 1.42 & 0.64 & 0.03 \\

\end{tabular}
\end{ruledtabular}

\end{table}


\newpage

\clearpage
\section{Cosine of the Opening Angle}

The material in this section provides additional information for the cross-section measurement as a function of the cosine of the opening angle: the total selected event rates and associated selection efficiencies, fractional systematic uncertainty breakdowns for the signal and $\bar{\nu}_e$ CC$0\pi Np$ background, response matrix, unfolded central-value cross-section values and covariance, and the additional smearing matrix $\bm{A_c}$ returned by the Wiener-SVD method. Here, detector systematic uncertainty is included in the gray error bands and the calculation of $\chi^2/n$. 


\begin{figure}[h]
\centering
\begin{tabular}{cc}
\begin{tikzpicture}
  \node[inner sep=0] {
    \includegraphics[width=0.48\textwidth]{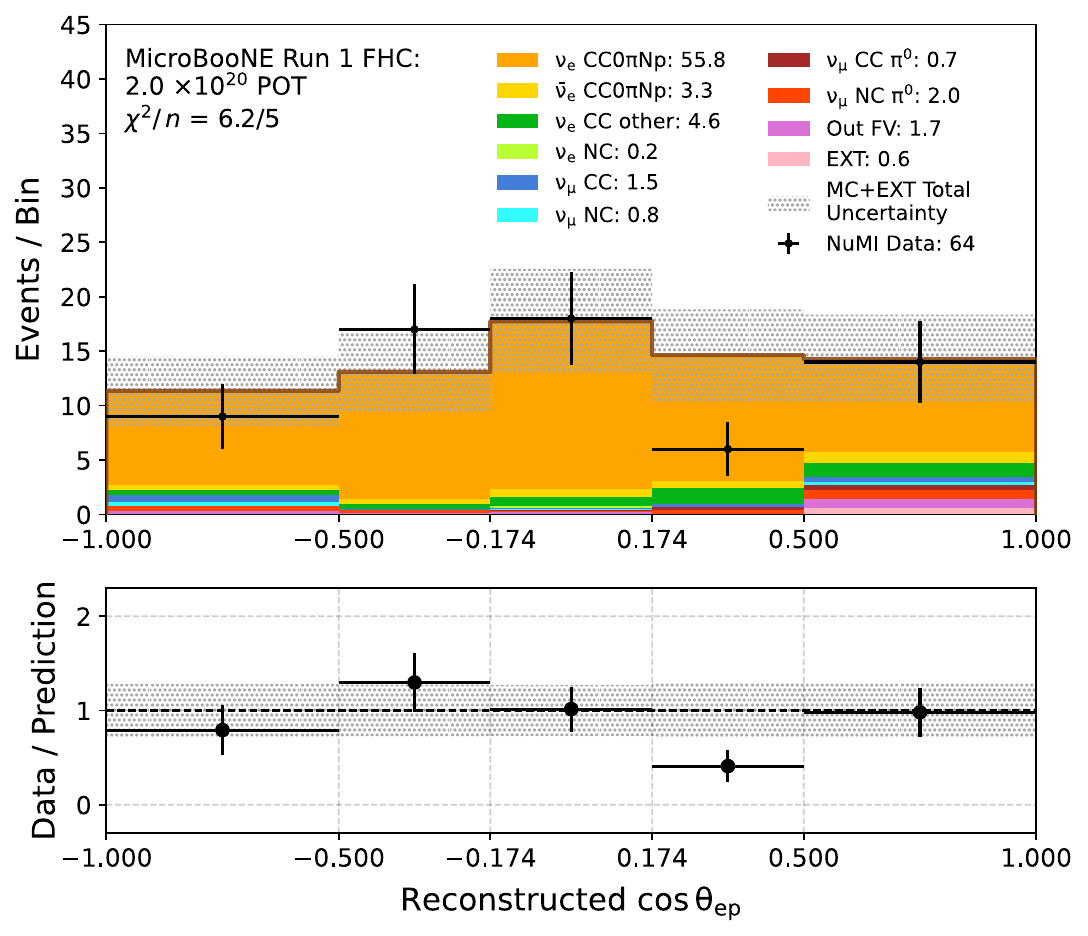}};
  \node at (3.6,-1.3) {\textbf{(a)}};
\end{tikzpicture}
& 
\begin{tikzpicture}
  \node[inner sep=0] {
    \includegraphics[width=0.48\textwidth]{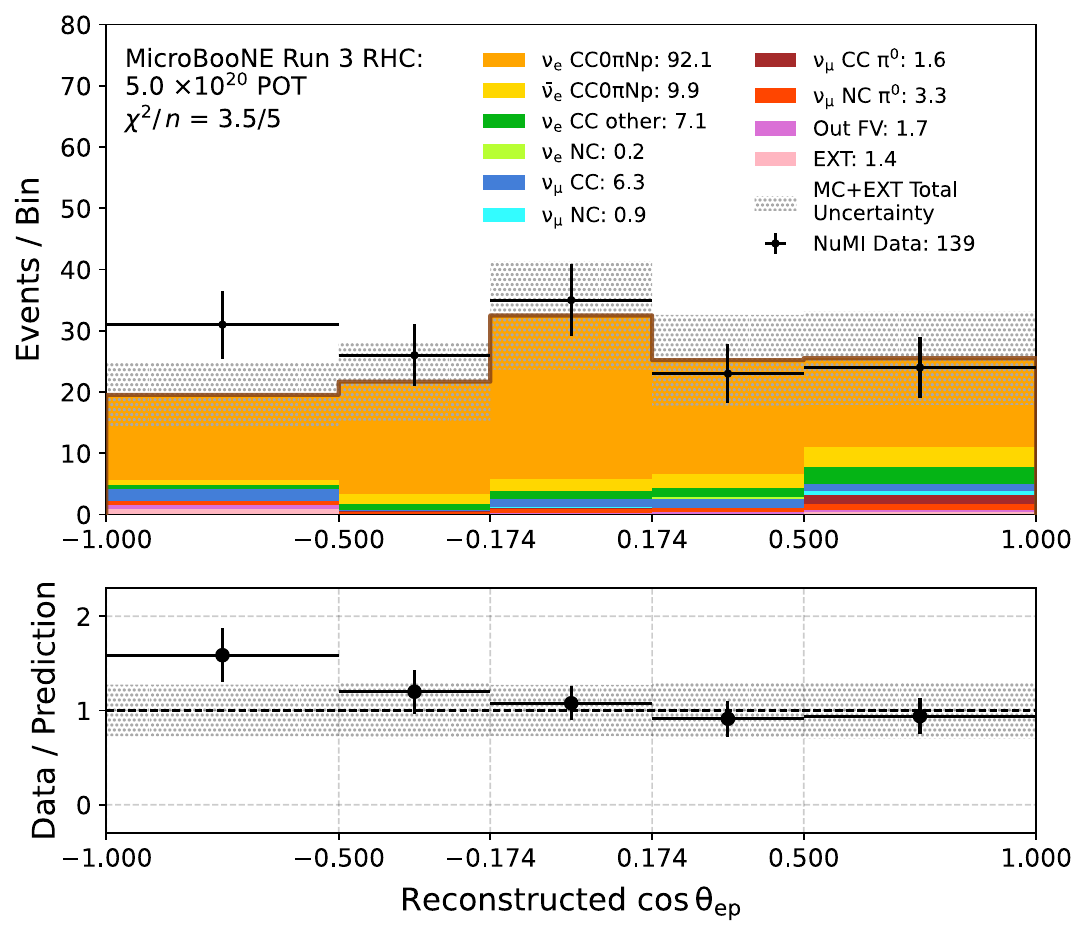}};
  \node at (3.6,-1.3) {\textbf{(b)}};
\end{tikzpicture}
\\[1.2em]
\multicolumn{2}{c}{
\begin{tikzpicture}
  \node[inner sep=0] {
    \includegraphics[width=0.48\textwidth]{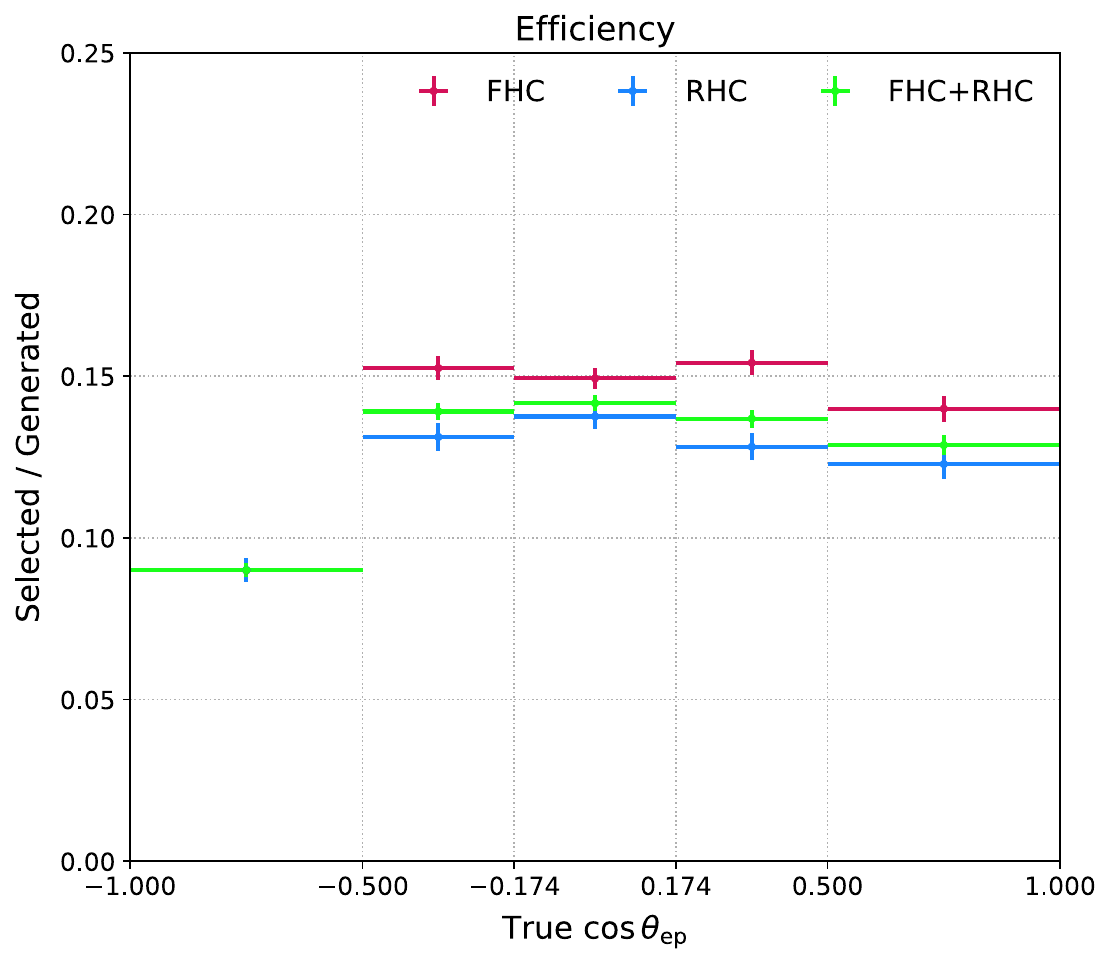}};
  \node at (3.6,-2.6) {\textbf{(c)}};
\end{tikzpicture}
}
\end{tabular}

\caption{Run~1 FHC (a) and Run~3 RHC (b) selected event rates as a function of the reconstructed cosine of the opening angle, and FHC-only, RHC-only, and FHC+RHC selected event rate efficiencies (c) after subtracting background predicted by the tuned \code{GENIE v3.0.6 G18\_10a\_02\_11a} model as a function of true cosine of the opening angle.}
\label{fig:opening_angle_abc}
\end{figure}


\begin{figure}[h]
\centering
\begin{tabular}{cc}
\begin{tikzpicture}
  \node[inner sep=0] {
    \includegraphics[width=0.48\textwidth]{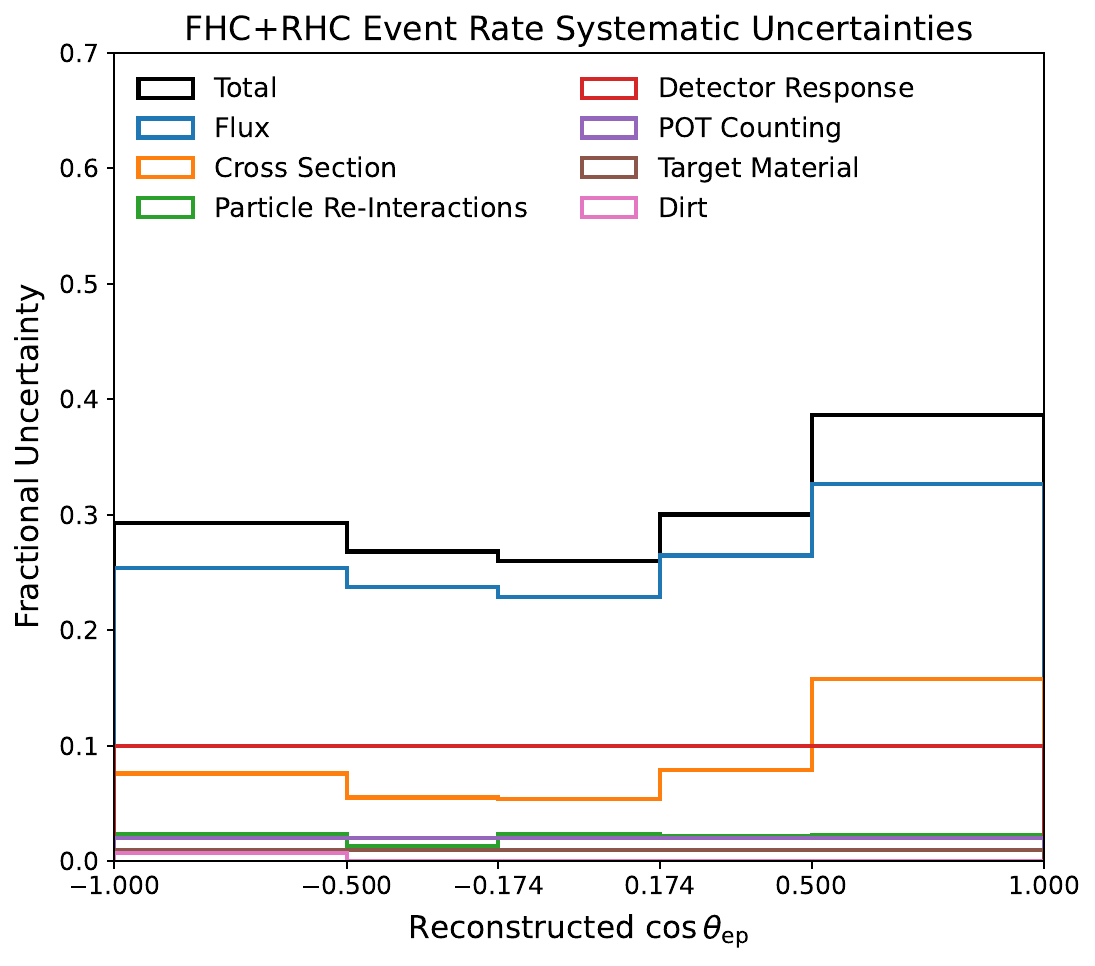}};
  \node at (3.55,3.0) {\textbf{(a)}};
\end{tikzpicture}
&
\begin{tikzpicture}
  \node[inner sep=0] {
    \includegraphics[width=0.48\textwidth]{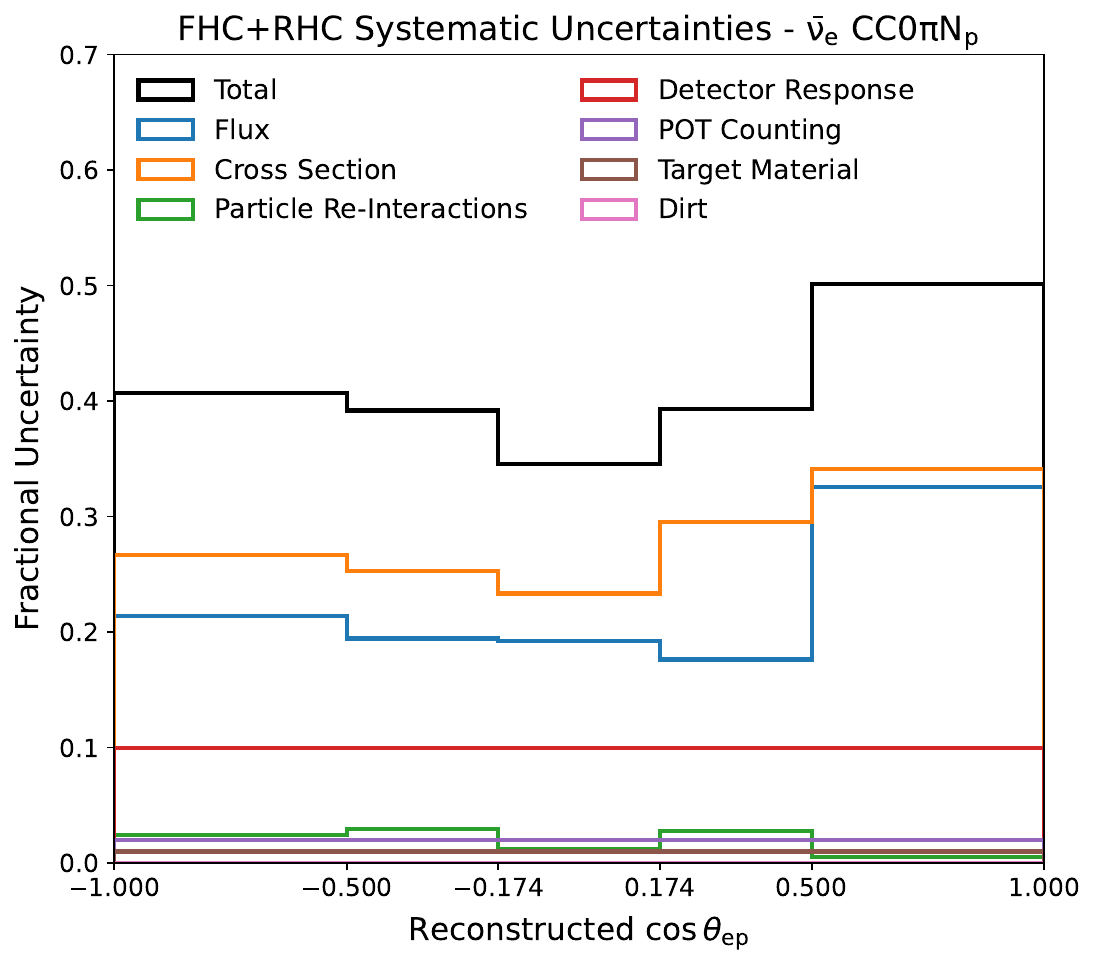}};
  \node at (3.55,3.0) {\textbf{(b)}};
\end{tikzpicture}
\end{tabular}

\caption{Fractional systematic uncertainties after subtracting background predicted by the tuned \code{GENIE v3.0.6 G18\_10a\_02\_11a} model (a) and on background $\bar{\nu}_{e}$ CC$0\pi Np$ events only (b) as a function of the reconstructed cosine of the opening angle.}
\label{fig:opening_angle_unc}
\end{figure}


\begin{figure}[h]
\centering

\includegraphics[width=.7\textwidth]{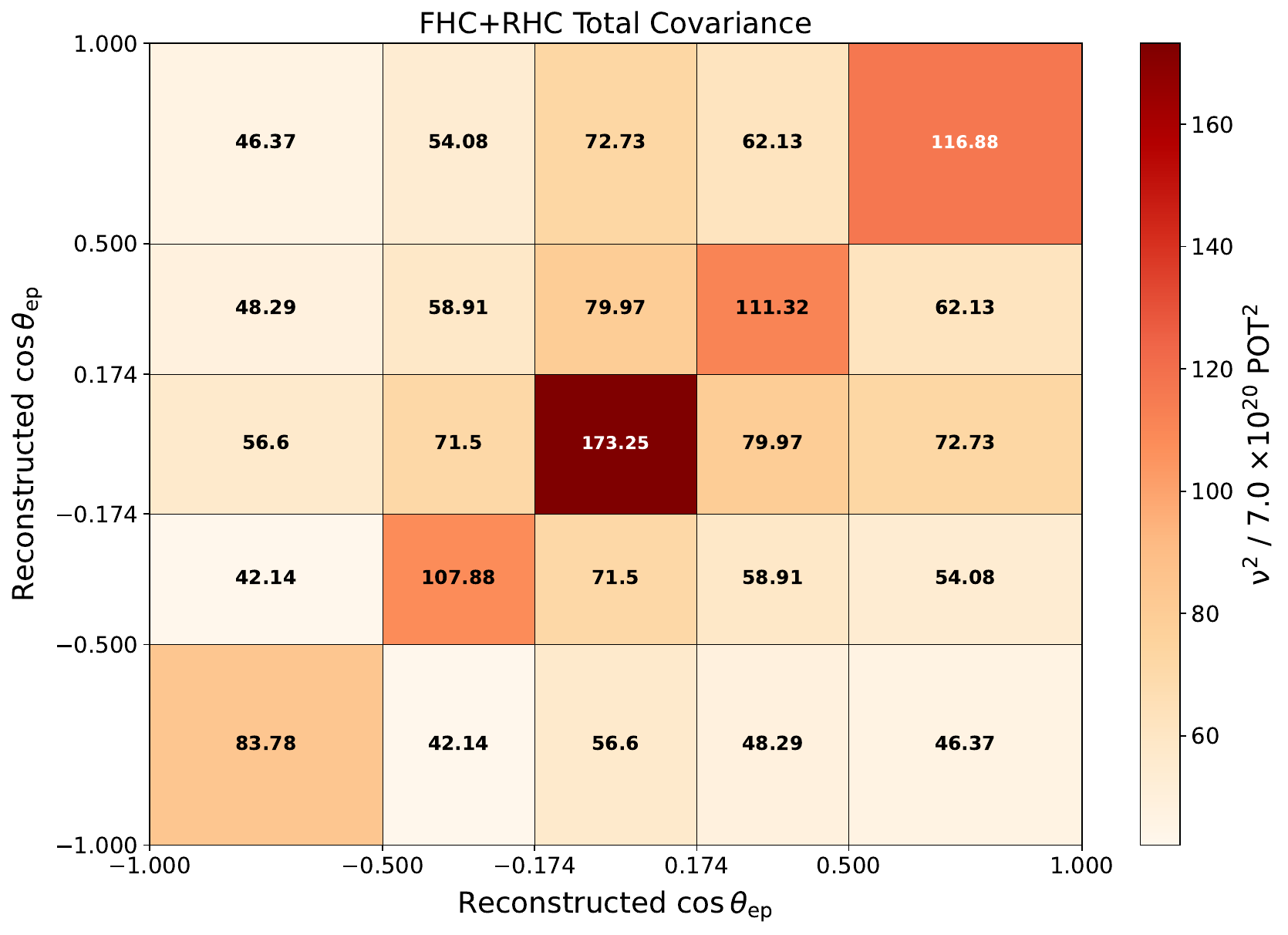}

\caption{{Total covariance (statistical + systematic) matrix for the FHC+RHC background-subtracted event rate as a function of the reconstructed cosine of the opening angle.}}
\end{figure}


\begin{figure}[h]
\centering
\begin{tabular}{cc}
\begin{tikzpicture}
  \node[inner sep=0] {
    \includegraphics[width=0.48\textwidth]{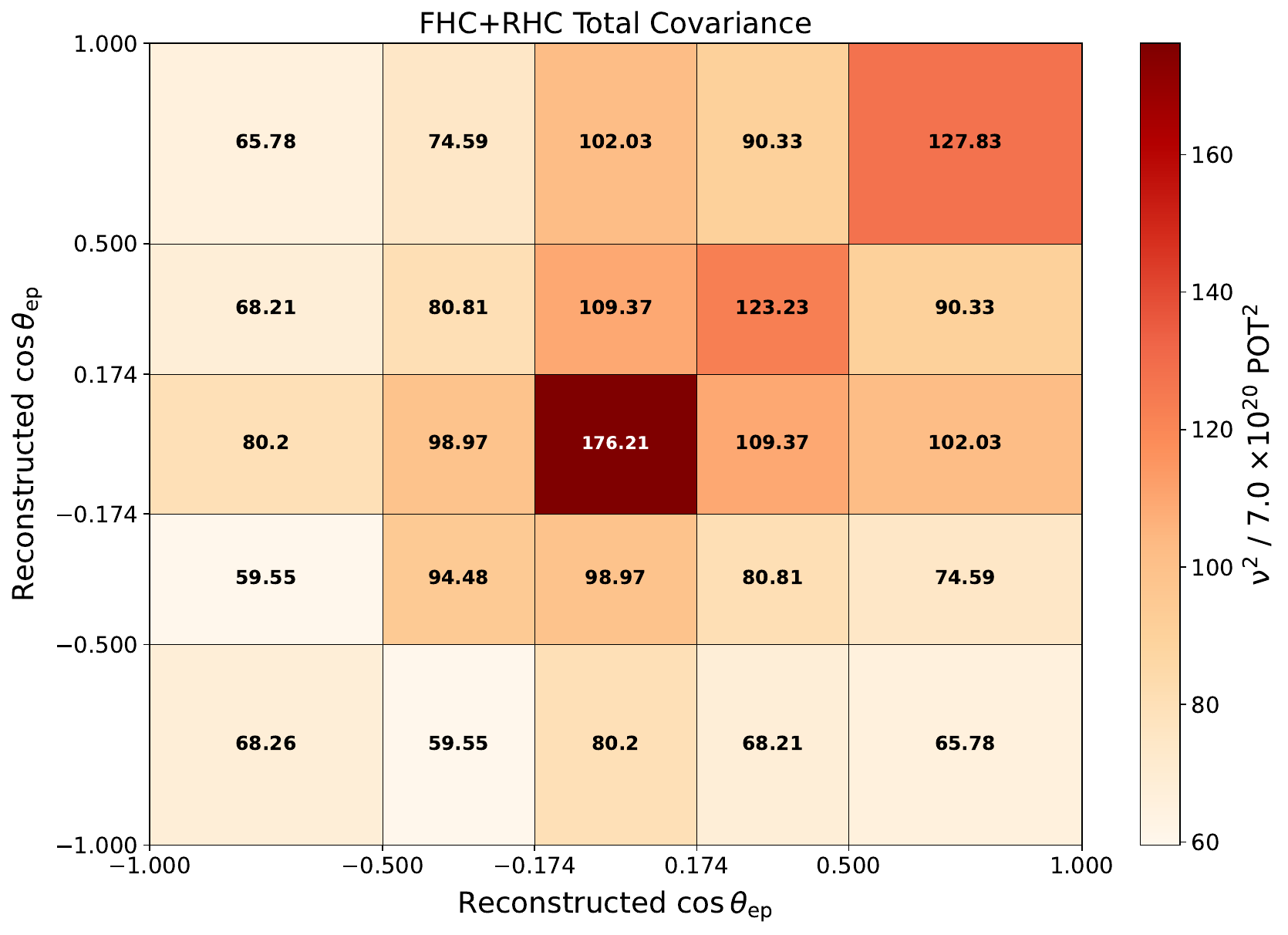}};
  \node at (-2.95,2.55) {\textbf{(a)}};
\end{tikzpicture}
&
\begin{tikzpicture}
  \node[inner sep=0] {
    \includegraphics[width=0.48\textwidth]{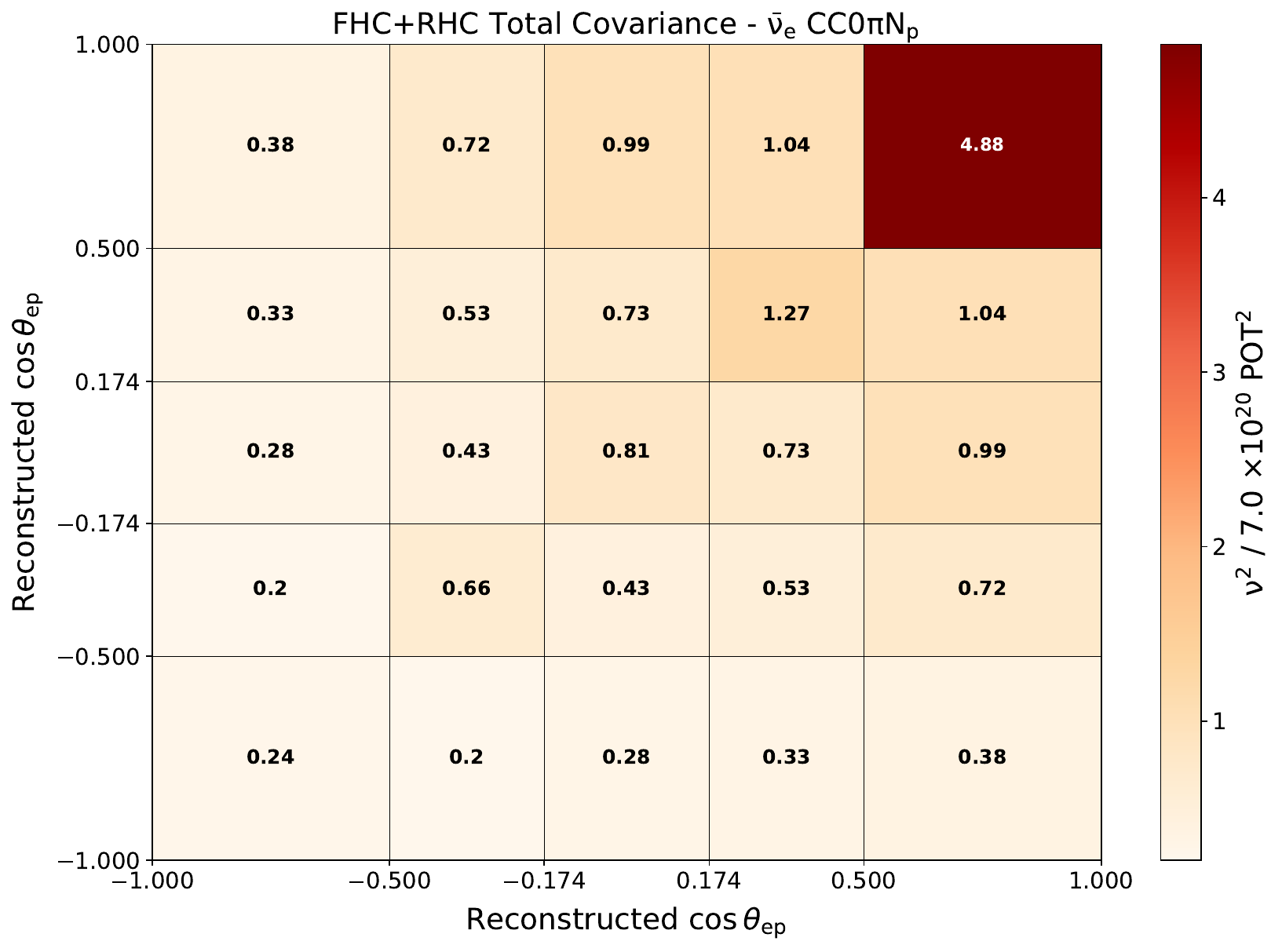}};
  \node at (-2.9,2.6) {\textbf{(b)}};
\end{tikzpicture}
\end{tabular}

\caption{Total covariance (statistical + systematic) matrix for the FHC+RHC selected event rate (a) and $\bar{\nu}_{e}$ CC$0\pi Np$ background events only (b) as a function of the reconstructed cosine of the opening angle.}
\end{figure}


\begin{figure}[h]
\centering

\includegraphics[width=.7\textwidth]{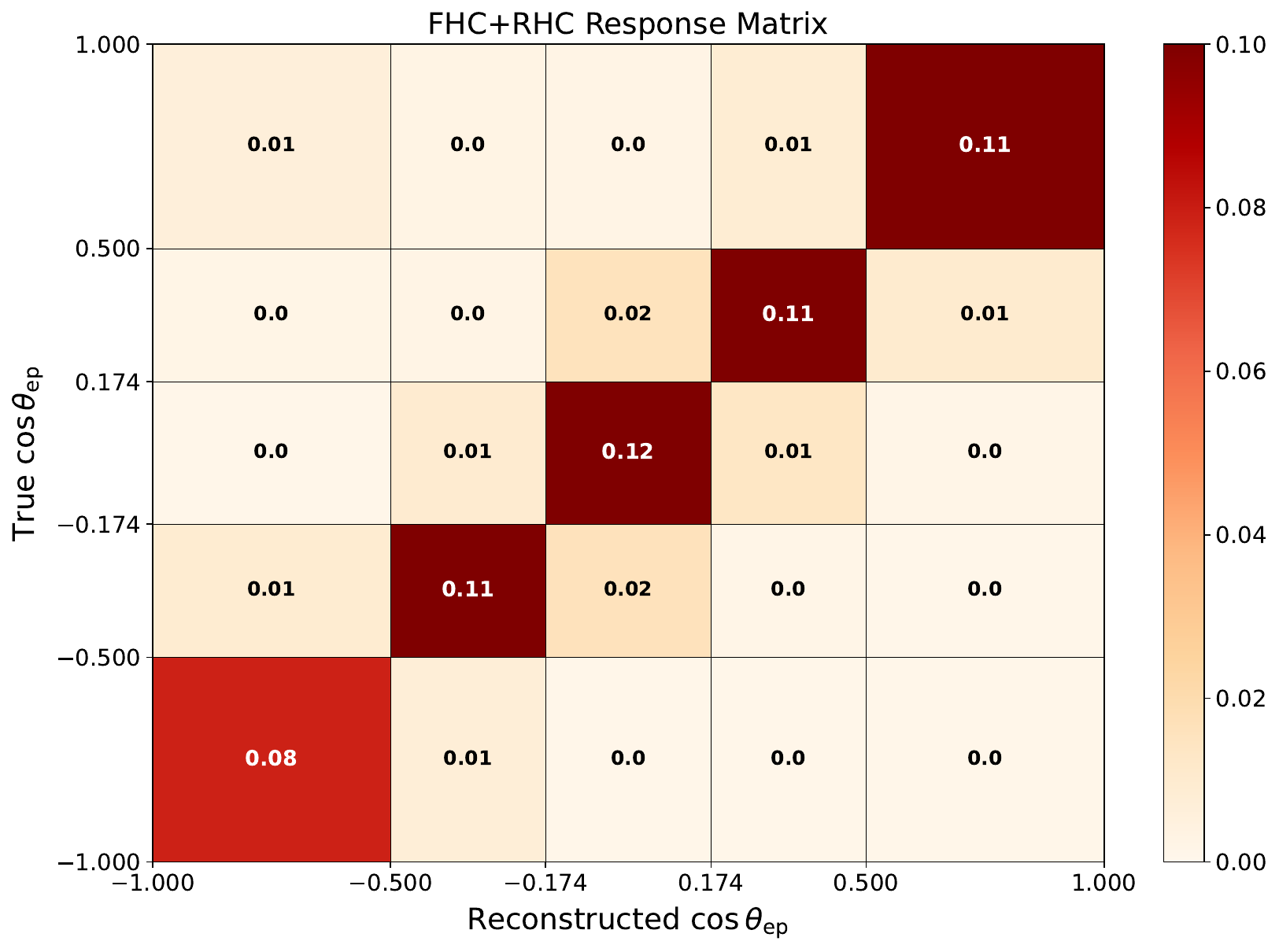}

\caption{{The response matrix, constructed using the \code{GENIE v3.0.6 G18\_10a\_02\_11a} selected signal prediction, as a function of the reconstructed and true cosine of the opening angle. Row $j$ of the matrix is normalized such that $\sum_i r_{ij} = \epsilon_j$, where $i$ is the column and $\epsilon_j$ is the estimated selection efficiency of true bin $j$. This histogram is used as input for the Wiener-SVD unfolding algorithm.}}
\end{figure}


\begin{figure}[h]
\centering

\includegraphics[width=.7\textwidth]{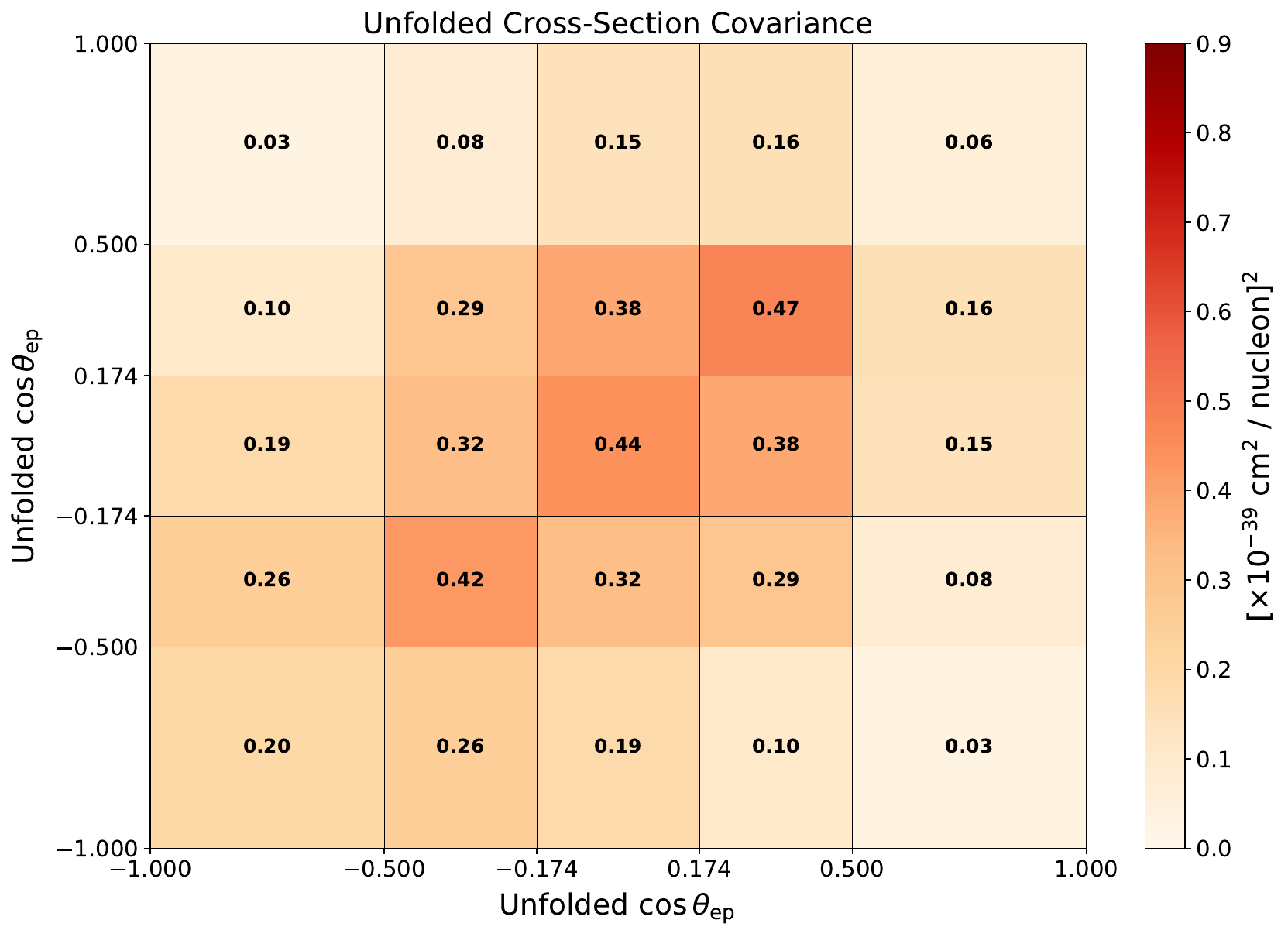}

\caption{{Unfolded covariance of the cross-section result as a function of the cosine of the opening angle, in units of $\left[10^{-39}\,\mathrm{cm}\,/\,\mathrm{nucleon}\right]^{2}$.}}
\end{figure}


\begin{figure}[h]
\centering

\includegraphics[width=.7\textwidth]{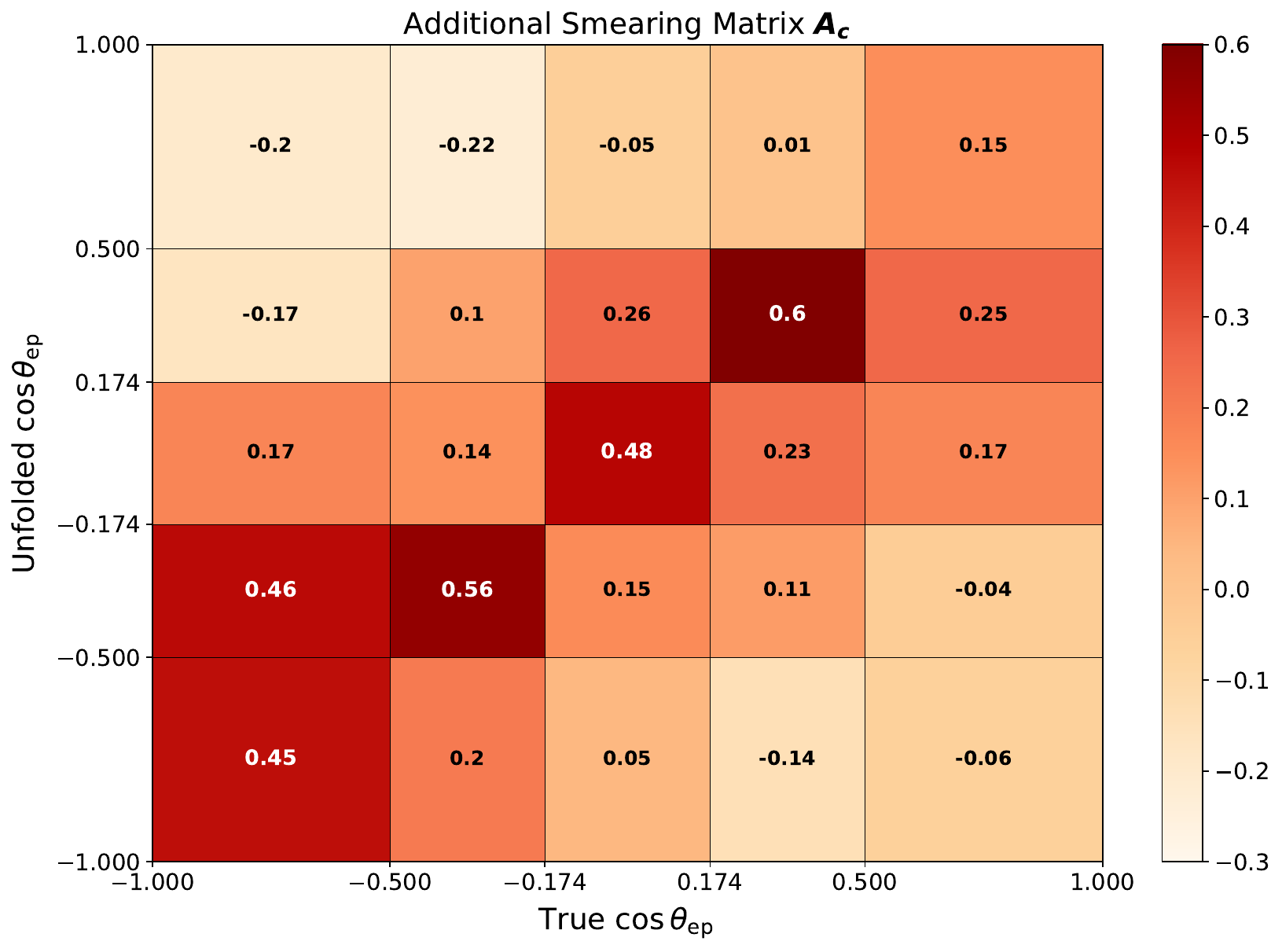}

\caption{{The additional smearing matrix $\bm{A_c}$ returned by the Wiener-SVD method for unfolded cosine of the opening angle. This is used to smear event rate generator predictions into unfolded space. }}
\end{figure}


\clearpage

\begin{table}
\caption{The background-subtracted event count and unfolded cross section as a function of the cosine of the opening angle. The unfolded cross section is in units of $10^{-39}\,\mathrm{cm}\,/\,\mathrm{nucleon}$.}
\begin{ruledtabular}
\begin{tabular}{llllll}
\textbf{$\cos\: \theta_{ep}$} & \textbf{[-1, -0.5]} & \textbf{[-0.5, -0.174]} & \textbf{[-0.174, 0.174]} & \textbf{[0.174, 0.5]} & \textbf{[0.5, 1]}  \\
\hline
Background-subtracted event count & 31.67 & 38.24 & 44.91 & 19.37 & 21.16  \\ 
Unfolded cross section & 1.97 & 2.65 & 2.36 & 1.38 & 0.54

\end{tabular}
\end{ruledtabular}

\end{table}


\normalsize 
\clearpage
\section{Proton Multiplicity}

The material in this section provides information about the distribution of $\nu_{e}+\,^{40}{\rm Ar} \rightarrow 1e+Np+0\pi$ interactions as a function of proton multiplicity needed to extract a cross section. This includes the total selected event rates and associated selection efficiencies, fractional systematic uncertainty breakdowns for the signal and $\bar{\nu}_e$ CC$0\pi Np$ background, and the response matrix. Detector systematic uncertainty is included in the gray error bands and the reported $\chi^2/n$ shown in this section.


\begin{figure}[h]
\centering
\begin{tabular}{cc}
\begin{tikzpicture}
  \node[inner sep=0] {
    \includegraphics[width=0.48\textwidth]{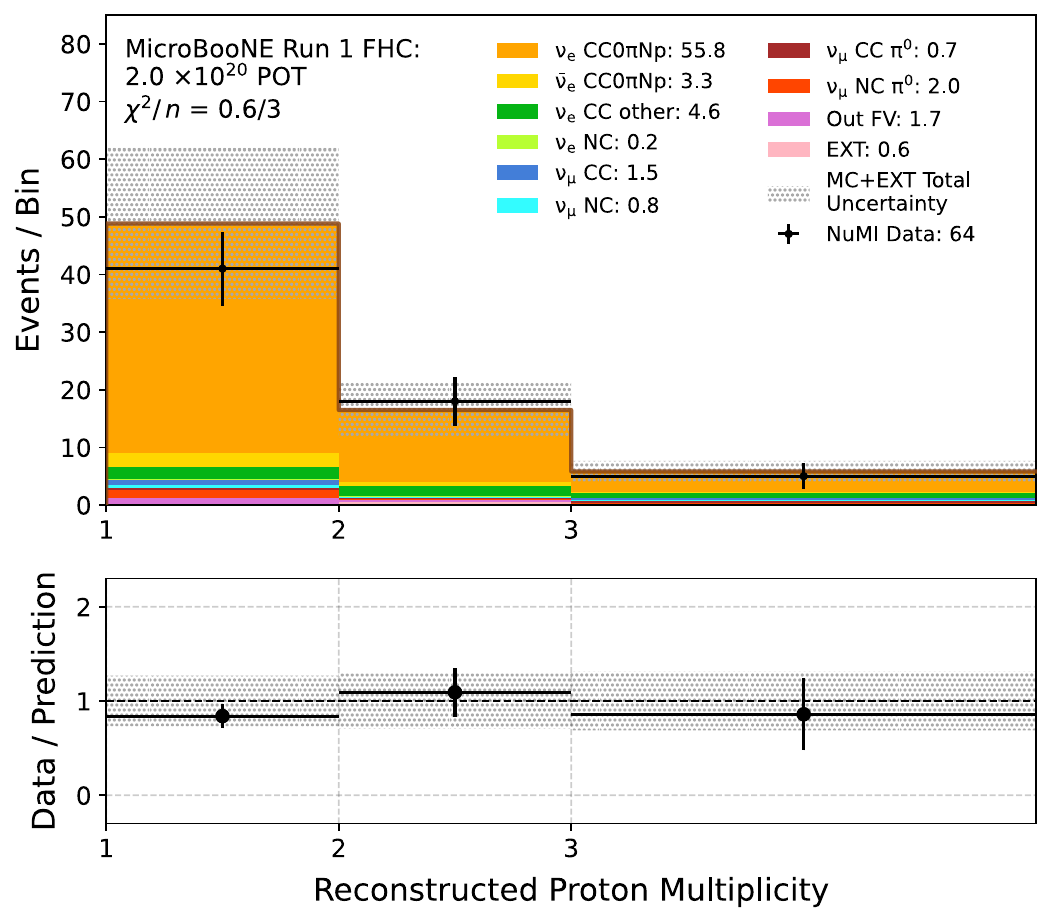}};
  \node at (3.6,-1.3) {\textbf{(a)}};
\end{tikzpicture}
& 
\begin{tikzpicture}
  \node[inner sep=0] {
    \includegraphics[width=0.48\textwidth]{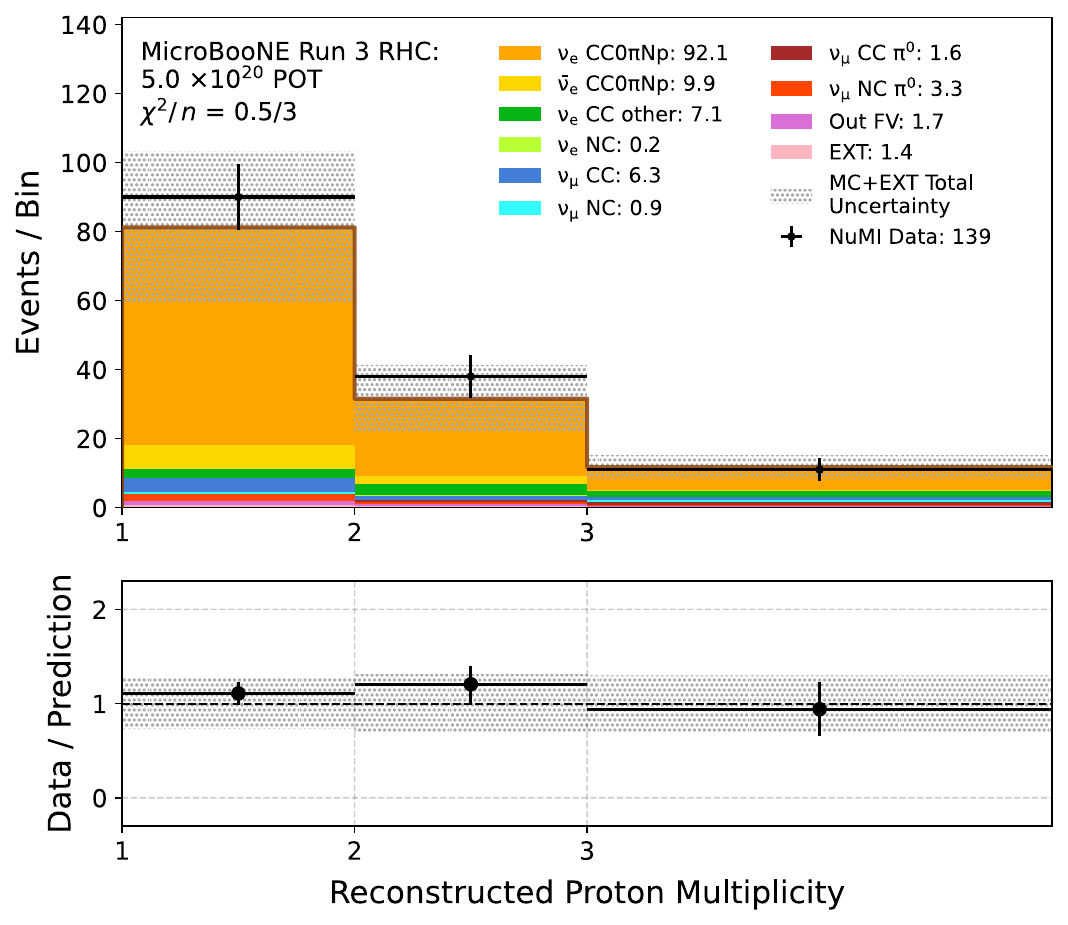}};
  \node at (3.6,-1.3) {\textbf{(b)}};
\end{tikzpicture}
\\[1.2em]
\multicolumn{2}{c}{
\begin{tikzpicture}
  \node[inner sep=0] {
    \includegraphics[width=0.48\textwidth]{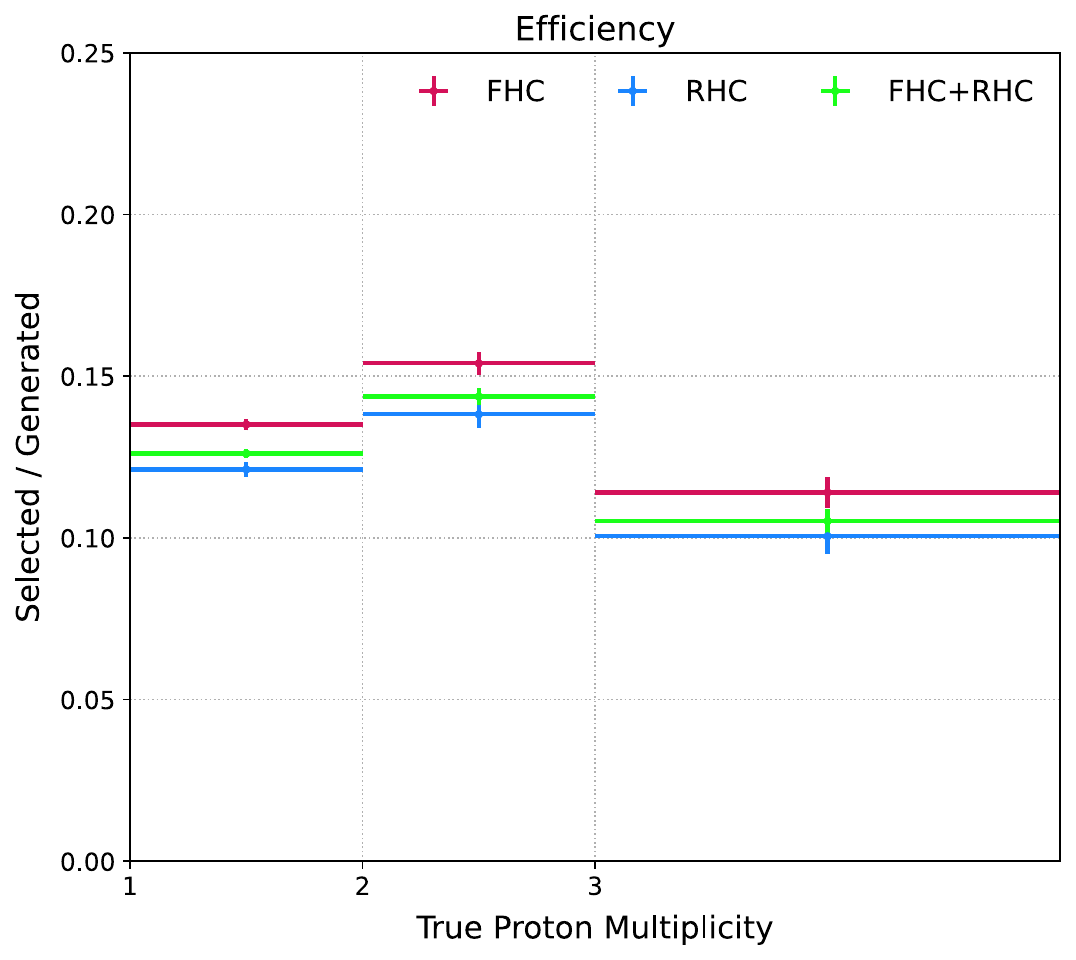}};
  \node at (3.6,-2.6) {\textbf{(c)}};
\end{tikzpicture}
}
\end{tabular}

\caption{Run~1 FHC (a) and Run~3 RHC (b) selected event rates as a function of reconstructed proton multiplicity, and FHC-only, RHC-only, and FHC+RHC selected event rate efficiencies (c) after subtracting background predicted by the tuned \code{GENIE v3.0.6 G18\_10a\_02\_11a} model as a function of true proton multiplicity.}
\label{fig:proton_multiplicity_abc}
\end{figure}


\begin{figure}[h]
\centering
\begin{tabular}{cc}
\begin{tikzpicture}
  \node[inner sep=0] {
    \includegraphics[width=0.48\textwidth]{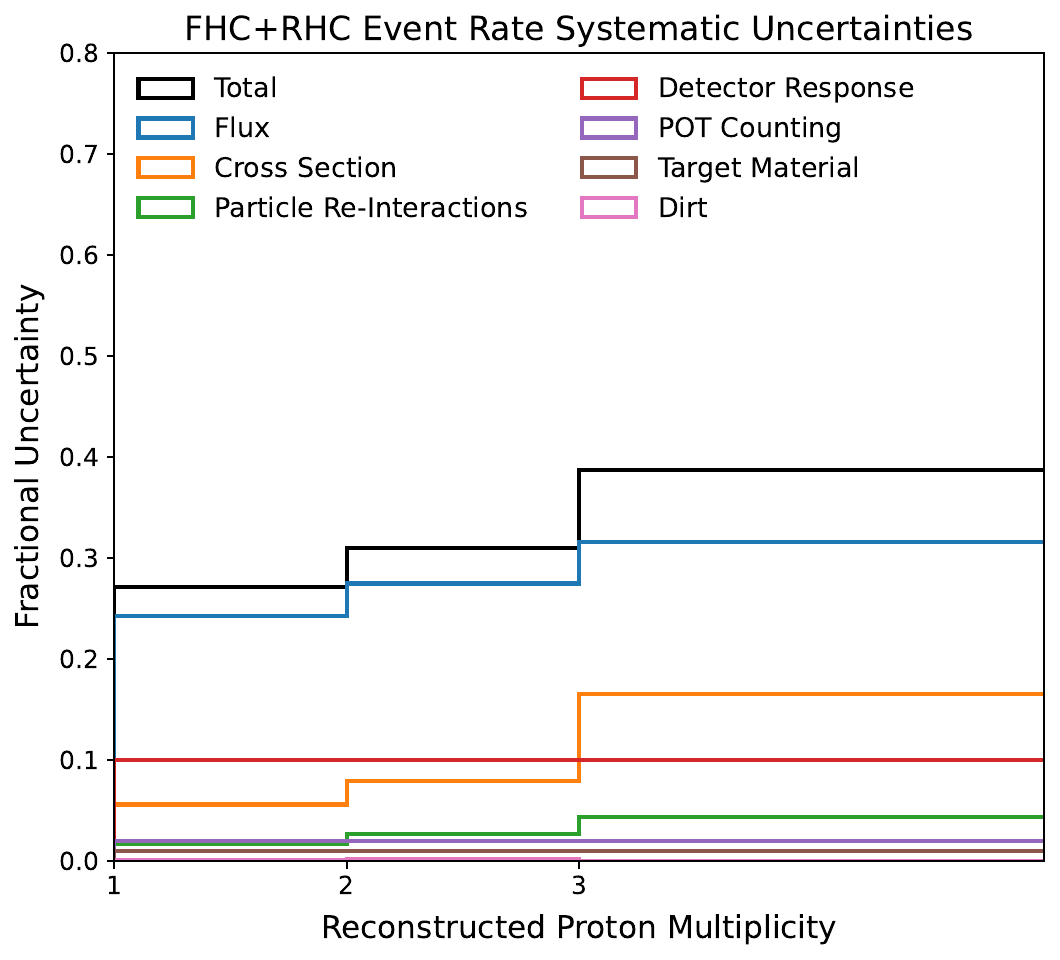}};
  \node at (3.75,3.1) {\textbf{(a)}};
\end{tikzpicture}
&
\begin{tikzpicture}
  \node[inner sep=0] {
    \includegraphics[width=0.48\textwidth]{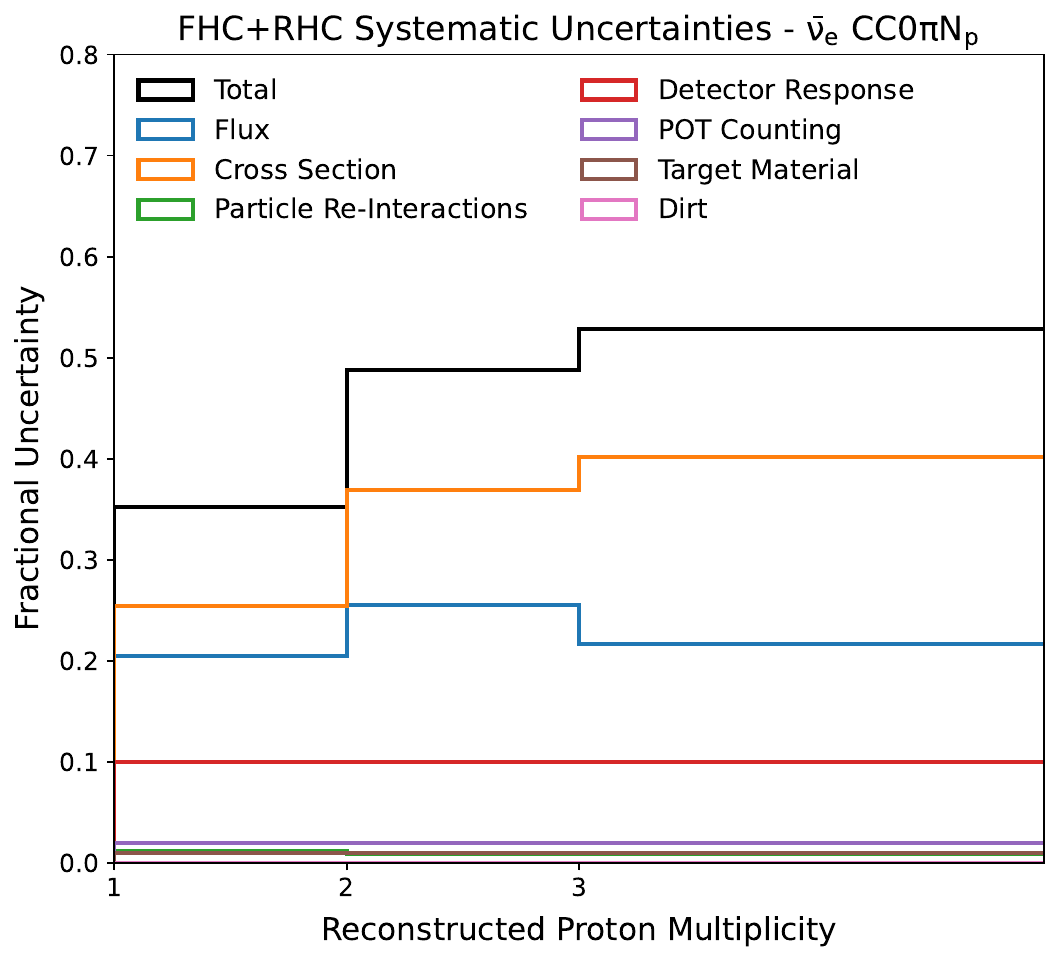}};
  \node at (3.75,3.1) {\textbf{(b)}};
\end{tikzpicture}
\end{tabular}

\caption{Fractional systematic uncertainties after subtracting background predicted by the tuned \code{GENIE v3.0.6 G18\_10a\_02\_11a} model (a) and on background $\bar{\nu}_{e}$ CC$0\pi Np$ events only (b) as a function of reconstructed proton multiplicity.}
\label{fig:proton_multiplicity_unc}
\end{figure}


\begin{figure}[h]
\centering

\includegraphics[width=.7\textwidth]{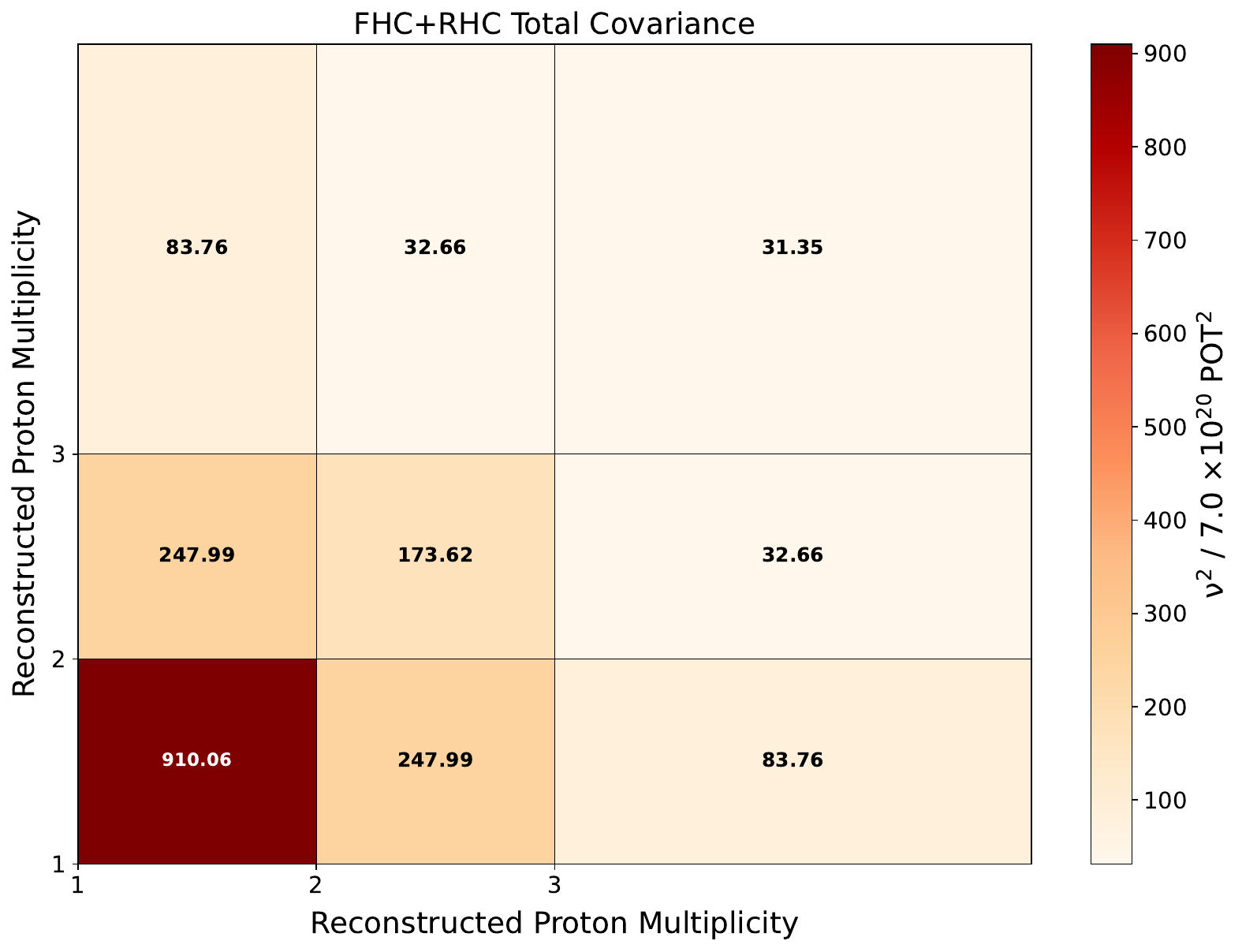}

\caption{{Total covariance (statistical + systematic) matrix for the FHC+RHC background-subtracted event rate as a function of reconstructed proton multiplicity.}}
\end{figure}


\begin{figure}[h]
\centering
\begin{tabular}{cc}
\begin{tikzpicture}
  \node[inner sep=0] {
    \includegraphics[width=0.48\textwidth]{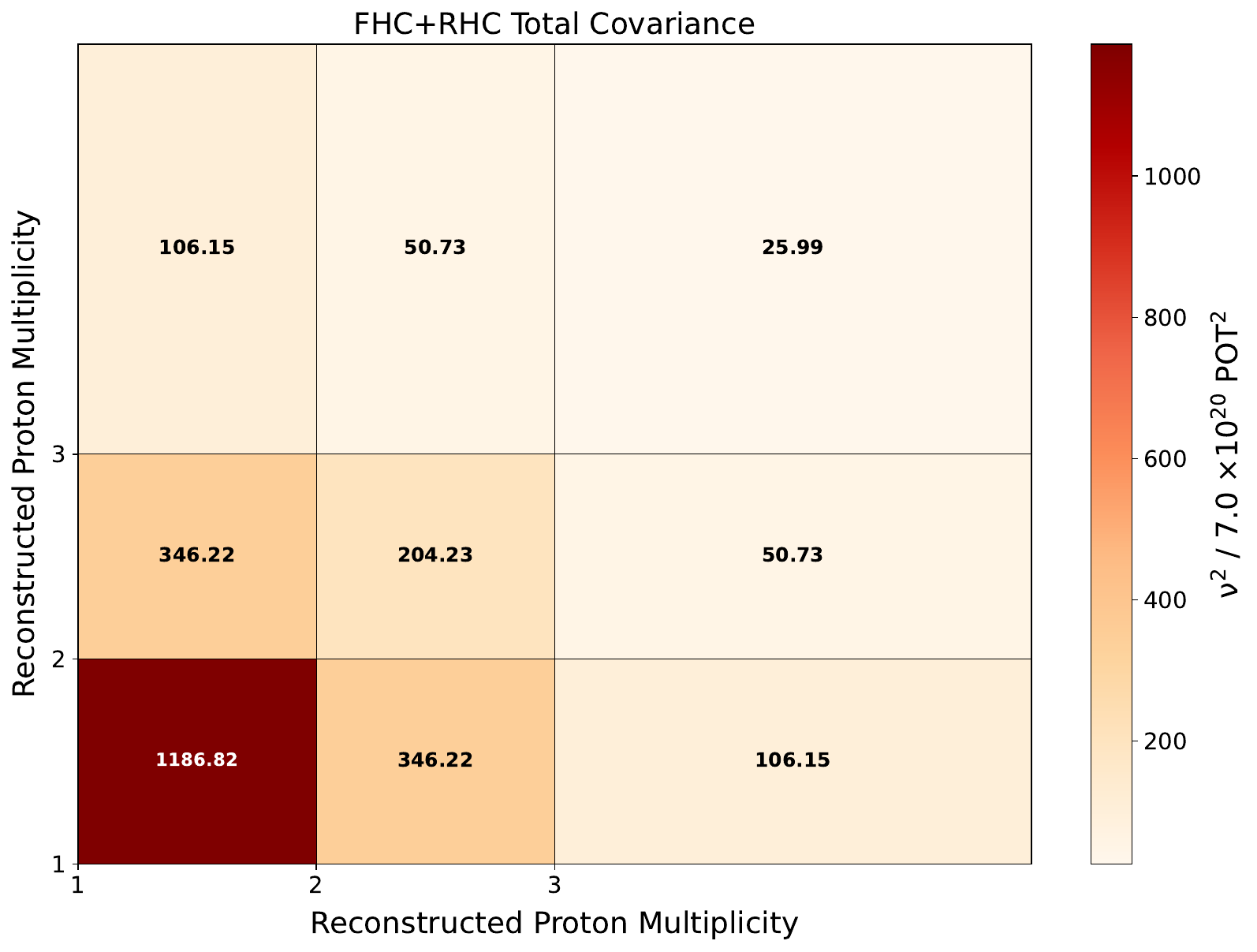}};
  \node at (2.35,2.6) {\textbf{(a)}};
\end{tikzpicture}
&
\begin{tikzpicture}
  \node[inner sep=0] {
    \includegraphics[width=0.48\textwidth]{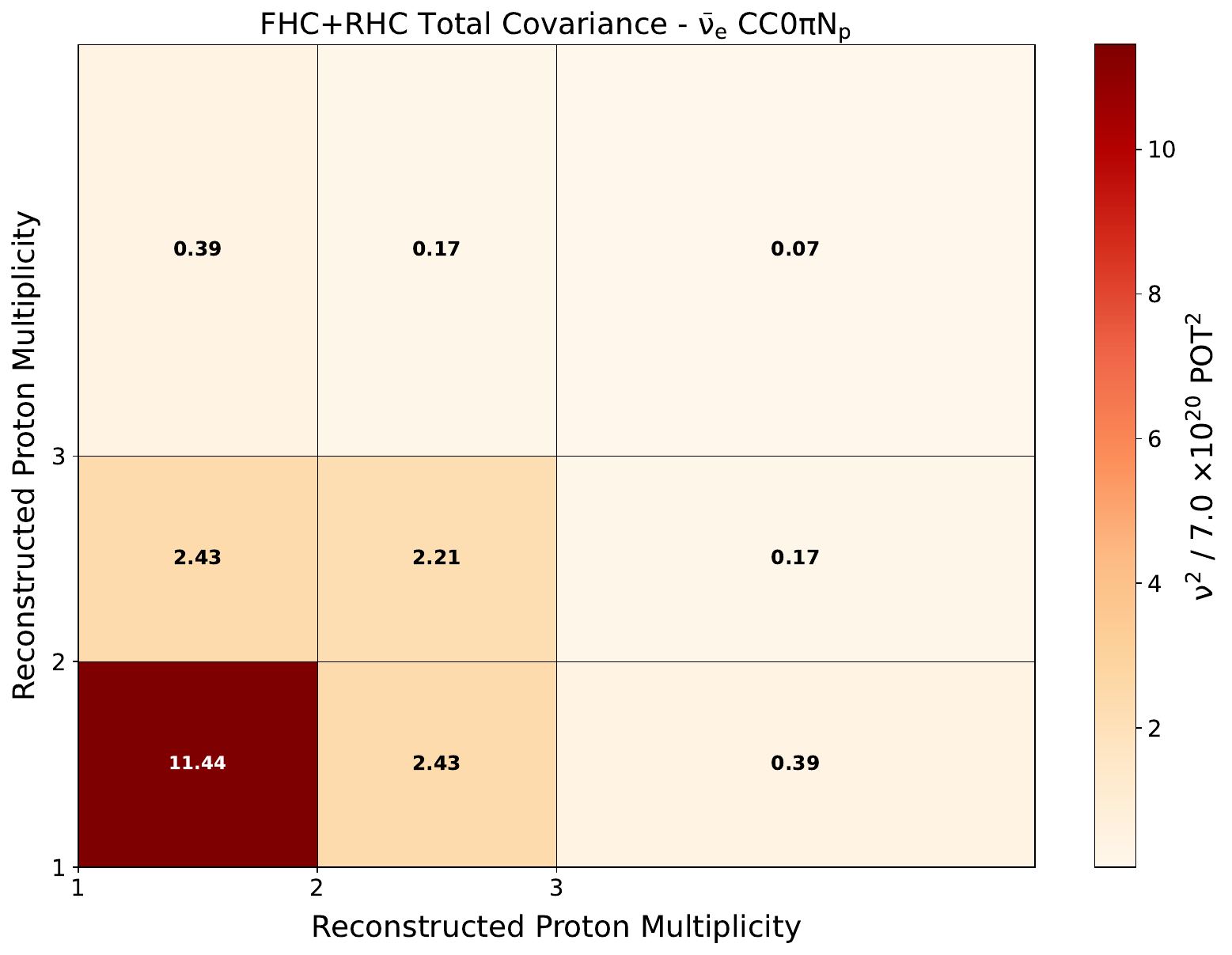}};
  \node at (2.55,2.65) {\textbf{(b)}};
\end{tikzpicture}
\end{tabular}

\caption{Total covariance (statistical + systematic) matrix for the FHC+RHC selected event rate (a) and $\bar{\nu}_{e}$ CC$0\pi Np$ background events only (b) as a function of reconstructed proton multiplicity.}
\end{figure}


\begin{figure}[h]
\centering

\includegraphics[width=.7\textwidth]{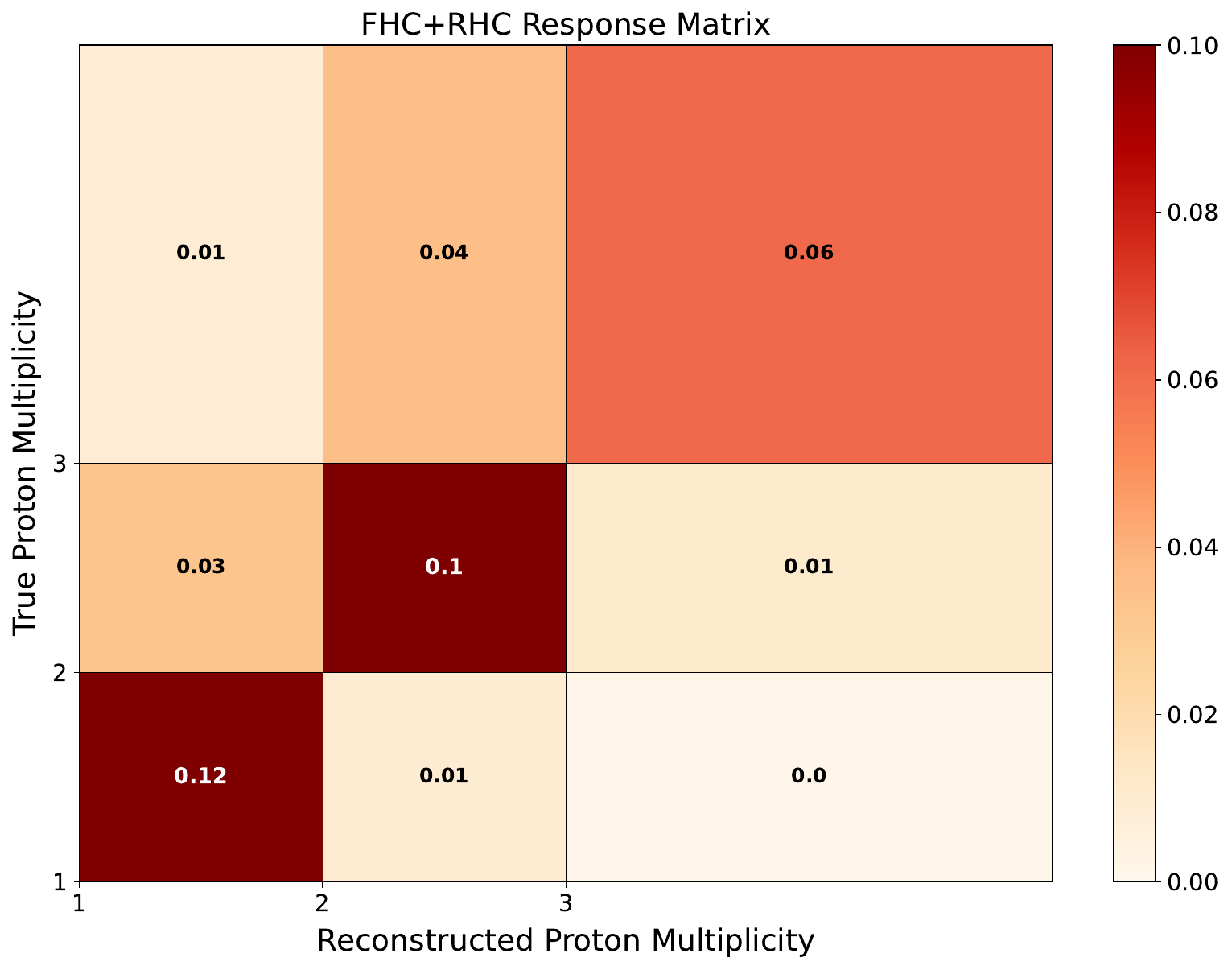}

\caption{{The response matrix, constructed using the \code{GENIE v3.0.6 G18\_10a\_02\_11a} selected signal prediction, as a function of true and reconstructed proton multiplicity. Row $j$ of the matrix is normalized such that $\sum_i r_{ij} = \epsilon_j$, where $i$ is the column and $\epsilon_j$ is the estimated selection efficiency of true bin $j$.}}
\end{figure}


\clearpage

\begin{table}
\caption{The background-subtracted event counts as a function of reconstructed proton multiplicity.}
\begin{ruledtabular}
\begin{tabular}{llll}
\textbf{Proton multiplicity} & \textbf{1} & \textbf{2} & \textbf{3+}  \\
\hline
Background-subtracted event count & 103.82 &  42.93 & 8.60 \\
\end{tabular}
\end{ruledtabular}

\end{table}


\normalsize

\clearpage 

\section{Wiener-SVD Smoothing}
\label{smoothing}

The Wiener-SVD unfolding method takes advantage of a technique traditionally used in digital signal processing, which requires deconvolution of the field and electronics response from a measured signal to recover the true signal. For measured time-series distributions, this can be achieved by executing a Fourier transformation of the signal into the frequency domain, and then introducing a filter function (often, the Wiener filter) to suppress large fluctuations arising from noise. 

This approach is adopted in Wiener-SVD unfolding by constructing an additional smearing matrix $\bm{A_c}$ that is dependent on the Wiener filter function $W$ \cite{wsvd}. The generalized Wiener-SVD method involves a free parameter matrix $\bm{C}$ that modifies the basis of the frequency domain. That is, 

\begin{equation}
    \bar{M} = \bm{R} \cdot \bm{C}^{-1} \cdot \bm{C} \cdot \bar{s}
\end{equation}
\newline
\noindent where $\bar{M}$ is the expectation of the measured signal distribution based on $\bar{s}$, the expectation of the true signal distribution. $\bm{C}$ is generally chosen to be a matrix that computes the derivative between the adjacent bins based on the shape of the spectrum to be regularized. This causes the effective frequency domain to be formed based on the curvature of the spectrum. Higher-order choices of $\bm{C}$ correspond to computing second or higher derivatives, which increasingly emphasize curvature in the spectrum.

Using singular value decomposition, the effective response matrix can be broken down as

\begin{equation}
\label{eqn:c_wsvd}
    \bm{R} \cdot \bm{C}^{-1} = \bm{U} \cdot \bm{D} \cdot \bm{V}^T
\end{equation}
\newline
\noindent where $\bm{U}$ and $\bm{V}$ are orthogonal matrices satisfying $\bm{U}^T \: \bm{U} = \bm{U} \: \bm{U}^T = \bm{V}^T \: \bm{V} = \bm{V} \: \bm{V}^T = \bm{I}$ (the identity matrix) and $\bm{D}$ is a diagonal matrix with non-negative elements. The final unfolded event rate, $\hat{s}$, can be expressed as

\begin{equation}
    \hat{s} = \bm{A_c} \cdot (\bm{R}^T \bm{R})^{-1} \cdot \bm{R}^T \cdot M
\end{equation}
\newline
\noindent where $M$ is the real measured spectrum and $\bm{A_c} = \bm{C}^{-1} \cdot \bm{V} \cdot W \cdot \bm{V}^T \cdot \bm{C}$. 

By design, the $\chi^2$ between data and prediction is invariant under the choice of $\bm{C}$. Results shown in the main paper are presented using an applied smoothing of $\bm{C} = 1$. This choice avoids over-smoothing and retains some of the shape differences between data and simulation in the background-subtracted event rates. Here, the shorthand $\bm{C} = 1$ refers to a first derivative matrix, and likewise for other orders.
\clearpage
\section{Unfolded Results}

The material in this section provides a comparison between the regularized and unregularized generator predictions as a function of electron energy, visible energy, and $\cos{\theta_{ep}}$.


\begin{figure}[h]
\centering

\begin{tikzpicture}
  \draw (0, 0) node[inner sep=0] {
    \includegraphics[width=0.49\textwidth]{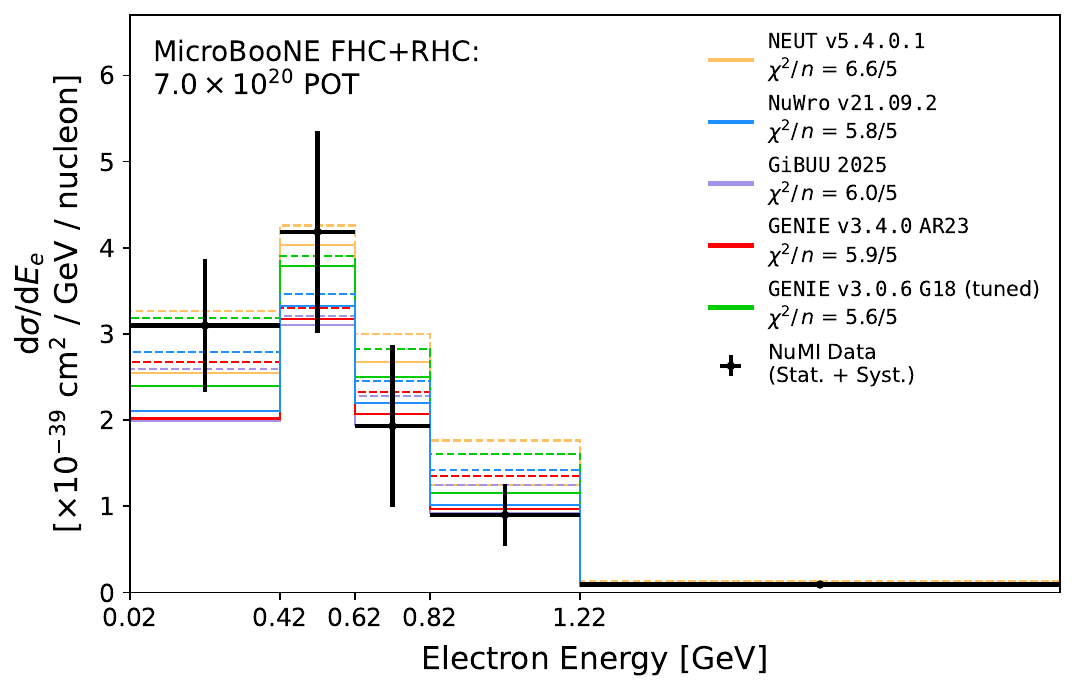}};
  \draw (3.9, -1.6) node {\textbf{(a)}};
\end{tikzpicture}
\hfill
\begin{tikzpicture}
  \draw (0, 0) node[inner sep=0] {
    \includegraphics[width=0.49\textwidth]{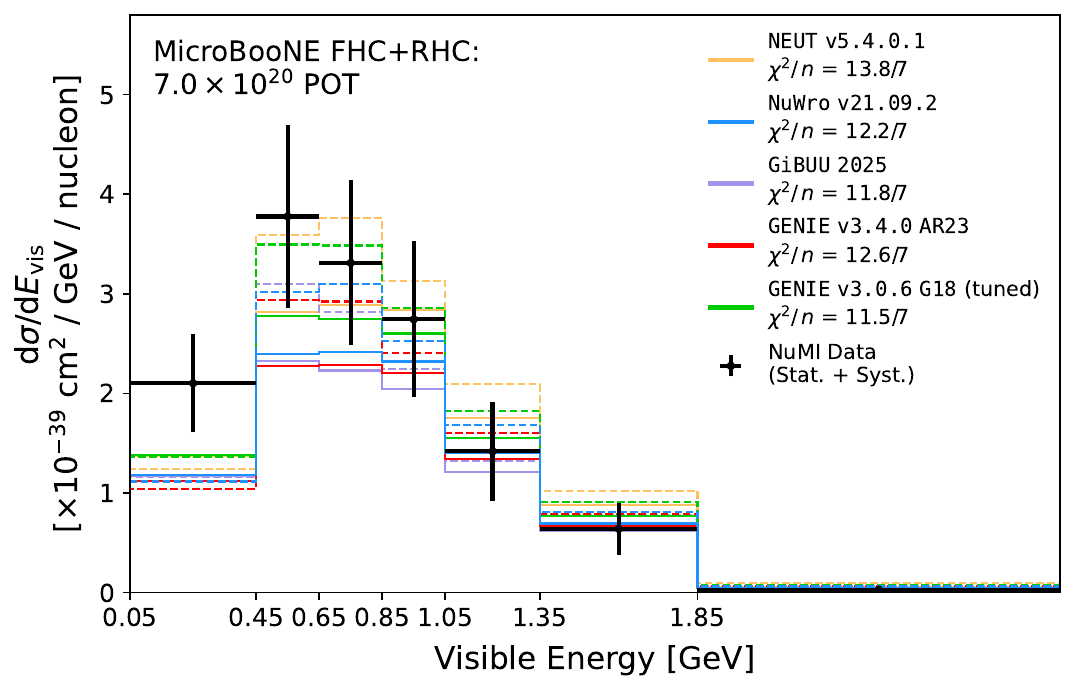}};
  \draw (3.9, -1.6) node {\textbf{(b)}};
\end{tikzpicture}

\vspace{1.2em}

\begin{tikzpicture}
  \draw (0, 0) node[inner sep=0] {
    \includegraphics[width=0.49\textwidth]{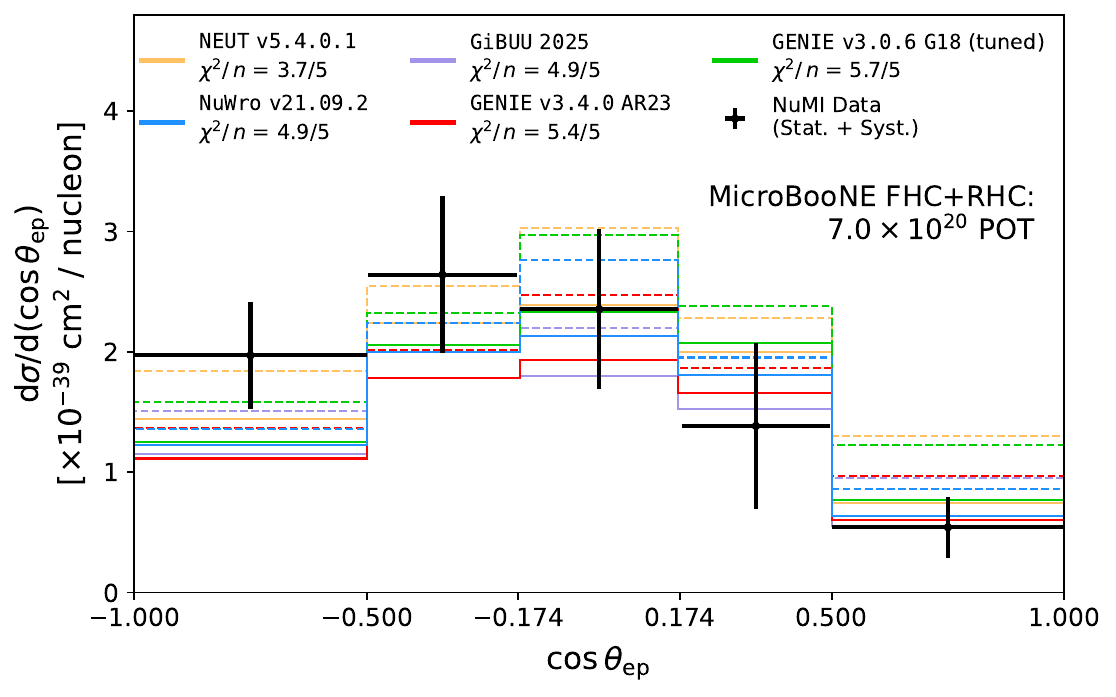}};
  \draw (-2.9, -1.6) node {\textbf{(c)}};
\end{tikzpicture}

\caption{{The unfolded NuMI data result (black) as a function of electron energy (a), visible energy (b), and $\cos{\theta_{ep}}$ (c), displayed with unfolded statistical and systematic uncertainties. Comparisons are made to regularized predictions from \code{NEUT v5.4.0.1} (solid orange), \code{NuWro v.21.09.2} (solid blue), \code{GiBUU 2025} (solid purple), \code{GENIE v03.4.2 AR23} (solid red), and \code{GENIE v03.0.6 G18} tuned (solid green). The unregularized generator predictions are shown by dashed lines.}}
\end{figure}

\bibliography{references_supplemental}